\documentclass{aastex63}

\received{5 August 2021}
\revised{15 October 2021}
\accepted{5 November 2021}

\submitjournal{ApJ}

\shorttitle{Pebbles \& Planetesimals to Planets \& Dust}
\shortauthors{Najita, Kenyon \& Bromley}

\graphicspath{{./}{figures/}}

\usepackage{epsf,amssymb,amsmath}
\usepackage{wrapfig}

\def\nbody{$n$-body}

\def\deg{\ifmmode {^\circ}\else {$^\circ$}\fi}
\def\degree{\ifmmode {^\circ}\else {$^\circ$}\fi}
\def\mum{\ifmmode {\rm \,\mu {\rm m}}\else $\rm \,\mu {\rm m}$\fi}
\def\arcsec{\ifmmode ^{\prime \prime}\else $^{\prime \prime}$\fi}
\def\secpoint{\mbox{$''\mskip-7.6mu.\,$}}

\def\inch{\ifmmode ^{\prime \prime}\else $^{\prime \prime}$\fi}
\def\gs{\ifmmode {{\rm g~s^{-1}}}\else ${\rm g~s^{-1}}$\fi}
\def\msunyr{\ifmmode {M_{\odot}~{\rm yr^{-1}}}\else $M_{\odot}~{\rm yr^{-1}}$\fi}
\def\msun{\ifmmode {M_{\odot}}\else $M_{\odot}$\fi}
\def\rsun{\ifmmode {R_{\odot}}\else $R_{\odot}$\fi}
\def\lsun{\ifmmode {L_{\odot}}\else $L_{\odot}$\fi}
\def\mstar{\ifmmode {M_{\star}}\else $M_{\star}$\fi}
\def\rstar{\ifmmode {R_{\star}}\else $R_{\star}$\fi}
\def\tstar{\ifmmode {T_{\star}}\else $T_{\star}$\fi}
\def\lstar{\ifmmode {L_{\star}}\else $L_{\star}$\fi}
\def\mwd{\ifmmode {M_{wd}}\else $M_{wd}$\fi}
\def\rwd{\ifmmode {R_{wd}}\else $R_{wd}$\fi}
\def\twd{\ifmmode {T_{wd}}\else $T_{wd}$\fi}
\def\lwd{\ifmmode {L_{wd}}\else $L_{wd}$\fi}
\def\md{\ifmmode {M_d}\else $M_d$\fi}
\def\ld{\ifmmode {L_d}\else $L_d$\fi}
\def\ad{\ifmmode A_d\else $A_d$\fi}
\def\ldlwd{\ifmmode L_d / L_{wd}\else $L_d / L_{wd}$\fi}
\def\ldlstar{\ifmmode L_d / L_\star\else $L_d / L_{\star}$\fi}
\def\rearth{\ifmmode {\rm R_{\oplus}}\else $\rm R_{\oplus}$\fi}
\def\mearth{\ifmmode {\rm M_{\oplus}}\else $\rm M_{\oplus}$\fi}
\def\qc{\ifmmode Q_c\else $Q_c$\fi}
\def\qdstar{\ifmmode Q_D^\star\else $Q_D^\star$\fi}
\def\rt{\ifmmode r_t\else $r_t$\fi}
\def\vc{\ifmmode v_c\else $v_c$\fi}
\def\vsqd{\ifmmode v^2 / Q_D^\star\else $v^2 / Q_D^\star$\fi}
\def\kms{\ifmmode {\rm km~s^{-1}}\else $\rm km~s^{-1}$\fi}
\def\ms{\ifmmode {\rm m~s^{-1}}\else $\rm m~s^{-1}$\fi}
\def\vrel{\ifmmode v_{rel}\else $v_{rel}$\fi}
\def\mdot{\ifmmode \dot{M}\else $\dot{M}$\fi}
\def\mdotz{\ifmmode \dot{M}_0\else $\dot{M}_0$\fi}
\def\mesc{\ifmmode m_{esc}\else $m_{esc}$\fi}
\def\rmin{\ifmmode r_{min}\else $r_{min}$\fi}
\def\rmax{\ifmmode r_{max}\else $r_{max}$\fi}
\def\xmax{\ifmmode x_{max}\else $x_{max}$\fi}
\def\mmin{\ifmmode m_{min}\else $m_{min}$\fi}
\def\mmax{\ifmmode m_{max}\else $m_{max}$\fi}
\def\rmind{\ifmmode r_{min,d}\else $r_{min,d}$\fi}
\def\rmaxd{\ifmmode r_{max,d}\else $r_{max,d}$\fi}
\def\mmaxd{\ifmmode m_{max,d}\else $m_{max,d}$\fi}
\def\vrad{\ifmmode v_{rad}\else $v_{rad}$\fi}
\def\qz{\ifmmode q_{0}\else $q_{0}$\fi}
\def\qi{\ifmmode q_{i}\else $q_{i}$\fi}
\def\ql{\ifmmode q_{l}\else $q_{l}$\fi}
\def\qs{\ifmmode q_{s}\else $q_{s}$\fi}
\def\vhill{\ifmmode v_H\else $r_H$\fi}
\def\rhill{\ifmmode r_H\else $r_H$\fi}
\def\Rhill{\ifmmode R_H\else $R_H$\fi}
\def\rbrk{\ifmmode r_{brk}\else $r_{brk}$\fi}
\def\rdamp{\ifmmode r_{damp}\else $r_{damp}$\fi}
\def\rin{\ifmmode r_{in}\else $r_{in}$\fi}
\def\rout{\ifmmode r_{out}\else $r_{out}$\fi}
\def\tin{\ifmmode t_{in}\else $t_{in}$\fi}
\def\tout{\ifmmode t_{out}\else $t_{out}$\fi}
\def\ain{\ifmmode a_{in}\else $a_{in}$\fi}
\def\aout{\ifmmode a_{out}\else $a_{out}$\fi}
\def\r0{\ifmmode r_{0}\else $r_{0}$\fi}
\def\R0{\ifmmode R_{0}\else $R_{0}$\fi}
\def\m0{\ifmmode m_{0}\else $m_{0}$\fi}
\def\mone{\ifmmode m_{1}\else $m_{1}$\fi}
\def\mtwo{\ifmmode m_{2}\else $m_{2}$\fi}
\def\atwo{\ifmmode a_{2}\else $a_{2}$\fi}
\def\etwo{\ifmmode e_{2}\else $e_{2}$\fi}
\def\mf{\ifmmode m_{f}\else $m_{f}$\fi}
\def\af{\ifmmode a_{f}\else $a_{f}$\fi}
\def\ef{\ifmmode e_{f}\else $e_{f}$\fi}
\def\M0{\ifmmode M_{0}\else $M_{0}$\fi}
\def\amax{\ifmmode a_{max}\else $a_{max}$\fi}
\def\a0{\ifmmode a_{0}\else $a_{0}$\fi}
\def\e0{\ifmmode e_{0}\else $e_{0}$\fi}
\def\v0{\ifmmode v_{0}\else $v_{0}$\fi}
\def\xm{\ifmmode x_{m}\else $x_{m}$\fi}
\def\sigz{\ifmmode \Sigma_0\else $\Sigma_0$\fi}
\def\ergg{\ifmmode {\rm erg~g^{-1}}\else ${\rm erg~g^{-1}}$\fi}
\def\ergs{\ifmmode {\rm erg~s^{-1}}\else ${\rm erg~s^{-1}}$\fi}
\def\gyr{\ifmmode {\rm g~yr^{-1}}\else ${\rm g~yr^{-1}}$\fi}
\def\cms{\ifmmode {\rm cm~s^{-1}}\else ${\rm cm~s^{-1}}$\fi}
\def\gcms{\ifmmode {\rm g~cm^{-2}}\else $\rm g~cm^{-2}$\fi}
\def\gcmc{\ifmmode {\rm g~cm^{-3}}\else $\rm g~cm^{-3}$\fi}
\def\atil{\ifmmode {\tilde{a}}\else $\tilde{a}$\fi}
\def\ttil{\ifmmode {\tilde{t}}\else $\tilde{t}$\fi}
\def\sqrttt{\ifmmode {\tilde{t}^{1/2}}\else $\tilde{t}^{1/2}$\fi}

\def\orch{{\it Orchestra}}

\def\mp{\ifmmode M_P\else $M_P$\fi}
\def\mc{\ifmmode m_C\else $m_C$\fi}
\def\mh{\ifmmode m_H\else $m_H$\fi}
\def\mk{\ifmmode m_K\else $m_K$\fi}
\def\ms{\ifmmode m_S\else $m_S$\fi}
\def\mn{\ifmmode m_N\else $m_N$\fi}
\def\rp{\ifmmode r_P\else $r_P$\fi}
\def\rc{\ifmmode r_C\else $r_C$\fi}
\def\apc{\ifmmode a_{PC}\else $a_{PC}$\fi}
\def\mpc{\ifmmode m_{PC}\else $m_{PC}$\fi}
\def\epc{\ifmmode e_{PC}\else $e_{PC}$\fi}

\begin{document}

\title{From Pebbles and Planetesimals to Planets and Dust: the Protoplanetary Disk--Debris Disk Connection}

\author{Joan R. Najita}
\affil{NSF’s NOIRLab, 
950 N. Cherry Avenue, Tucson, AZ 85719, USA}

\author[0000-0003-0214-609X]{Scott J. Kenyon}
\affil{Smithsonian Astrophysical Observatory,
60 Garden Street,
Cambridge, MA 02138, USA}

\author[0000-0001-7558-343X]{Benjamin C. Bromley}
\affil{Department of Physics \& Astronomy,
University of Utah, 201 JFB,
Salt Lake City, DC 20006, USA}

\begin{abstract}
The similar orbital distances and 
detection rates
of debris disks
and the prominent rings observed in protoplanetary disks suggest a
potential connection between these structures.  We explore this
connection with new calculations that follow the evolution of rings
of pebbles and planetesimals as they grow into planets and generate
dusty debris.  Depending on the initial solid mass and planetesimal
formation efficiency, the calculations predict diverse outcomes for
the resulting planet masses and accompanying debris signature.  When
compared with debris disk incidence rates as a function of luminosity
and time, the model results indicate that the known population of
bright cold debris disks can be explained by rings of solids with
the (high) initial masses inferred for protoplanetary disk rings
and modest planetesimal formation efficiencies that are consistent
with current theories of planetesimal formation.  These results
support the possibility that large protoplanetary disk rings evolve
into the known cold debris disks.  The inferred strong evolutionary
connection between protoplanetary disks with large rings and mature
stars with cold debris disks implies that the remaining majority
population of low-mass stars with compact protoplanetary disks leave
behind only modest masses of residual solids at large radii and
evolve primarily into mature stars without detectable debris beyond
30 au.  The approach outlined here illustrates how combining
observations with detailed evolutionary models of solids strongly
constrains the global evolution of disk solids and underlying
physical parameters such as the efficiency of planetesimal formation
and the possible existence of invisible reservoirs of solids in
protoplanetary disks.

\end{abstract}


\section{Introduction} \label{sec: intro}

Stars form surrounded by disks, the material from which planets form. Over the first Myr (or so) in the life of a disk, its solids are aggregated into planetesimals and protoplanets, and eventually into planets. 
The first step in this transformation may be hastened as disk solids collect in inhomogeneities---ring-like pressure bumps and other features---triggering processes such as the streaming instability, which create planetesimals.  
The planetesimals eventually grow into planets, which shape the disk gas and dust, producing rings, gaps, inner holes, and other structures.  
The concentration of solids in rings may spur further planetesimal (and planet) formation. After the disk gas  dissipates, residual planetesimals ``left behind'' in the planet formation process may eventually reveal themselves, as they collide and produce disks of debris 
that glow in the reprocessed light from the central star. 

Some of the most dramatic evidence in support of this picture comes from the millimeter continuum morphologies of Class II (protoplanetary) disks, many of which show the substructure that forming planets are expected to induce. 
When imaged at millimeter wavelengths 
and in scattered light, 
large disks (radii $\gtrsim$ 25--30~au) show rings, gaps, central cavities, and other features at distances of 20--200 au from the star 
\citep[e.g.,][and references therein]{avenhaus2018,huang2018a,
long2019,cieza2019}. 
While central cavities have been imaged for more than a decade 
\citep[e.g.,][]{andrews2010,andrews2015}, the realization that 
narrow rings and gaps are common features of disks at large radii is a discovery of the ALMA era \citep{alma2015}.
Whereas disk central cavities are thought to be created by a high mass giant planet that orbits within the cavity, the narrow rings and gaps can be created by lower mass (approximately Neptune-mass) planets 
\citep[e.g.,][]{bae2018,lodato2019}.

Emission associated with orbiting gas giant planets has also been detected in disks with central cavities through 
direct imaging techniques \citep{sallum2015,keppler2018,haffert2019,
zurlo2021} and spectroastrometry \citep[e.g.,][]{brittain2019}. 
However, the lower mass ice giants thought to be responsible for narrow rings and gaps at large radial distances ($\gtrsim$ 25--30 au) remain beyond our ability to detect directly. 
Because gas and ice giants at such large orbital radii have no counterpart in the Solar System, these protoplanetary disk structures appear to point to an even greater diversity of planet formation outcomes than previously contemplated. 
 
Another valuable clue in support of this picture,  that planetesimals---the hypothesized building blocks of planets---commonly form in protoplanetary disks, comes from debris disks,  the dusty debris that is found around some post-T Tauri and main sequence stars.  
Best explained as the result of collisions between large parent bodies, 
and identified by their infrared and millimeter excesses, debris disks accompany stars over wide range of ages. At ages 10 Myr to beyond 1 Gyr, 
approximately
20\% to 25\% of FGK stars have detected cold excesses at $\sim 100$~\mum\ \citep[e.g.,][]{carp2009a,bryden2009,eiroa2013,sibthorpe2018}, corresponding to debris at distances of $\gtrsim 40$ au. When imaged at high angular resolution, the debris also commonly shows substructure such as rings and gaps \citep[e.g.,][]{marino2018,hughes2018,marino2020,
nederlander2021}. 

Here we explore the possible evolutionary connection between the rings and gaps observed in protoplanetary disks and those in debris disks. 
Given that both protoplanetary disks and debris disks show structured continuum emission in the form of rings and gaps over similar radial distances (20--200 au) around approximately solar-mass stars, it seems plausible that large protoplanetary disks evolve into the known cold debris disk population. 
We complement related work on this topic \citep[e.g.,][]{michel2021} using new models of the evolution of rings of solids on Myr to Gyr timescales. 
The new ring models contrast with earlier generations of models that explored extended disks of solids as the origin of debris disk emission
\citep[e.g.,][]{kb2008,kb2010}.

Using the new models, we examine how the efficiency of planetesimal formation affects not only the outcome of planet formation, but also its associated debris production. 
By comparing the model results to observations of debris disk populations,
we explore the disk conditions (total initial mass of solids, planetesimal formation efficiency) that can reproduce the properties of debris disks (their incidence rate and luminosity) as a function of stellar age.
We also use the comparison to explore questions such as whether cold debris disks feature in the evolutionary histories of all disks or only a special subset.  

In \S\ref{sec:diskobs}, we review observations that link protostellar disks
to debris disks. After a brief review of planetesimal formation in 
\S\ref{sec:ringmodelback}, we set up a suite of numerical calculations
in rings with populations of small and large solids
(\S\ref{sec:ringmodelinit}) and describe the formation of 
planets (\S\ref{sec: ringmodelgrowth}), 
debris disks (\S\ref{sec: ringmodeldebris}),
and gaps (\S\ref{sec: ringmodelgaps}). 
After discussing the calculations in the context of observations and
other models (\S\ref{sec:discussion}), we conclude with a brief summary
(\S\ref{sec: summary}).

\section{Properties of Protoplanetary Disks and Debris Disks} \label{sec:diskobs}

\subsection{Protoplanetary Disks}

The millimeter continuum emission from large, bright protoplanetary disks 
(continuum sizes $\gtrsim 50$ au) is often highly structured. In deep, 
high angular resolution ALMA images 
of the brightest disks in 
nearby star-forming regions, the continuum emission typically arises from 
multiple concentric rings with radii of $\sim 10-150$ au, widths of a few
au to tens of au, and dust masses of 10--70~\mearth\ \citep{alma2015,
huang2018a,long2018,dullemond2018}.
Within larger samples of protoplanetary disks that span a wider range in disk properties, including lower
millimeter continuum flux, large resolved structures are less common. 
Of 147 disks in Ophiuchus studied by the ODISEA survey, the great majority have millimeter continuum 
emission restricted to radial distances
$< 15$ au. Only $\sim 20$\% of the sources show continuum emission larger than 20 au in radius
\citep[see Fig.\ 12 of][]{cieza2019}. In a sample of 32
disks in Taurus--Auriga that cover a broad range of continuum brightness, only a modest fraction of disks show large rings
\citep{long2019}. Nine of the 27  sources studied with stellar masses $< 1.6\msun$
have rings with effective millimeter continuum sizes $> 40$ au; the 
remainder have more compact emission. 

To 
extrapolate the results of Long et al.\ (2018, 2019) and
estimate the fraction of Taurus sources with large rings, we 
consider what is known about the entire Taurus disk population 
\citep[e.g.,][]{luhman2010,akeson2019}.
We follow the rough selection criteria of \citet{long2019} and 
select the Taurus Class II sources with spectral types M3 or 
earlier and no companions at angular separations of 
0\secpoint14--1\arcsec. 
We also impose an upper stellar mass limit 
of $1.6\msun$ to mirror the upper mass stellar limit of the debris disk samples. With these criteria,
there is a parent 
sample of 77 sources. The fraction with large rings in this sample is at
least 9/77 = 12\%. 

The actual fraction is likely to be larger. 
Beyond the 32 disks studied by \citet{long2019}, 
17 additional Taurus 
T Tauri stars
were known to have ALMA observations at the time the Long et al.\ paper was written.
Several of these sources have rings on large scales, e.g.,  
AA~Tau \citep{loomis2017}, LkCa~15 \citep{facchini2020}, and
DM~Tau \citep{hashimoto2021}. Others show smooth emission without rings, e.g., CW~Tau, CY~Tau, and 
DG~Tau \citep{simon2017,bacciotti2018}. At least one source is compact without large rings 
\citep[CX~Tau;][]{facchini2019}, and 5 other sources have binary companions
within $1\arcsec$ 
and would be excluded from our sample. 
The morphology of the remaining 5 sources is unknown (unpublished). 

As a result, of the 17 additional sources, 3 definitely have large rings; the remaining 5 sources of unknown morphology may also have large rings.
If all of 8 of these sources do in fact have large rings, the fraction of 
Taurus Class II sources with large rings 
could be as high as 17/77 or $\sim 22$\%. The true fraction of disks with large rings is larger if 
other (as yet unobserved) sources among the parent sample of 77 sources also have rings. 

At the same time, a robust estimate of the fraction of young stars 
with large rings requires that we also account for the lack of rings 
among the population of (diskless) Class III sources with similar ages, 
stellar masses, and companion properties as the Class II sources.
As a group, the Class III sources 
represent 25\% of all T Tauri stars in Taurus \citep[e.g.,][]{luhman2010,
luhman2018,esplin2019}. 
Accounting for these diskless sources reduces the fraction of young stars  
with large rings by a factor of $\sim 3/4$. 
In summary, the fraction of young stars in Taurus with large rings is plausibly 
$\sim 17$\%, 
although with significant uncertainty 
(J.\ Bae and A.\ Isella, private communication).
Further ALMA imaging of a larger sample is needed for a robust estimate.
Taking a similar approach, van der Marel \& Mulders 2021 estimated a fraction of $\sim 16$\% of structured disks in a sample study of almost 700 disks in nearby star forming regions.

\begin{figure}[t]
\begin{center}
\includegraphics[width=5.0in]{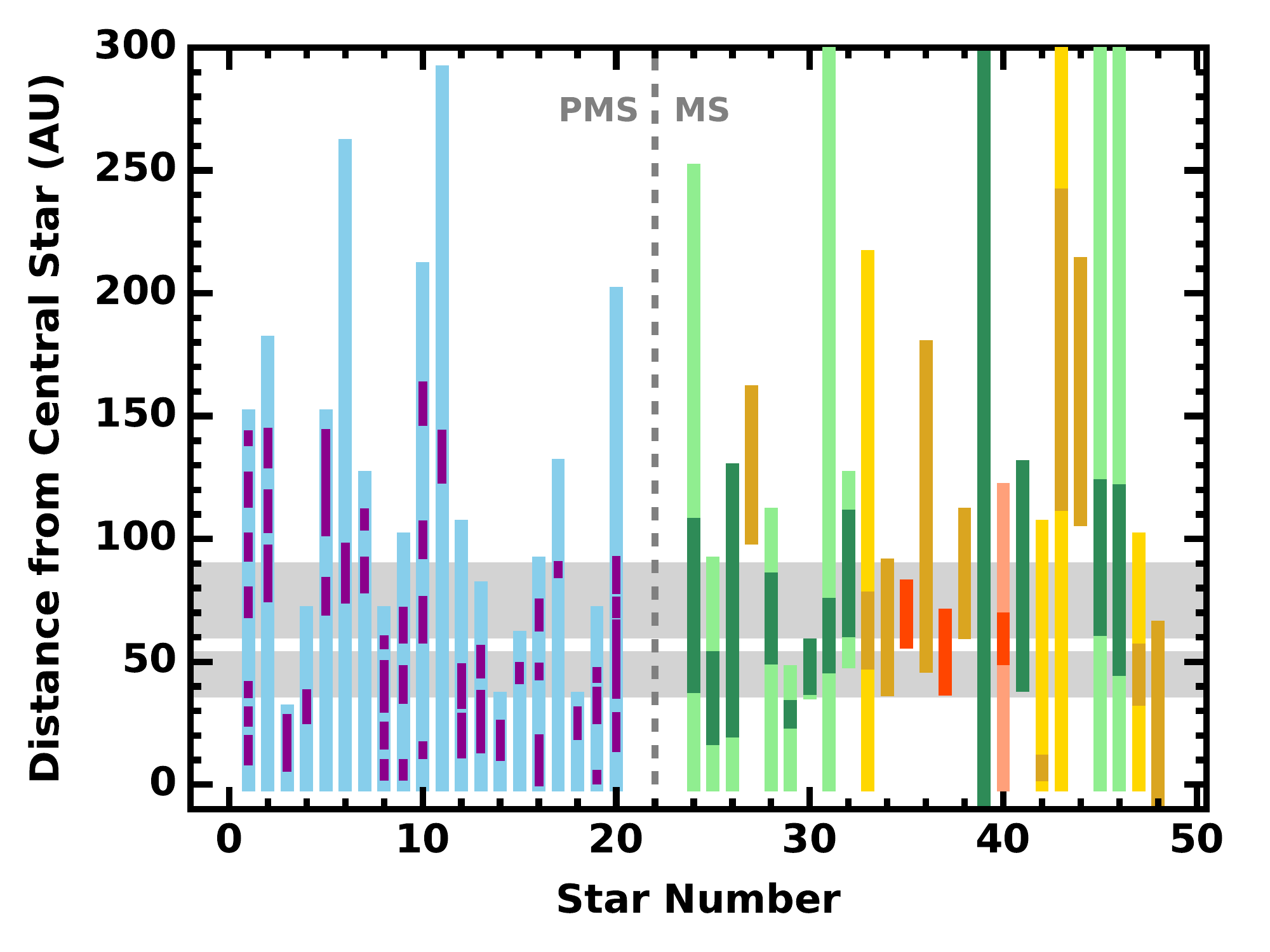}
\vskip -2ex
\caption{
\label{fig: extent}
Comparison of the observed positions of bright rings in the continuum emission from protoplanetary disks around T Tauri stars \citep[left side;][]{huang2018a} to bright rings within 
extended dusty debris disks surrounding FGK main sequence stars  \citep[right side;][]{hughes2018}. 
For each vertical bar, light (dark) regions indicate the extent of the
disk (rings). Disks around main sequence stars are ordered by age, from 12~Myr for star 24 (HD~146897)
to 8.2~Gyr for star 48 ($\tau$~Cet). Colors denote the spectral type---F (green), 
G (gold), or K (orange)---of the central star. The vertical grey dashed line separates pre-main-sequence (PMS) from main sequence (MS) stars. The horizontal grey bands represent the
two grids used in the numerical calculations described in section 3.
}
\end{center}
\end{figure}

Fig.~\ref{fig: extent} illustrates the range of sizes and widths of continuum emission rings observed in T Tauri disks (primarily from the DSHARP sample; Table 1 of Huang et al.\ 2018). In the left side of the diagram, the purple bars indicate the radial extent of bright rings; the blue bars indicate the rough radial extent of the dust continuum emission from the disk. The high angular resolution of the DSHARP observations (30--60 mas; equivalent to 5-8 au at the distance of the targets) probed structures at much smaller angular scales than was possible with the lower resolution, snapshot ALMA observations of Long et al.\ ($\sim 120$ mas or $\sim 16$ au).

\subsection{Debris Disks}

Spatially resolved images of debris disks also show rings  
in 
scattered light and thermal emission \citep[e.g.,][]{hughes2018}. 
Fig.~\ref{fig: extent} shows the radial extents of debris disks (light-colored bars) and their rings (dark-colored bars), as measured by direct imaging  
\citep{hughes2018}. 
As shown in the Figure, much of the emission from debris arises from within $a = 10$--150 au of the star, similar to the orbital distances from which the millimeter continuum arises from protoplanetary disks. 
The fractional widths of the rings are typically $\Delta a/a = 0.1 - 0.6,$ with $\Delta a$ typically ranging from a few au to 30 au, overlapping the range of widths of protoplanetary disk rings.

While fine substructure (i.e., narrow rings) is reported more commonly for protoplanetary disks than debris disks, 
only a few debris disks have been imaged with sufficient sensitivity and angular resolution to detect such substructure \citep[see discussion
in][]{marino2020,nederlander2021}. Of the 6 sources studied to date at high sensitivity and angular resolution, 4 show gaps in their continuum emission, suggesting that finer substructure may be common among large debris disks, 
as in protoplanetary disks
\footnote{The gaps are all located at $\sim 70$ au; these are HD~107146, HD~92945, HD~15115, HD~206893. The other two sources without reported gaps are beta Pic and AU Mic.}. Future ALMA imaging is needed to explore this possibility.

\begin{figure}[t]
\begin{center}
\includegraphics[width=5.0in]{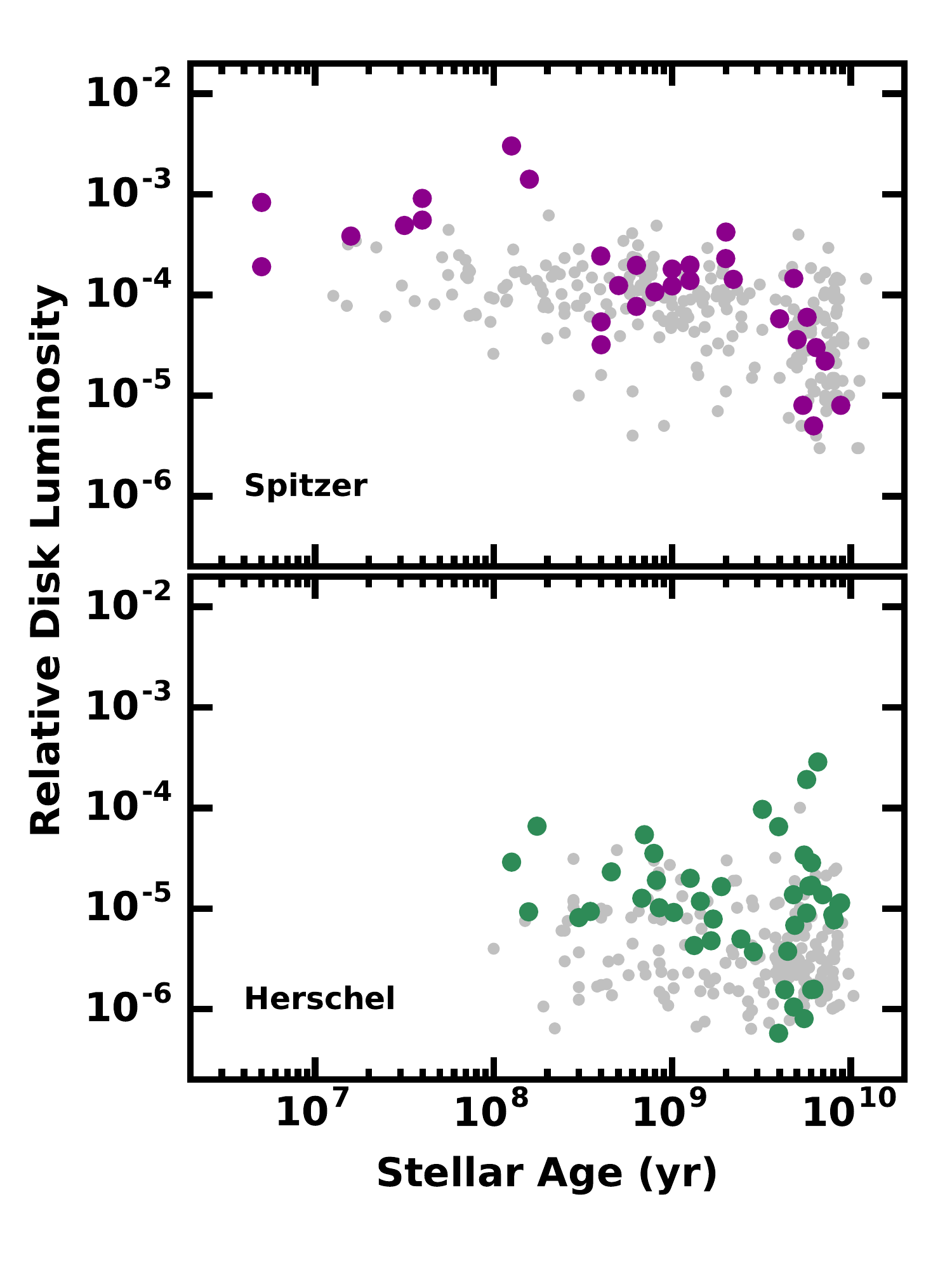}
\vskip -8ex
\caption{
\label{fig: excess}
Relative disk luminosity for debris disks from {\it Spitzer}
\citep[upper panel;][]{carp2009a,bryden2009} and {\it Herschel}
\citep[lower panel;][]{eiroa2013,sibthorpe2018} observations. 
Large filled circles indicate detections; small symbols indicate 
upper limits for all sources ({\it Herschel}) or those with 
$d \le$ 80~pc ({\it Spitzer}).
}
\end{center}
\end{figure}
   
Surveys of varying sensitivities to dust temperature and fractional luminosity have reported detection rates of debris disk emission.
As described in \citet{hillen2008}, the 
FEPS Survey used the 
{\it Spitzer Space Telescope} to carry out a census of cool dust surrounding 328 solar-type stars in the age range 3 Myr -- 3 Gyr located at distances of 10 pc -- 200 pc. 
Over the age range 30 Myr to 3 Gyr, $\sim 10$\% of solar-type stars show evidence for cold debris, with their rising spectral energy distributions toward 70~\mum\ indicating dust temperatures of $\sim 45$--85 K for dust in equilibrium with the stellar radiation field. 
The detected excesses are bright at young ages, with fractional excess luminosities of $L_d/L_* \sim 10^{-3}-10^{-4}$ at 30 Myr, declining with increasing stellar age to $L_d/L_* \sim 10^{-4}$ at 3 Gyr. 

While FEPS focused on the excess properties of FGK stars younger than 3 Gyr, other {\it Spitzer} surveys  
investigated the excess properties of older stars. \citet{bryden2009} reported the excess properties of planet-bearing stars with spectral types F5--K5 that are known from radial velocity studies to harbor one or more planets; the majority of the stars are 4--10 Gyr old where, as in the case of most of the FEPS sources, the ages are based on chromospheric activity and the calibration of \citet{mamajek2008}. The excess properties of the sample are statistically indistinguishable from those of a comparison sample of comparable nearby stars (i.e., similar in spectral type and age) without known planetary companions. 

To illustrate the limits placed by these  
{\it Spitzer} results on possible evolutionary paths for disk solids, 
the upper panel of Fig.~\ref{fig: excess}
shows detections (purple dots) and upper limits 
\citep[smaller gray dots;][]{carp2009a,bryden2009} as a function of stellar age.
Protoplanetary disk sources in the FEPS sample are excluded.
Only the FEPS sources within 80 pc are shown to highlight the value of the constraints placed by the upper limits on the nearby sample.
The full FEPS sample spans a large range in distance (out to $\sim 150$ pc) and upper limits on the distant sources are weak. 
For the few sources in common between the FEPS and Bryden et al.\ samples (HD38529 and HD150706), we adopted the excess properties reported by FEPS. A few sources with unusually high flux uncertainties and upper limits were removed from the Bryden et al.\ sample (HD 4203, HD 46375, HD168746, and HD330075).

To obtain the FEPS upper limits shown, we converted the 70 micron flux upper limits (typically $\sim 10$ mJy, 1-$\sigma$) to upper limits on $L_d/L_*$ assuming a typical ratio of $r= (L_d/L_*)/(F_d/F_*)= 10^{-5},$ a factor appropriate for the typical temperature of detected FIR excesses 
\citep[60 K;][]{hillen2008}; the conversion factor is insensitive to temperature in the range 40--80 K \citep[e.g., Figure 9 of][]{hillen2008}.
Following \citep{carp2009a} for both the FEPS sources discussed here and the 
{\it Herschel} sources discussed below, the plotted upper limits in 
Fig.~\ref{fig: excess} of  $L_d/L_* = r (F_d/F_* + 3\sigma),$ use the reported value of $F_d/F_*$ when its value is $> 0$ and 0 otherwise. The upper limits shown for the Bryden et al.\ survey are from their paper. 

Following the {\it Spitzer} studies, the DEBRIS and 
DUNES surveys searched for infrared excess emission at 
100~\mum\ and 160~\mum\ with the 
{\it Herschel Space Observatory} 
\citep{eiroa2013,sibthorpe2018}. Both programs observed 
several sources at 70~\mum; DUNES acquired additional 
data at 250--500~\mum. The surveys targeted 275 (DEBRIS) 
and 133 (DUNES) unique FGK stars with distances 
within
25~pc \citep[DEBRIS;][]{sibthorpe2018} and 
20--25~pc \citep[DUNES;][]{eiroa2013}. Stellar
activity ages range from 1~Myr to 11~Gyr (100~Myr to 
10~Gyr) for DEBRIS (DUNES), with median ages of $\sim$ 
3~Gyr. 

The DEBRIS (DUNES) survey detected excess emission from 
debris disks around 47 (31) stars for a nominal 
detection rate of 17\% (23\%). In DEBRIS, the detection 
rate is similar across the age bins 0.1--1~Gyr, 1--3~Gyr, and 3--10~Gyr. 
Corrected for incompleteness, the incidence rate for 
FGK stars is $\sim$ 28\%. For DUNES, the volume-limited
detection rate
is $\sim$ 20\% and is independent of
spectral type for FGK stars \citep[see
also][]{montesinos2016}. 

For the combined set of DEBRIS and DUNES detections, 
the median \ldlstar\ is roughly an order of magnitude 
smaller than the median dust luminosity of the
{\it Spitzer} detections. For both {\it Herschel} 
programs, blackbody dust temperatures have a broad 
range, 16--300~K for DEBRIS and 20--100~K for DUNES; 
the median dust temperatures are 48~K (DUNES) and 63~K (DEBRIS). 
Inferred radii for the dust are 
1--300~au with a median $\sim$ 20~au for DEBRIS and 
7--200~au with a median of $\sim$ 30~au for DUNES. 
Assuming realistic dust properties would place the 
emission at larger radii. 

The lower panel of Fig.~\ref{fig: excess} shows the {\it Herschel} 
detections (green dots) and upper limits (smaller gray dots) for the 
43 sources with FGK spectral types
(effective temperatures of 4000--7200 K) and ages as 
estimated from  stellar activity 
\citep[e.g.,][]{vican2012}. 
To translate flux detection upper limits to $L_d/L_*$, the reported
3-$\sigma$ upper limit on $F_d/F_*$ at 100~\mum\ was converted to
$L_d/L_*$ assuming a ratio $r = (L_d/L_*)/(F_d/F_*) \simeq 10^{-5.4},$ 
the value appropriate for an excess at 100~\mum\ that has a temperature
of $\sim 55$ K, the median temperature of detected excesses. The actual
adopted value of $r$ is appropriate for the stellar temperature of each
source.

\begin{figure}[t]
\begin{center}
\includegraphics[width=5.0in]{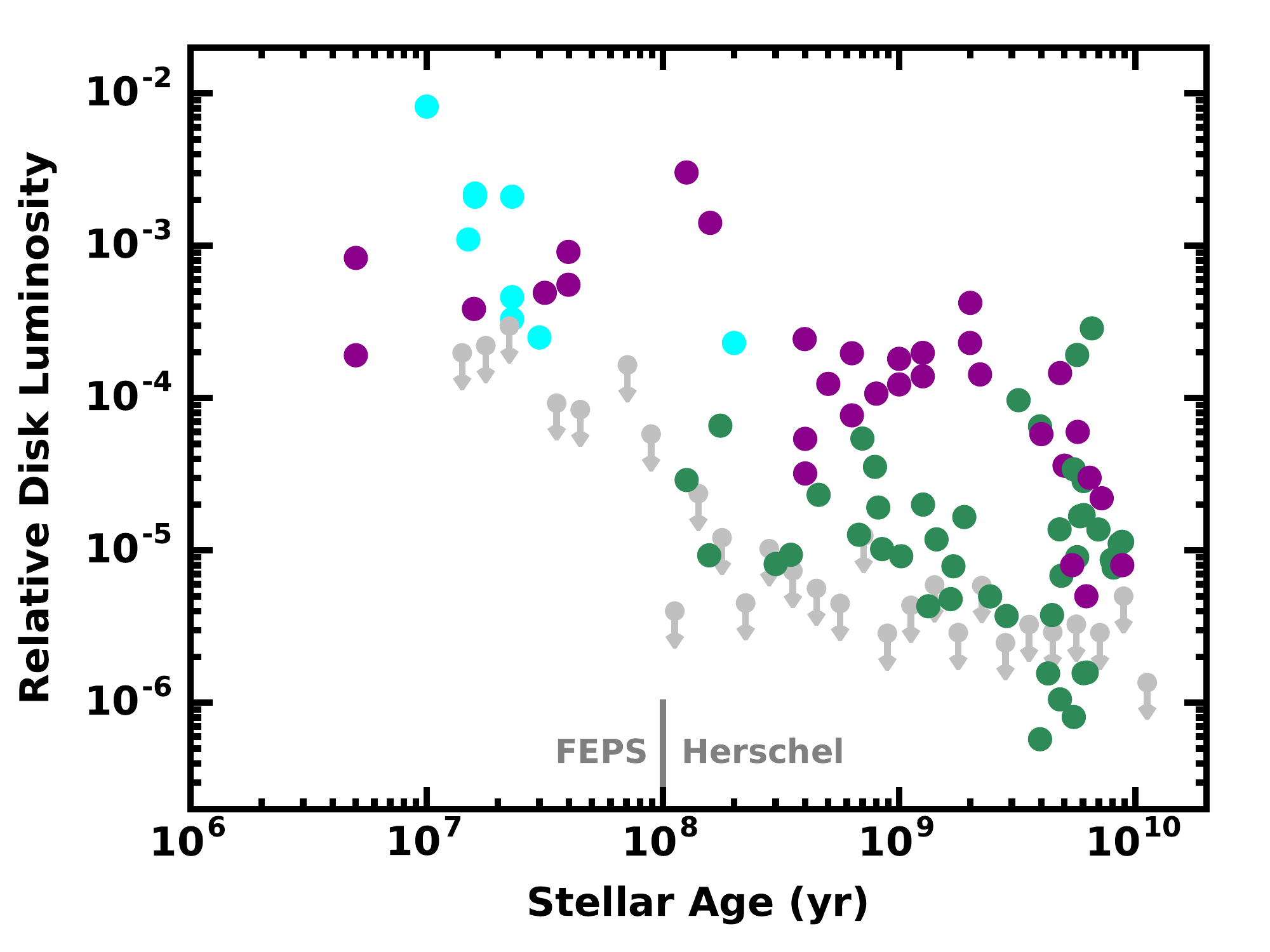}
\vskip -2ex
\caption{
\label{fig: alldata}
Relative disk luminosity for the full sample of debris disks summarized in 
the text. Filled circles indicate detections from 
\citet[][cyan]{matra2018}, 
{\it Spitzer} (purple, Fig.~\ref{fig: excess}, upper panel), and
{\it Herschel} (green, Fig.~\ref{fig: excess}, lower panel). 
Grey symbols indicate the median of upper limits in 0.1 intervals of log 
stellar age from FEPS (age $\le$ 100~Myr) or 
{\it Herschel} (age $\ge$ 100~Myr) data.
}
\end{center}
\end{figure}

To illustrate a larger range of excess properties observed among 
debris disks, Fig.~\ref{fig: alldata} supplements the {\it Spitzer} and 
{\it Herschel} detections (purple and green dots, respectively) with those 
from \citet{matra2018}, which compiles properties of (bright) debris disks 
that have been spatially resolved at millimeter wavelengths. Sources with 
FGK spectral types not included in the {\it Spitzer} \citep{carp2009a,
bryden2009} and {\it Herschel} \citep{eiroa2013,sibthorpe2018} samples are shown. 
The grey arrows show the median upper limits from FEPS at ages $\le 100$ Myr and from {\it Herschel} at ages $\ge 100$ Myr 
in 0.1 intervals of log stellar age.

Limits on \ldlstar\ for younger stars with ages $\lesssim$ 4--5~Myr are
rare.
\citet{lovell2021} detected one source with cold debris in an ALMA survey of 6 class III T Tauri stars that are likely members of the 1-3 Myr Lupus association; the other sources in the \citet{lovell2021} study are likely to be members of the Sco-Cen Association \citep{michel2021}.
Future observations of class III T Tauri stars in
other star-forming regions would improve links between class II sources 
with $\ldlstar \gtrsim 3 \times 10^{-2}$ \citep[e.g.,][]{michel2021}
and debris disks with $\ldlstar \lesssim 10^{-2}$.

Finally, parallax data from the {\it Gaia} satellite have revolutionized our knowledge of nearby moving groups of young stars \citep[e.g.,][]{faherty2018,gagne2018,gagne2020,ujjwal2020}, leading to new membership catalogs that allow better probes of the frequency of cold debris disks in the 20--150~Myr age range where the {\it Spitzer} and 
{\it Herschel} surveys have poor statistics.
Among the $\sim 10$ F stars 
within the 20--25~Myr old $\beta$ Pic moving group, 
50\% (75\%) have dusty 
material at $a \gtrsim$ 40~au \citep[$a \gtrsim$ 1~au;][]{pawellek2021}.
For F stars in older moving groups (20 stars in the
Tucana/Horologium association and the Columba and Carina groups), the 
cold debris disk frequency declines to $\sim$ 30\% at 45~Myr and $\sim$ 
15\% at 150~Myr\footnote{At 45~Myr (150~Myr), 6 of 20 (1 of 7) stars in
these samples have blackbody radii larger than 10~au.}. However, few of the new 45--150~Myr old debris disks  
have $a \gtrsim$ 30~au. 

With $\ldlstar \approx 2 \times 10^{-4}$ to $2 \times 10^{-3}$, the cold 
debris disks in the $\beta$ Pic moving group have dust luminosities 
similar to the \ldlstar\ of stars with ages of 10--40~Myr in 
Fig.~\ref{fig: alldata}. The new systems with ages of 40--50~Myr from
other moving groups have lower
dust luminosities, \ldlstar $\approx 2 \times 10^{-5}$ to $10^{-4}$, which
roughly coincides with the FEPS upper limits in Fig.~\ref{fig: excess}.
These data thus follow the observed trend of decreasing \ldlstar\ with 
stellar age. Continued analysis of cold debris in 30--100~Myr old stars 
would provide essential connections between the younger more luminous 
debris disks and those 
much older 
than 100~Myr.

To summarize, debris disk detections fall along a broad swath in $L_d/L_*$ 
and decline with time (Fig.~\ref{fig: alldata}). The most stringent upper 
limits in $L_d/L_*$ are $\lesssim 10^{-5}$ at ages of 0.1--10 Gyr (from 
the {\it Herschel} surveys), and $\sim 10^{-4}$ at 10--100 Myr (from FEPS). 
Current data suggest the frequency of cold debris disks 
is roughly constant at $\sim$ 25\% for stellar ages of $\sim$ 50~Myr to 
10~Gyr. Within the much younger $\beta$ Pic moving group ($\sim$ 
20--25~Myr), the frequency may be higher, $\sim$ 50\%, 
based on the study of a small sample of stars. 
As discussed below, 
future observations that lead to greater certainty in 
the debris disk frequency and \ldlstar\ for stars with ages of 20--150~Myr 
will bear on the variety of ways in which  protoplanetary disks evolve into debris disks (Section 4).

\section{Evolution of Rings of Solids} \label{sec:ringmodel}

To understand whether the rings of solids observed at 
$a \approx$ 20--200 au in young stars could plausibly
evolve into the rings of debris detected at similar $a$ 
in much older stars, we
perform a suite of multiannulus coagulation calculations. For the 
geometry of the rings, we rely on the observed properties outlined in
\S\ref{sec:diskobs}. In previous calculations, \citet{kb2008,kb2010}
considered the evolution of swarms of 1--100~km planetesimals in disks
extending from 30--150~au. These calculations matched the time evolution 
of available data for \ldlstar\ rather well. However, for disks with the largest \ldlstar, the models predict that dust is produced at increasingly large distances from the central star at late times. This trend is 
not observed among known debris disks \citep[e.g.,][see also 
Fig.~\ref{fig: extent}]{najita2005,kennedy2010,matthews2014,hughes2018}.  Moreover, 
theory currently favors scenarios where planets grow in seas of small 
and large solids (see below). Together, these observational and
theoretical developments motivate an updated set of models of 
debris production. To support our choices for the initial mix of pebbles 
and planetesimals, we briefly review recent theoretical results. Following 
this summary, we outline the numerical procedures and then describe 
results of new calculations.

\subsection{Background} \label{sec:ringmodelback}

In the core accretion model, planet formation is a three-step process. 
Within a circumstellar disk of gas and dust, micron-sized dust grains
grow into cm-sized pebbles then km-sized or larger planetesimals 
then planets. Within each step, various uncertainties in the initial
conditions, the physical properties of the gas and solids, and the 
important chemical and physical processes prevent a robust understanding
of the path from grains to planets. For this discussion, we 
summarize the current picture of planet formation, highlighting several 
areas of significant uncertainty that we explore in our models.

Initially, the gas and dust are well-mixed \citep[e.g.,][]{chiang2010,
youdin2010,youdin2013,liu2020}. As the disk evolves, small grains collide 
slowly, stick together, and grow into larger and larger aggregates 
\citep[e.g.,][]{dominik1997,wurm1998,blum2008,birnstiel2016,nimmo2018}.
As growth proceeds, collisions compactify particles
significantly \citep[e.g.,][]{weidling2009}. Particles with 
larger filling factors are less well-coupled to the gas. When grains 
uncouple from the gas, they settle to the midplane and
collide at higher velocities. Interactions between particles become more 
elastic, which limits additional growth \citep[e.g.,][]{zsom2010,
kelling2014,kruss2017}. Detailed studies suggest that particles 
experience a `bouncing barrier' at sizes $\sim$ 1--10~cm beyond 
which agglomeration effectively ceases \citep[see also][]{brauer2008,
windmark2012,grundlach2015,kruss2020,teiser2021}.

Although the presence of charged or organic grains may circumvent 
the bouncing barrier \citep[e.g.,][]{homma2019,steinpilz2019}, 
recent analyses have concentrated
on the `streaming instability' as a way to generate km-sized or
larger planetesimals from ensembles of cm-sized `pebbles'
\citep{youdin2005}. In this mechanism, aerodynamic drag concentrates 
pebbles into clumps with large overdensities compared to the typical
solid-to-gas ratio throughout the disk \citep[e.g.,][]{johansen2007a,
johansen2007b,johansen2009}. Continued concentration of pebbles within
the clumps enables the formation of planetesimals with radii 
$r \approx$ 100--1000~km \citep[e.g.,][]{birnstiel2016,simon2016,
schafer2017,yang2017,li2018,sekiya2018,lenz2019,lei2019,li2019,
chen2020,umurhan2020,pan2020,gerbig2020,squire2020}.

In recent numerical studies of the streaming instability, the size
distribution of the largest planetesimals and the efficiency of 
planetesimal formation depend on the physical conditions of the 
gaseous disk and the size distribution of pebbles 
\citep[e.g.,][]{simon2016,li2018,abod2019,carrera2020,klahr2020,
gole2020,rucska2021}.  Large solid-to-gas ratios generated by
radial drift and low turbulence ($\alpha \lesssim 10^{-4}$) favor 
efficient concentration of mono-disperse (i.e., single-sized)
sets of pebbles into 
much larger solids. High turbulence ($\alpha \gtrsim10^{-3}$), 
smaller solid-to-gas ratios, and broader size distributions of 
pebbles appear to limit the ability of the streaming instability 
to form large planetesimals \citep[however, see also][]{mcnally2021}. 
Among calculations with identical starting conditions, local 
fluctuations in these and other physical conditions within the disk 
lead to variations in the maximum size \rmax\ of a planetesimal and 
the fraction $f$ of the initial solid mass in pebbles that is 
concentrated into massive planetesimals.

Among other options for planetesimal formation, such as turbulent 
clustering \citep[e.g.,][]{cuzzi2008,pan2011,hartlep2020} and the 
settling instability \citep{squire2018}, outcomes for $f$ and \rmax\ are 
also uncertain. For any instability mechanism, local chemistry, 
radial diffusion, and sublimation modify the growth of pebbles and 
the concentration of pebbles into planetesimals \citep[e.g.,][]{ida2016,
hyodo2019}. Prior to the onset of instability, the porosity and 
compactness of pebbles are also uncertain \citep[e.g.,][]{okuzumi2012,
kataoka2013}

Once planetesimals form, the path to protoplanets is more certain.
In systems with $f \approx$ 1 (completely efficient planetesimal 
formation), the growth of massive protoplanets may be too slow to
form super-Earth mass and larger planets during the likely lifetime of 
the gaseous disk \citep[e.g.,][]{kb2008,liss2009,kb2009,kb2010,koba2010b,
levison2010,dangelo2014,mordasini2015,boden2018,dangelo2021}. From 
analytical considerations, \citet{gold2004} and \citet{raf2005} 
demonstrated that a few large planetesimals in a sea of pebbles grow 
rapidly due to the small scale height of the pebbles. Subsequent numerical 
calculations of `pebble accretion' yield 10~\mearth\ and larger ice 
giants on time scales of a few Myr \citep{kb2009,bk2011a,lamb2012}. 
More recent studies with $\rmax \gtrsim$ 100~km and $f \lesssim10^{-2}$ 
illustrate the ability of pebble accretion to form ice and gas giants 
in many circumstances on short time scales \citep[e.g.,][]{matsumura2017,
alibert2018,lin2018,bitsch2019,johansen2019,lambrechts2019,morbidelli2020,
voelkel2020,klahr2020,chambers2021}. 

In contrast to the many studies of giant planet formation via pebble 
accretion, there have been few attempts to investigate the long-term 
evolution of the debris signatures produced by growing protoplanets in a 
sea of pebbles. Formation scenarios for a planet nine at 
$a \gtrsim$ 200~au in the Solar System illustrate how systems with an 
initial \rmax\ = 100~km and various $f$ generate super-Earth mass planets 
and very luminous debris disks at 200--750~au around solar-type stars
on time scales of 0.1--1~Gyr \citep{kb2015a,kb2016b}. The debris disks 
in some model systems have properties similar to those observed in the 
bright debris disks orbiting HD~107146, HD~202628, and HD~207129 
\citep{corder2009,krist2010,krist2012,marshall2011,ricci2015,marino2018}. 
In the next sections, we consider whether swarms of pebbles and planetesimals 
can produce debris disks similar to those in the {\it Herschel} and
{\it Spitzer} samples described in section 2. 

\subsection{Initial Conditions}\label{sec:ringmodelinit}

To follow the evolution of pebbles and planetesimals in a ring, we use 
the multiannulus coagulation routine within \orch, an ensemble of codes
for planet formation. As outlined in the Appendix, the code uses a 
particle-in-a-box algorithm for collision rates and energy scaling for
collision outcomes. Particles evolve dynamically with Fokker-Planck 
routines. To avoid the extra free parameters 
associated with the gaseous component of the ring, we ignore radial drift 
and circularization of solids by the gas.
\citet{kb2008,kb2010,kb2012} describe the formulation and procedures in 
more detail.

For this study, we perform calculations in two separate grids, each with 
28 concentric annuli. With an inner radius of 36~au (60~au) and an outer 
radius of 54~au (90~au), the grids cover a reasonable subset of the rings
observed in protoplanetary and debris disks (Fig.~\ref{fig: extent}). 
Within each annulus, particles occupy distinct mass bins with sizes 
ranging from 1~\mum\ to $10^5$~km, orbital eccentricity $e$, and 
inclination $\imath$. Initially, solids have sizes of 1~cm (residual 
pebbles) and 100~km (planetesimals produced by the streaming 
instability), eccentricity $e_0 = 10^{-3}$, and inclination $\imath_0 = 
e/2$. These parameters are appropriate for solids recently liberated 
from a protostellar disk with turbulence parameter $\alpha \sim 10^{-3}$.

\begin{figure}[t]
\begin{center}
\includegraphics[width=5.0in]{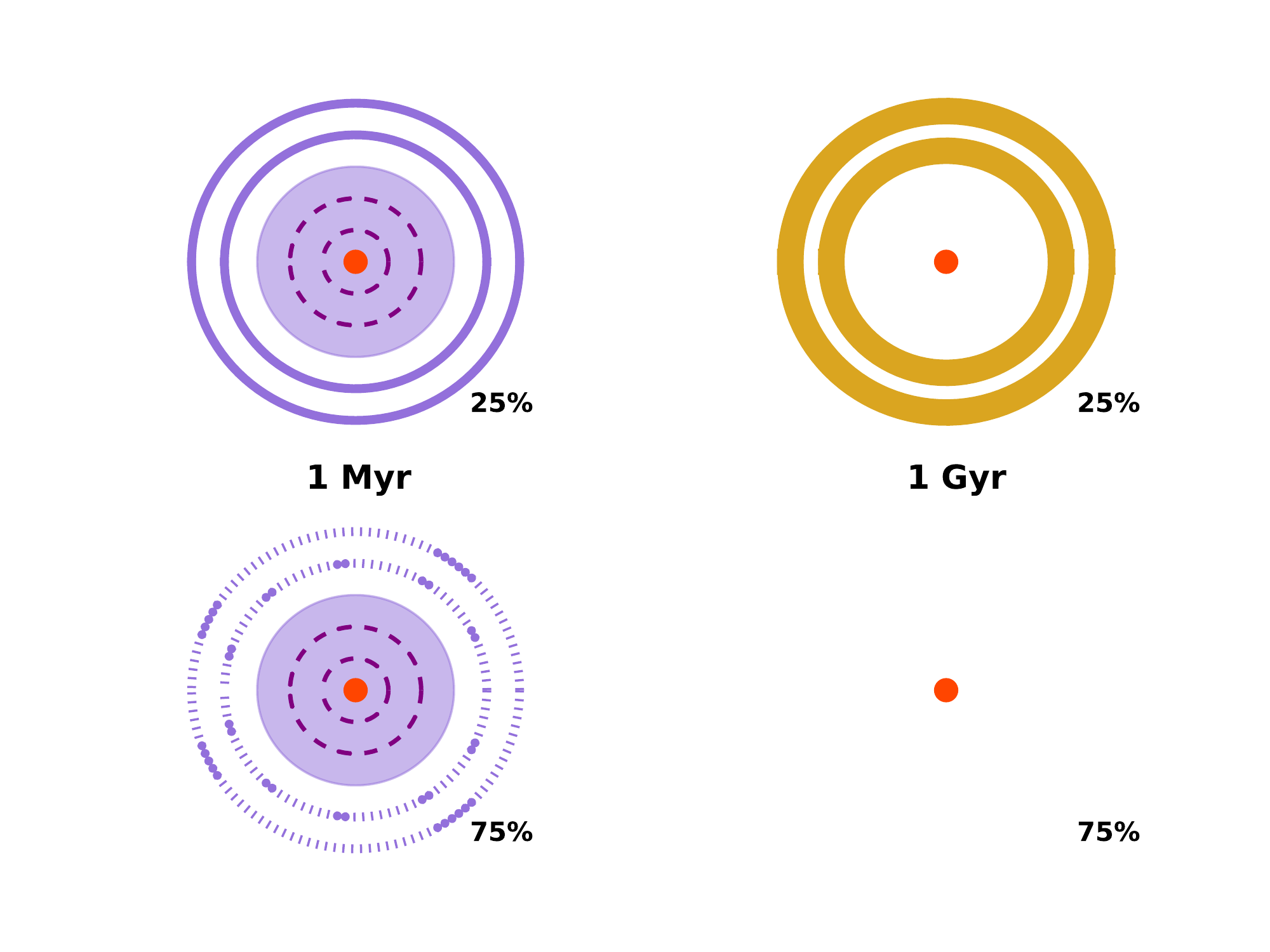}
\vskip -2ex
\caption{
\label{fig: schema}
Schematic architectures for protoplanetary disks (left diagrams) and
older main sequence stars (right diagrams). Each illustration includes
the approximate percentage of stars with the sketched architecture.
On the left, light purple filled circles indicate compact gaseous disks; 
dashed circles show possible locations of dense rings within the disk. 
Outside the compact disks, solid (dotted) circles illustrate locations 
of dense (invisble) rings of solids. On the right, the gold bands in 
the upper diagram mark the location of a debris disk.
}
\end{center}
\end{figure}

Solids initially have a total mass \M0, a surface density distribution
$\Sigma(a)$, and a fraction $f$ of mass in large planetesimals (see 
section 3.1). To select
these parameters, we rely on the observational constraints described 
above and shown schematically in Fig.~\ref{fig: schema}. Among class II 
sources with ages $\sim$ 1~Myr, roughly a quarter have a compact disk 
with an outer radius of $\sim$ 30~au and bright rings of solids at 
larger distances (Fig.~\ref{fig: schema}, top left; see also
Fig.~\ref{fig: extent}). The rings are well-fit with gaussian 
distributions of pebbles having total masses $\sim$ 10--60~\mearth\ and
dispersions $\sigma \approx$ 4--8~au \citep{dullemond2018}. To span this 
range, we adopt upper limits of \M0\ = 10~\mearth\ at 45~au and 
\M0\ = 45~\mearth\ at 75~au. For $\Sigma(a)$, we adopt a fixed ratio
$\sigma / a$ = 1/15; solids then have a gaussian $\Sigma(a)$ centered at 
45~au (75~au) with $\sigma$ = 3~au (5~au).
Defining $P$ as the orbital period around the 
central star, the maximum \M0\ and the gaussian $\Sigma(a)$ yield
a similar ratio for $P / \Sigma$ at 45~au and 75~au and therefore similar
time scales for the growth of planets \citep[e.g.,][]{liss1987}.

In the majority of class II sources with compact disks and no bright rings
(Fig.~\ref{fig: schema}, bottom left), the lack of millimeter emission at
large radii is consistent with the solids outside the compact disk having 
either (i) low mass if composed primarily of pebbles or (ii) higher mass 
if composed primarily of large planetesimals. These possibilities motivate 
a broad range in \M0 (from 0.01~\mearth\ to the adopted upper limits) and 
$f$ (0--1).
\footnote{The lower mass limit is somewhat smaller 
than current mass
estimates for the Kuiper belt \citep[0.02--0.06~\mearth;][]{pitjeva2018,
diruscio2020}, which has a dust luminosity well below {\it Herschel} 
sensitivity limits \citep[e.g.,][]{backman1995,vitense2012}.}
A plausible third option for the lack of millimeter emission in
these sources is a small vertical scale height which prevents solids from
intercepting a detectable fraction of light from the central star. For the
calculations reported here, we do not consider this possibility but return 
to it in the discussion.

The observed properties of debris disks also motivate a range in \M0\ and
$f$. Among solar-mass stars with ages of $\sim$ 1~Gyr, roughly 25\% have
bright rings of cold debris at $a \gtrsim$ 40~au (Fig.~\ref{fig: schema}, 
top right). In the remaining 75\%, cold debris is absent or undetectable 
(Fig.~\ref{fig: schema}, bottom right). The goal of the calculations is 
to identify the initial conditions and evolutionary paths that connect 
the protoplanetary disks (left) to the debris disks (right), while 
satisfying the constraints on the incidence rates and 
dust luminosities of known debris disks as a function of stellar age 
(e.g., Fig.~\ref{fig: extent} and Fig.~\ref{fig: alldata}). 
One simple hypothesis we explore is whether the massive rings of 
protoplanetary solids (upper left of Fig.~\ref{fig: schema}) evolve into
rings of debris at similar distances at Gyr ages (upper right of 
Fig.~\ref{fig: schema}). 

Other evolutionary scenarios are also plausible. If some of the 75\% of 
class II sources that appear as compact disks at millimeter wavelengths
possess 
substantial rings of solids that are invisible at millimeter wavelengths 
as a result of their lower
mass or lack of pebbles, these may evolve into bright debris disks at 
late  times \citep[e.g.,][and references therein]{wyatt2008,kb2010,
matthews2014,krivov2021}. To explore the evolution of these systems, we 
also follow the evolution of low mass rings of pebbles and more massive 
rings composed of large planetesimals. 

Based on these considerations, we consider initial masses 
\M0\ = 0.01--45~\mearth\ for eight values of $f$ between 0 and 1. 
As summarized in Appendix Tables~\ref{tab: tab1} and \ref{tab: tab2}, some combinations of \M0\ and $f$ are not physically realizable in a ring with pebbles and large planetesimals. For example, with \M0\ = 0.01~\mearth\ and $f = 10^{-3}$), the mass in a single 100~km planetesimal exceeds $\M0 \times f$, the total mass allocated for all larger bodies. 

In identifying the relevant evolutionary path (or combination of paths) 
for protoplanetary and debris disks, we aim to ensure continuity between 
the starting and ending points outlined in Fig.~\ref{fig: schema}. That is,
if young stars simply move horizontally in Fig.~\ref{fig: schema}, from 
bright rings at 1~Myr to bright rings at 1~Gyr {\it and} from compact 
disks into diskless main sequence stars, we are guaranteed to match the 
observed fraction of debris disks at every epoch from $\sim$ 50~Myr to 
10~Gyr. However, if some sources with compact disks and invisible rings 
of solids at 1~Myr evolve into observable debris disks at 1~Gyr, then a 
similar fraction of class II sources with bright rings must evolve into 
diskless main sequence stars. In this case, we need to ensure that the 
fraction of class II sources that evolve diagonally upward in the figure 
matches the fraction that evolve diagonally downward.

If the high fraction of cold debris disks in the $\beta$ Pic moving group is typical of all 20--25~Myr-old stars \citep[50\%;][]{pawellek2021}, then a large fraction of class II sources
without bright rings need to evolve into bright debris disks at 20--25~Myr
and then fade below current detection limits for stellar ages of 50--100~Myr.
An additional goal of this study is to understand the physical properties
of protoplanetary disks that lead to this type of evolution.

The following sections describe the growth of the largest objects 
(\S\ref{sec: ringmodelgrowth}), the evolution of the dust luminosity
from collisional debris (\S\ref{sec: ringmodeldebris}), and the impact 
of the largest objects on the radial distribution of the debris and gap
formation (\S\ref{sec: ringmodelgaps}). 

\subsection{Growth of Planets}
\label{sec: ringmodelgrowth}

As each calculation proceeds, pebbles and planetesimals collide and merge
into larger objects. Along with mergers, debris from collisions deposits 
mass into other mass bins. At first, the low initial $e$ and $\imath$
limits collisions between particles within different annuli. Evolution of
$e$ and $\imath$ allows more interactions among all particles; these 
interactions disperse collision debris throughout the grid. Most 
collisions yield some debris with particle sizes smaller than 1~\mum.
This material is assumed ejected by radiation pressure from the central
star. Over 10~Gyr of evolution, the amount of lost material ranges from 
less than 1\% to more than 90\% 
(see Appendix Tables~\ref{tab: tab1} and \ref{tab: tab2}). Ejected
solids do not interact with other solids in the grid.

To set some expectations for the calculations, we estimate the time 
scale for the growth of planets in rings at 45~au and 75~au. In 
\citet{kb2010}, 100~km planetesimals grow into Pluto-mass planets on
a time scale $t_{1k} \sim$ 10~Gyr at 80~au in a disk with the solid
surface density distribution of a minimum mass solar nebula. In these
calculations, $t_{1k} \propto P / \Sigma$, where $P$ is the orbital
period and $\Sigma$ is the initial surface density \citep[see 
also][]{liss1987,gold2004,raf2005,kb2008}. Scaling these results to 45~au and 75~au,
\begin{equation}
\label{eq: t1kp}
t_{1k} \approx 
2~{\rm Gyr} 
\left ( \frac{M_{ref}}{M_0} \right ) \\
\end{equation}
where $f$ = 1 and $M_{ref}$ = 10~\mearth\ (45~\mearth) is the mass
contained in rings at 45~au (75~au). When $f < 1$, analytical estimates 
suggest $t_{1k} \propto f^n$, with $n \approx$ 0.5--1.0 \citep{gold2004,
raf2005}. The calculations here provide good tests of these estimates, 
as discussed in the Appendix.

\begin{figure}[t]
\begin{center}
\includegraphics[width=5.5in]{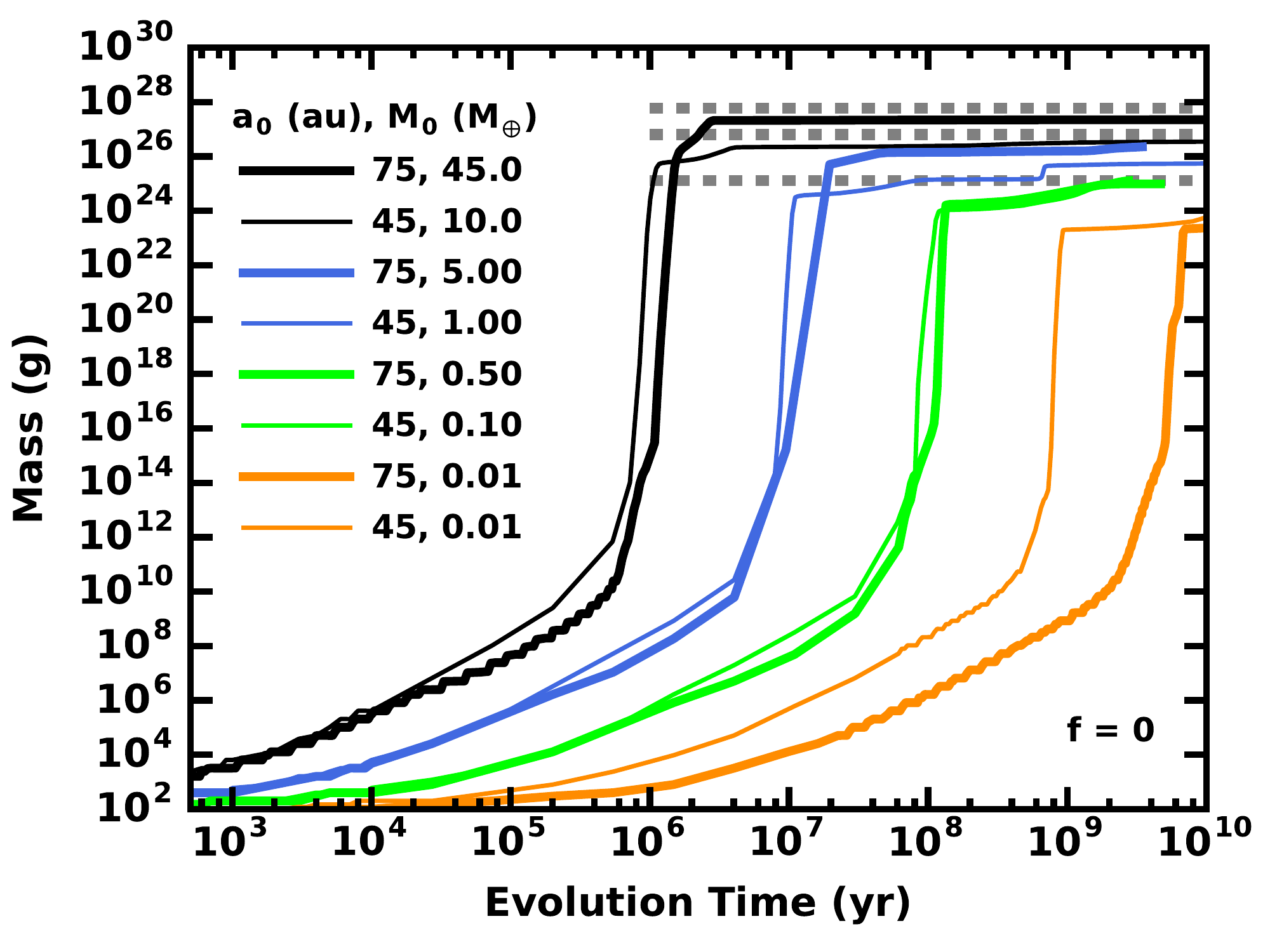}
\vskip -2ex
\caption{
\label{fig: mass0}
Time evolution of the mass of the largest object in rings of solids with 
initial $f$ = 0 and the values for $a_0$ and $M_0$ indicated in the legend.
Dashed horizontal lines indicate the mass of Pluto (lower), Mars 
(middle), and Earth (upper). At 45--75~au, massive rings of solids produce 
super-Earths in 1~Myr (\M0\ = 45~\mearth\ at 75~au, thick purple curve), 
Mars mass planets in 10--100~Myr (\M0\ = 5--10~\mearth\ at 45--75~au, thin
purple and thick dark green curves), and Plutos in 100~Myr to 1~Gyr 
(\M0\ = 0.01--0.5~\mearth\ at 45--75~au, light green and orange curves).
}
\end{center}
\end{figure}

In the full set of calculations, outcomes are sensitive to \M0\ and $f$.
Systems with more mass in solids tend to evolve more rapidly and form the
largest planets and the brightest debris disks. For fixed initial mass,
rings with more of their initial mass in planetesimals evolve more slowly.
When $f$ is large, planetesimals compete for pebbles. As the initial 
population of
pebbles declines, growth depends on infrequent planetesimal--planetesimal 
collisions and the evolution stalls. When $f$ is small, planetesimals do
not compete for pebbles; growth is then rapid.

Fig.~\ref{fig: mass0} illustrates the growth of the largest object in 
systems with $f$ = 0. At the start of each calculation, gravitational 
focusing is negligible; pebbles grow roughly linearly with time. Slow
pebble growth allows collisional damping to reduce $e$ and $\imath$ by 
an order of magnitude. As the largest particles approach km sizes, 
collision velocities remain low despite increased dynamical friction and 
viscous stirring by the largest objects. Larger gravitational focusing 
factors initiate an explosive phase of runaway growth, where planetesimals
with maximum radii $\rmax\ \sim$ 1~km rapidly reach much larger sizes, 
$\rmax\ \gtrsim$ 1000~km. After the runaway, a few protoplanets continue
to accrete more and more material from the smaller objects and to stir up 
the velocities of the leftovers. At first, collisional damping among the 
pebbles counters stirring by protoplanets. Eventually, stirring overcomes 
damping and initiates a collisional cascade, where km-sized and smaller 
objects are ground into smaller and smaller particles. The cascade robs 
protoplanets of remaining small solids; growth ceases. 

Aside from the time scale and the final mass, the evolution of the mass
of the largest protoplanet is amazingly independent of $a$ and \M0. All
of the curves in Fig.~\ref{fig: mass0} have the same linear phase for 
$M \lesssim 10^{12}$~g followed by a nearly vertical rise to a 
maximum planet mass that scales with \M0. During runaway growth, 
planets accrete nearly all of the mass in their annulus (and sometimes 
in adjacent annuli). With more mass in protoplanets than pebbles, stirring
reduces gravitational focusing factors. The evolution is then oligarchic,
where the largest protoplanets slowly accumulate leftover pebbles
\citep[e.g.,][]{kok1998,ormel2010c}. Once
pebbles are exhausted, growth ceases. In the most massive rings,
Earth mass planets form in 3--10~Myr. When \M0\ is smaller, 
growth takes longer and results in lower mass protoplanets. In the 
lowest mass rings we consider, Pluto mass planets form on Gyr 
or longer time scales. The progression of longer time scales to 
produce lower mass planets when \M0\ is small is characteristic of 
coagulation calculations (eq.~\ref{eq: t1kp}).

\begin{figure}[t]
\begin{center}
\includegraphics[width=5.5in]{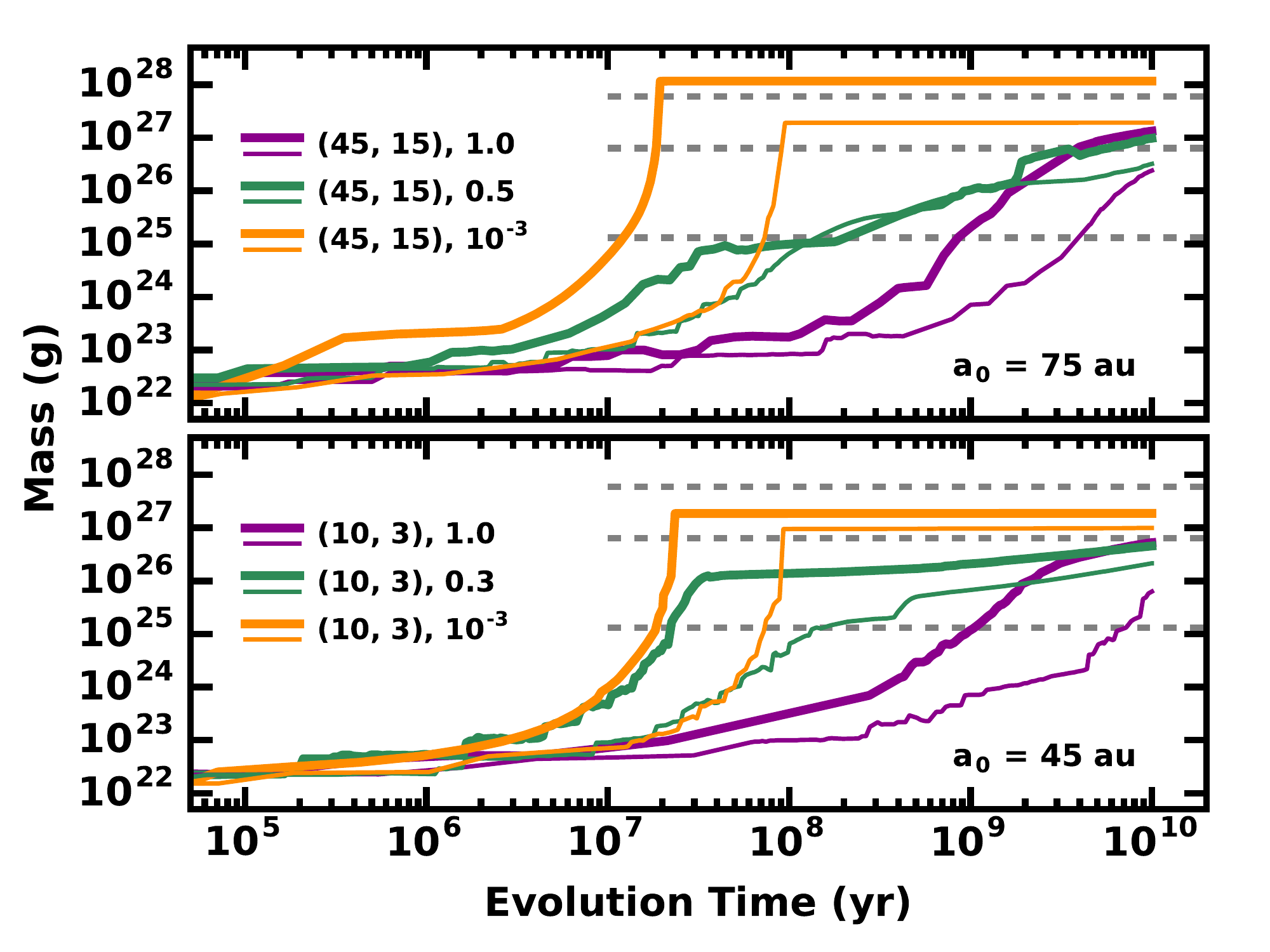}
\vskip -2ex
\caption{
\label{fig: mass1}
As in Fig.~\ref{fig: mass0} for calculations at 45~au (lower panel) and 
at 75~au (top panel) with various \M0\ and $f$. The legend in the upper 
left of each panel indicates a pair of \M0\ for thick (first \M0) and 
thin (second \M0) lines and the value of $f$ for each curve. Dashed 
horizontal lines indicate the masses of Pluto (lower), Mars (middle), 
and Earth (upper). Runaway growth is more dramatic in rings
with small $f$. At 75~au, massive rings with small $f$ produce 
super-Earths in 20--30~Myr (thick orange curve). Rings with smaller $f$ 
generate Mars-mass planets on much longer times scales, 1--10~Gyr (thick 
purple, thick green, and thin green curves) At 45~au, rings with
\M0\ = 3--10~\mearth\ and $f \approx 10^{-1} - 10^{-4}$ produce Mars mass 
planets in 20--100~Myr. Models with $f \approx$ 0.3--1.0 yield Plutos to 
Mars-mass planets in 1--10~Gyr..
}
\end{center}
\end{figure}

When $f > 0$, the final planet mass correlates well with \M0\ and $f$. 
At 75~au (Fig.~\ref{fig: mass1}, upper panel), rings with 
$f \approx 10^{-2} -10^{-5}$ have a 
similar explosive growth phase as systems with $f$ = 0. Initially, 100~km
planetesimals slowly accrete pebbles and the occasional planetesimal. Once
their masses exceed $10^{23} - 10^{24}$~g, the runaway quickly carries
them to super-Earth (\M0\ = 45~\mearth, thick orange curve) to super-Mars 
(\M0\ = 15~\mearth, thin orange curve) masses. Unlike the systems in 
Fig.~\ref{fig: mass0}, the lack of solids with masses intermediate between
planetesimals and pebbles allows protoplanets to accrete a large fraction
of the mass in their respective annuli. Oligarchic growth is limited.
Protoplanets maintain roughly constant masses for the rest of the 
calculation.

As the initial mass in pebbles drops, protoplanets grow more slowly and
just manage to reach the mass of Mars (Fig.~\ref{fig: mass1}, green and 
purple curves in the upper panel). Systems with $f \approx$ 0.5 produce 
many Pluto-mass planets in 10--100~Myr. 
Continued accretion of pebbles and leftover planetesimals during an 
extended oligarchic growth phase enables protoplanets to match the mass 
of Mars. When $f \approx$ 1, it takes 1--3~Gyr for several protoplanets 
to exceed the mass of Pluto. This time scale is close to the estimate 
derived from scaling previous calculations in 30--150~au disks of solids
(eq.~\ref{eq: t1kp}). Oligarchic growth eventually carries some of these 
to the mass of Mars.

At 45~au, the behavior of rings with 3--10~\mearth\ and $f = 10^{-5}$ to 
$f$ = 1 closely follows the evolution at 75~au (Fig.~\ref{fig: mass1},
lower panel). 
Although time scales are similar, the lower mass rings at 45~au prevent 
planets from reaching the same masses as the higher mass rings at 75~au.
Thus, Mars-mass (Earth-mass) planets form in 3--10~\mearth\ 
(15--45~\mearth) rings at 45~au (75~au) in 30--100~Myr when 
$f \approx 10^{-4}$--0.1, For $f$ = 0.5, time scales to reach Pluto-mass 
planets at 45~au are somewhat longer than at 75~au; Mars-mass planets take 
much longer, $\sim$ 3--10~Gyr (green curves). Calculations with $f$ = 1 
(purple curves) yield a few Mars-mass planets only at the highest mass 
considered; lower mass systems manage to form planets with 3--5 $\mp$ 
(Pluto masses). 

\begin{figure}[t]
\begin{center}
\includegraphics[width=5.5in]{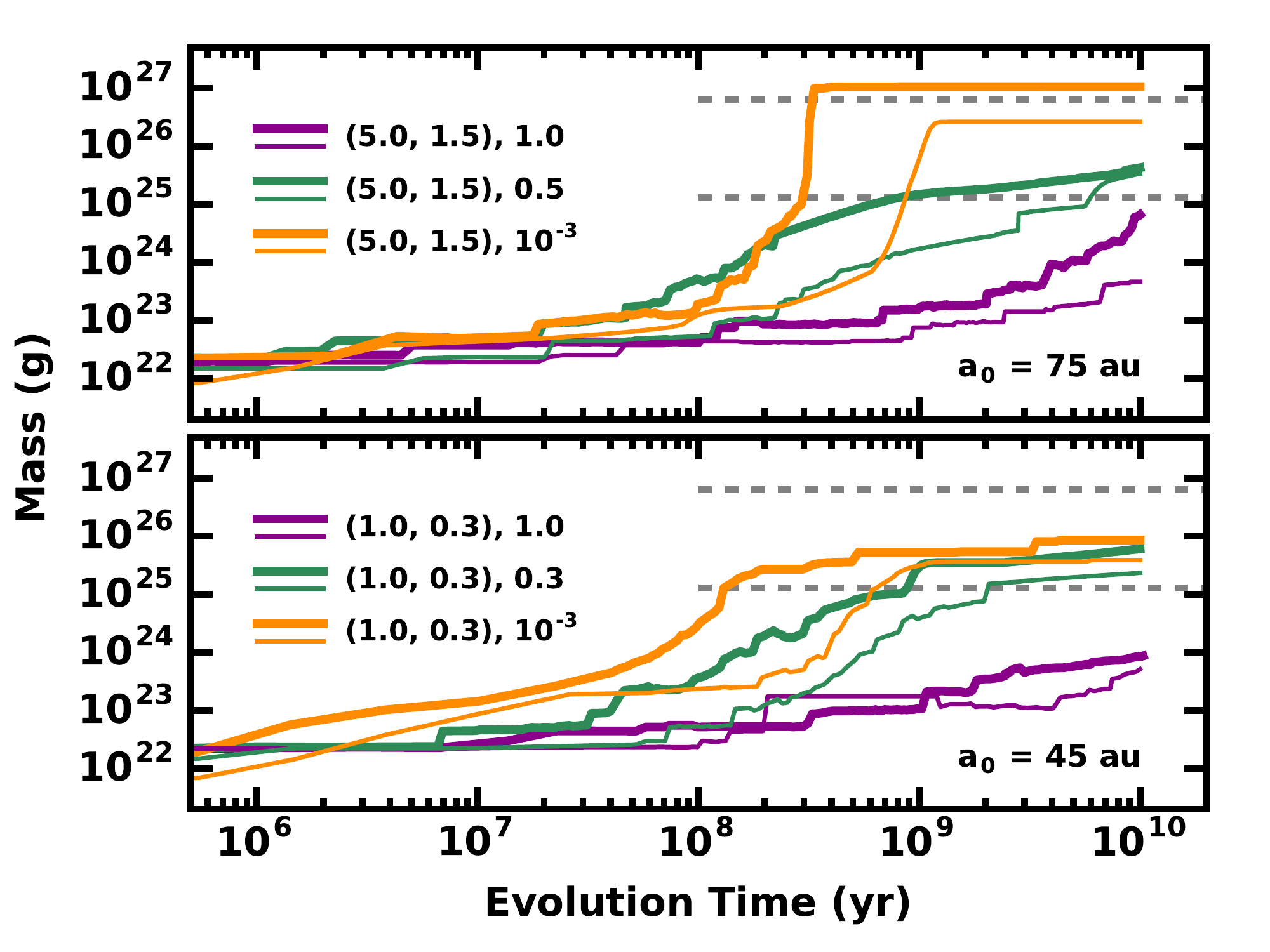}
\vskip -2ex
\caption{
\label{fig: mass2}
Growth of the largest objects in medium mass rings of solids at 45~au
(lower panel) and at 75~au (upper panel). The legend associates \M0\ and 
$f$ for each curve. Numbers in parentheses indicate initial mass in
\mearth\ for thick (first number) and thin (second number) lines. 
Horizontal dashed lines indicate the mass of Pluto (lower) and Mars 
(upper). The final mass of the largest object and the time scale to 
reach this mass scale inversely with \M0: lower mass rings produce 
lower mass planets. 
}
\end{center}
\end{figure}

When rings have less mass than the examples in Fig.~\ref{fig: mass1}, 
lower-mass planets form on longer time scales (Fig.~\ref{fig: mass2}).
At 45~au (lower panel), rings with \M0\ = 1~\mearth\ (0.3~\mearth) and 
$0 < f < 1$ produce Pluto-mass planets on time scales of 100~Myr to 
1~Gyr (green and orange curves). After 10~Gyr, the planets in these 
systems have masses of 2--10~\mp. Growth is negligible when $f$ = 1 
(purple curves); maximum masses are $\sim$ 0.1~\mp. The more massive 
systems at 75~au easily generate Mars mass planets in 300~Myr to 1~Gyr 
when $f$ is very small (Fig.~\ref{fig: mass2}, upper panel, orange 
curves). In each of these systems, planetesimals accrete nearly all of 
the available pebbles during runaway growth and then maintain a roughly
constant mass for the rest of the calculation. As $f$ increases, growth 
slows; the mass of the largest planets drops by several orders of 
magnitude (green and purple curves). When $f$ = 0.5, Pluto mass
planets form on 1--10~Gyr time scales. Very efficient planetesimal 
formation ($f$ = 1) leads to the growth of Pluto mass planets at late
times when \M0\ = 5~\mearth; in lower mass rings, large planets fail to
form after 10~Gyr.

Very low mass rings produce much smaller planets. 
Mixes of solids and pebbles with \M0\ = 0.01--0.1~\mearth\ in rings at
45~au struggle to make a single Pluto.
For any $f \ne$ 1, the \M0\ = 0.1~\mearth\ models have a 
modest runaway that converts 200~km planetesimals into 900~km dwarf 
planets. Rings with a factor of three (ten) less mass produce several 
Charon-mass objects (Kuiper belt objects) with radii of 500~km (250~km). 
Rings with no pebbles barely evolve over 10~Gyr.

At 75~au, the lowest mass rings in our study show a little more activity. 
Models with \M0\ = 0.5~\mearth\ and $f \lesssim$ 0.1 begin a robust 
runaway at 1~Gyr. After 3~Gyr, these rings have a modest set of dwarf
planets with masses 5--10 times the mass of Pluto. When $f \approx$ 0.5,
the delay in the runaway to 3--6~Gyr fails to produce a Pluto-mass dwarf
planet, but generates a few Charon-mass objects. As \M0\ drops to
0.01--0.1~\mearth, growth is modest, yielding a few 200--300~km objects
in 0.1~\mearth\ rings and many 150~km planetesimals in 0.01~\mearth\ rings.
Once again, systems with $f$ = 1 barely evolve over the age of the 
universe.

To collate some results from the calculations in a convenient form, 
Tables~\ref{tab: tab1}--\ref{tab: tab2} in the Appendix list the time
scale to form at least one Pluto-mass planet, $t_{1k}$; the maximum
radius \rmax\ of the largest planet at 10~Gyr, the final mass in solids
$M_f$ at the end of each calculation, and other useful parameters. In
addition to examining the relation between $t_{1k}$, \M0, and $f$, the
discussion in the Appendix includes several examples of collisional
damping and some comparisons with previous calculations.

\subsection{Evolution of Dust Luminosity}
\label{sec: ringmodeldebris}

Throughout each calculation, the dust luminosity depends on the vertical
scale height and the surface area of small solids. Initially, pebbles and
planetesimals have the same typical orbital inclination, 
$\imath_0 = 5 \times 10^{-4}$. 
At large distances $a$ = 30--90~au, the upper and lower surfaces of the 
swarm intercept a negligible fraction of stellar flux 
\citep[e.g.,][]{kh1987,chiang1997}. The maximum stellar luminosity 
intercepted by pebbles is then the fraction of solid angle subtended by 
a ring with vertical scale height $H$ above and below the midplane
at distance $a$ from the central star: 
$L_{d,max} = H /a = 5 \times 10^{-4}$ for the adopted 
$\imath_0 = 5 \times 10^{-4}$. Ignoring the negligible contribution from 
large planetesimals for rings with any $f$ the initial dust luminosity is:
\begin{equation}
\label{eq: ldlstar0}
L_{d,0} / \lstar \approx \begin{cases}
                7.3 \times 10^{-4} ~ (1 - f) \left ( \frac{M_0}{\mearth} \right ) & a = 45~{\rm au}, \M0\ \lesssim 0.7~\mearth\ \\
                1.9 \times 10^{-4} ~ (1 - f) \left ( \frac{M_0}{\mearth} \right ) & a = 75~{\rm au}, \M0\ \lesssim 2.6~\mearth\ \\
                \end{cases}
\end{equation}
When \M0\ exceeds the limits quoted in eq.~\ref{eq: ldlstar0}, 
$L_{d,0} / \lstar\ = 5 \times 10^{-4}$. 

These expressions for \ldlstar\ establish clear constraints on plausible 
evolutionary paths for rings of pebbles and planetesimals. To exceed the 
FEPS upper limits of \ldlstar $\approx 10^{-4}$ for young stars with ages 
$\lesssim$ 100~Myr, rings at 45~au (75~au) must have initial masses $\M0 
\gtrsim$ 0.15~\mearth\ (0.5~\mearth). Lower mass disks require additional 
dust production from a collisional cascade. Among older stars, exceeding 
the {\it Herschel} upper limits of $\ldlstar \approx 3 \times 10^{-6}$ 
requires much less mass, $\M0 \gtrsim$ 0.005~\mearth\ (0.02~\mearth) at
45~au (75~au). If the time evolution of \ldlstar\ is slow, the small 
{\it Herschel} upper limits allow a broader set of rings to match observed
systems.

As each calculation proceeds, the dust luminosity responds to changes in
$H$ and the surface area of small solids. When $f \lesssim 10^{-2}$, 
collisional damping initially reduces $H$ by a factor of 2--5. Debris
production is minimal; \ldlstar\ drops as $H$ falls. As planetesimals 
accrete more and more pebbles, stirring (damping) becomes more (less)
efficient. Debris production and $H$ begin to grow; \ldlstar\ rises. Rings
of solids with larger $f$ have more stirring from growing planetesimals 
and less damping. The dust luminosity then rises at the start of each 
calculation. In all systems, $H$ eventually reaches a maximum; debris
generated from collisions of protoplanets and planetesimals falls below 
the loss  of small particles from the collisional cascade. The dust 
luminosity then fades with time.

\begin{figure}[t]
\begin{center}
\includegraphics[width=5in]{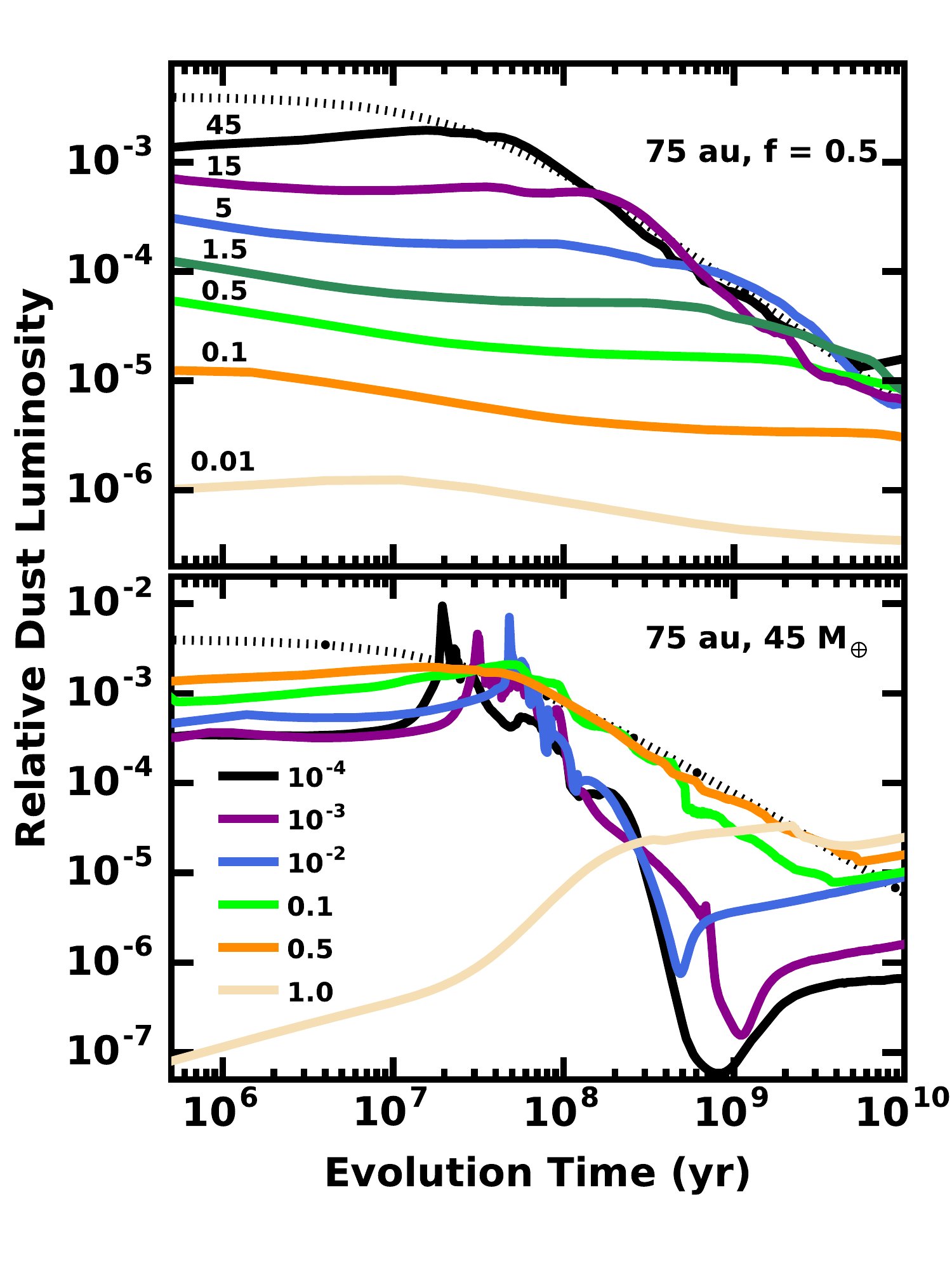}
\vskip -2ex
\caption{
\label{fig: ldust0}
Evolution of the relative dust luminosity \ldlstar\ for rings at 75~au. 
In each panel, the dotted line shows the evolution predicted by an
analytical model where $ \ldlstar\ \propto L_0 (1 + t/t_0)^{-\alpha}$ with
$\alpha$ = 1.12 \citep{kb2017a}.
{\it Upper panel:} For models with $f$ = 0.5, the value for \M0\ is listed
to the left of each curve. Systems with \M0\ $\gtrsim$ 1.5~\mearth\ have a 
period of constant \ldlstar\ followed by a $t^{-1}$ decline. In lower mass 
systems, \ldlstar\ is roughly constant over 10~Gyr. 
{\it Lower panel:} For models with \M0\ = 45~\mearth, the legend lists $f$ 
for each curve. Systems with $f$ = 1 evolve slowly to a maximum \ldlstar\ 
$\sim 10^{-5}$. When $f \approx$ 0.1--0.5, \ldlstar\ gradually drops from 
$\sim 10^{-3}$ to $10^{-5}$. Rings of solids with $f \lesssim 10^{-2}$ 
start fainter, abruptly rise to \ldlstar $\approx 10^{-2}$ and then 
dramatically fall to \ldlstar $\lesssim 10^{-6}$. Collisions between pairs 
of protoplanets occasionally produce spikes in \ldlstar.
}
\end{center}
\end{figure}

The top panel of Fig.~\ref{fig: ldust0} illustrates the smooth evolution of the relative
dust luminosity \ldlstar\ for a set of calculations with $f$ = 0.5 at 
75~au. When half of the initial mass is in pebbles, the initial
\ldlstar\ ranges from $\ldlstar\ \sim 10^{-6}$ for \M0\ = 0.01~\mearth\ to
$5 \times 10^{-4}$ for \M0\ = 2.6--45~\mearth\ (eq.~\ref{eq: ldlstar0}). As the
most massive rings evolve (15--45~\mearth, black and purple curves),
stirring overcomes collisional damping; $H$ increases. Despite the lack of 
debris production from planetesimal growth, \ldlstar\ grows with time. 
Over the next 10--30~Myr, increased debris production roughly balances 
losses from the collisional cascade; \ldlstar\ remains roughly constant. 
As protoplanet growth slows considerably at 
100-200~Myr, stirring initiates a stronger cascade that removes 
more and more of the remaining small
solids. The dust luminosity then begins a roughly linear decline. 
With more massive protoplanets that form earlier and stir solids more 
strongly, the 45~\mearth\ rings begin to decline before the 15~\mearth\ 
rings. Once the strong cascade begins, \ldlstar\ follows the same 
roughly linear decline in both systems.

Lower mass models have a similar behavior. In rings with 0.5--5~\mearth\ 
(light green, dark green, and blue curves), the dust luminosity slowly
declines with time during most of the first 100~Myr to 1~Gyr of evolution. 
During this decline, protoplanets gradually reach the mass of Pluto.
Stirring from these large solids increases $H$. Larger relative velocities
enhance debris production, but mass removal from the cascade keeps 
\ldlstar\ falling. Eventually, stirring becomes more effective; the 
cascade strengthens and removes more and more small solids from the ring. 
The dust luminosity then declines more rapidly, joining the roughly linear 
decline of the higher mass models.

In the lowest mass models (0.01--0.1~\mearth, beige and orange curves), the
evolution is extremely slow. Over 10~Gyr, planetesimal masses grow by a
factor of 3 (0.01~\mearth) to 30 (0.1~\mearth). The collisional cascade is
equally weak; the dust luminosity changes little over 10~Gyr. Early in the
evolution (1--10~Myr), these systems show a small rise in \ldlstar\ due to
the extra production of small dust grains relative to the loss of pebbles.
The systems then begin a protracted decline. Were the calculations extended
to 20--30~Gyr, the decline in \ldlstar\ would speed up and eventually
join the other models on a roughly linear decline of \ldlstar\ with 
time. 

Compared to the models with $f$ = 0.5, systems with small $f$ undergo a more 
erratic evolution (Fig.~\ref{fig: ldust0}, lower panel). When massive 
rings have $f \approx 10^{-4} - 10^{-2}$, runaway growth is explosive; 
protoplanets rapidly reach super-Earth masses 
(Figs.~\ref{fig: mass0}--\ref{fig: mass1}). The dust luminosity echoes 
this behavior (black, purple, and blue curves). Initially, \ldlstar\ 
drops
by a factor of $\sim$ 2--5 as collisional damping among pebbles reduces 
the 
scale height of the smallest solids. After the runaway begins, 
protoplanets stir up smaller solids; the cascade starts to produce 
copious 
amounts of debris. With more debris and a larger scale height, the dust 
luminosity rises dramatically to $\sim 10^{-2}$ and then drops 
precipitously as the cascade depletes the rings of pebbles and smaller
solids. During the steep decline, protoplanet--protoplanet collisions 
replenish the debris and produce additional brief rises in \ldlstar. After 
a nearly Gyr-long decline, a few large collisions create enough 
debris to fuel a final, modest rise in the dust luminosity, 
$\ldlstar\ \sim 10^{-6} - 10^{-5}$ that lasts until $\sim$ 10~Gyr.

Calculations with $f \lesssim 10^{-4}$ behave in a similar fashion. Early on,
damping is more important; \ldlstar\ declines by a factor of 10--20. As
km-sized planetesimals evolve into super-Earths, rapid growth of $H$ 
allows \ldlstar\ to reach $\sim 10^{-2}$ somewhat earlier than models with
$f = 10^{-4} - 10^{-2}$. In these systems, the drop in \ldlstar\ is more
dramatic, falling well below $10^{-7}$, and the recovery is smaller than
displayed by the black, purple, and blue curves in the lower panel of 
Fig.~\ref{fig: ldust0}.

Rings with a larger initial mass in planetesimals, $f$ = 0.1--0.5, evolve
more smoothly 
(light green and orange curves). 
In these systems, \ldlstar\ rises by a factor of four during the first
10--20~Myr when planetesimals grow into Pluto-mass planets (Fig.~\ref{fig: 
mass1}). While protoplanets continue to accrete pebbles, the dust 
luminosity levels off and begins to decline. The systems then enter an
extended oligarchic growth phase from $\sim$ 100~Myr to 10~Gyr, where the 
dust luminosity declines roughly linearly with time, from  
$\ldlstar \sim 10^{-3}$ to $\ldlstar \sim 10^{-5}$. Near the end of this 
epoch, both calculations exhibit a small rise in \ldlstar\ from several 
collisions among the remaining planetesimals and protoplanets.

When $f$ = 1, the ponderous evolution of \ldlstar\ closely parallels the
slow growth of planetesimals into Mars-mass planets
(Fig.~\ref{fig: ldust0}, beige curve). During the first 
30~Myr, \ldlstar\ rises slowly from $\sim 10^{-9}$ to $\sim 10^{-6}$. 
Growth of planetesimals and a rise in debris production powers a more 
rapid rise to \ldlstar\ = a few $ \times 10^{-5}$ at 200--300~Myr. The
system then enters oligarchic growth, where protoplanets continue to grow
and the cascade grinds leftovers to dust. Unlike other systems with smaller
$f$, these rings maintain a roughly constant luminosity from 300~Myr to
10~Gyr. Eventually, the dust luminosity will decline; however, the decline
will occur after the central star leaves the main sequence.

When the initial mass in solids is smaller than 45~\mearth, the evolution of the dust
luminosity in models with $f \lesssim 10^{-2}$ is slower and less 
dramatic. The overall shape in the $L_d(t) / \lstar$ curve follows the 
examples in the top panel of Fig.~\ref{fig: ldust0}, with a fairly constant \ldlstar\
at the start of the calculation followed by a nearly linear decline. 
Superimposed
on this generic evolution is a series of spikes in \ldlstar\ generated by
debris from occasional giant collisions between protoplanets and 
planetesimals. The amplitudes of these spikes decline with decreasing
initial mass in solids. Rings with \M0\ = 10~\mearth\ at 45~au and
\M0\ = 15~\mearth\ at 75~au have large spikes 
as in the black, purple, and blue curves in the lower panel of 
Fig.~\ref{fig: ldust0}, while rings with
0.3--1~\mearth at 45~au and 0.5--1.5~\mearth\ at 75~au have modest spikes.
Lower mass systems have insignificant spikes in \ldlstar\ during the 
overall decline.

In systems with $f$ = 1, the evolution in \ldlstar\ becomes less and less
interesting with decreasing \M0. Rings with \M0\ $\gtrsim$ 1~\mearth\ 
(5~\mearth) at 45~au (75~au) have $\ldlstar \gtrsim 10^{-6}$ at 5--10~Gyr.
As \M0\ decreases from these limits, the maximum dust luminosity also
drops. At 45~au, the maximum dust luminosity falls to $3 \times 10^{-7}$ 
for 0.3~\mearth\ rings to less than $10^{-7}$ for 0.01~\mearth\ rings.
Low mass rings at 75~au are equally invisible with current technology;
all systems with \M0\ $\lesssim$ 0.5~\mearth\ have a maximum \ldlstar\
smaller than $10^{-7}$. The tables in the Appendix include the maximum
\ldlstar\ for all calculations and allow for a more extensive comparison
among the calculations.

To connect these results to previous studies, we compare
with published analytical models of collisional cascades
\cite[e.g.,][]{wyatt2002,dom2003,krivov2008,lohne2008}. In a cascade at
75~au where the radius $r_c$ of the largest object participating in the 
cascade does not change, the dust luminosity is constant at early times 
and then falls linearly with time, $\ldlstar = L_0 / (1 + t / \tau_0)$. 
However, numerical models demonstrate that $r_c$ also declines with time 
\citep[e.g.,][]{kb2017a}. Including this behavior in the analytical model 
yields a steeper decline in the dust luminosity, 
$\ldlstar = L_0 / (1 + t / \tau_0)^{1.12}$, 
where $\tau_0 = 1.12 \alpha t_0$ and $t_0 = r_c \rho P / 12 \pi \Sigma$ 
is the collision time \citep{kb2017a}. The $\alpha$ term is a function of
the ratio of the collision energy to the binding energy of planetesimals.

The dotted line in each panel of Fig.~\ref{fig: ldust0}
shows the luminosity evolution for $L_0 = 4 \times 10^{-3}$ and 
$t_0$ = 30~Myr. The dust luminosity is nearly constant for 5--10~Myr and
then declines. Although the analytical model has a somewhat larger 
\ldlstar\ at early times, it matches numerical models for massive disks 
with $f$ = 0.1--0.5 at late times. Tracking the behavior of the numerical 
models for less massive disks requires smaller $L_0$ and larger $t_0$. 

In the \citet{kb2017a} analytical model, the reference 
luminosity $L_0$ and the time scale $\tau_0$ depend primarily on $M_0$ and 
$r_c$: $L_0 \propto M_0 r_c^{-1/2}$ and $\tau_0 \propto r_c M_0^{-1}$. For 
the cascade in Fig.~\ref{fig: ldust0}, the collision velocity is just large
enough to shatter objects with $r \approx r_c$; then, $r_c \approx$
0.2--0.5~km and $M_0 \approx$ 50--55~\mearth. For fixed $r_c$, reducing 
$L_0$ by a factor of 100 yields $M_0 \approx$ 0.5~\mearth\ and 
$t_0 \approx$ 3~Gyr. This model provides a reasonable match to numerical 
calculations with $M_0 \approx$ 0.1--0.5~\mearth.

Analytic cascade models with the larger planetesimals 
expected from streaming instability models, e.g., $r_c \approx$ 
100--200~km, require unreasonably large \M0\ to approximate the evolution 
in Fig.~\ref{fig: ldust0} \citep[see also][]{shannon2011,krivov2021}. 
Adopting $r_c \approx$ 100~km instead of $r_c \approx$ 0.3~km requires 
$\sim$ 17 times more mass to achieve $L_0 \approx 4 \times 10^{-3}$. To 
match the short collision time, $\tau_0 \approx$ 30~Myr, collision 
velocities must be $\sim$ 3--4 times larger than the minimum required to 
shatter 100~km objects. While this model provides a reasonable match to 
the simulation, the high mass in solids, $M_0 \approx$ 800--900~\mearth,
makes this solution unattractive compared to the analytical model with 
small planetesimals or the numerical model with a mix of pebbles and large
planetesimals.

Although not shown in Fig.~\ref{fig: ldust0}, analytic 
models also match the results of numerical models at 45~au. 
For \M0\ = 10~\mearth\ and $r_c$ = 0.2~km, an analytical model with
$L_0 \approx 2 \times 10^{-3}$ and $\tau_0 \approx$ 25~Myr tracks the
numerical models of disks with $f$ = 0.1--0.3 for evolution times 
exceeding 30--40~Myr. Lower mass disks with larger $\tau_0$ also match the 
simulations. Compared to calculations at 75~au, the analytic model prefers 
a somewhat smaller $r_c$ and shorter $\tau_0$ to track the numerical 
results adequately at 45~au. Once again, much larger $r_c$ requires 
unreasonably large \M0.

\subsection{Gap Formation}
\label{sec: ringmodelgaps}

In addition to the dust luminosity, the coagulation calculations provide
a quantitative measure of the radial distribution of dust as a function of 
time. For low mass rings that generate little dust luminosity, 
$\ldlstar \lesssim 10^{-4}$, final protoplanet masses are typically less
than the mass of Mars. These planets tend to follow the initial surface
density distribution and cluster near the center of the ring. With little
mass in planets, the dust also has a gausssian distribution in surface 
density with a peak in the center of the ring, e.g., at 45~au and at 75~au.

More massive rings have more obvious features in the radial surface density
of the dust. The planets in these calculations do not follow the initial
surface density distribution and are more evenly distributed among the 28
annuli in each ring. Super-Earth mass planets tend to eliminate all solids
from their annuli and sometimes from adjacent annuli. These systems thus 
have 1--3~au gaps in the surface density of dust along the orbits of the
super-Earths. Rings with several super-Earths have multiple gaps.

Predicting the structure of these gaps requires a parallel set of 
coagulation and \nbody\ calculations to allow the gravity of planets 
to open gaps with sizes that depend on the mass of the planet and the 
remaining mass in smaller solids within the ring  \citep[e.g.,][and 
references therein]{kok1995,raf2001,bk2011b,bk2013}. These calculations 
are computationally expensive \citep[e.g.,][]{bk2020,kb2021a}; we defer 
them to a later study. Here, we use previous results to infer the likely
structure of gaps in the dust distribution of rings with massive planets.

For this initial exploration of gap formation, we compare the 
relative separation of protoplanet orbits to the Hill radius, 
$\rhill\ = a (M / \mstar)^{1/3}$. From analytic and numerical 
calculations, planets clear out a `ring of influence' with a radial 
extent $\delta a \approx 2 \sqrt{3} \rhill$ on either side of their 
orbits \citep[e.g.,][]{gladman1993,kok1995,raf2001}. Among pairs of 
protoplanets, those separated by more than 4~\rhill\ in semimajor axis 
do not interact dynamically \citep[e.g.,][]{kok1995,chambers1996,
weiden1997b,chambers2001a,bk2006,
kb2006}; each carves out its own gap with an extent comparable to 
$\delta a$. Protoplanets on closer orbits interact dynamically and 
produce large gaps in an `interaction region' defined by the extent of 
their chaotic orbits prior to a merger or ejection event. This region 
has a width of 2--3$\delta a$.

\begin{figure}[t]
\begin{center}
\includegraphics[width=5.5in]{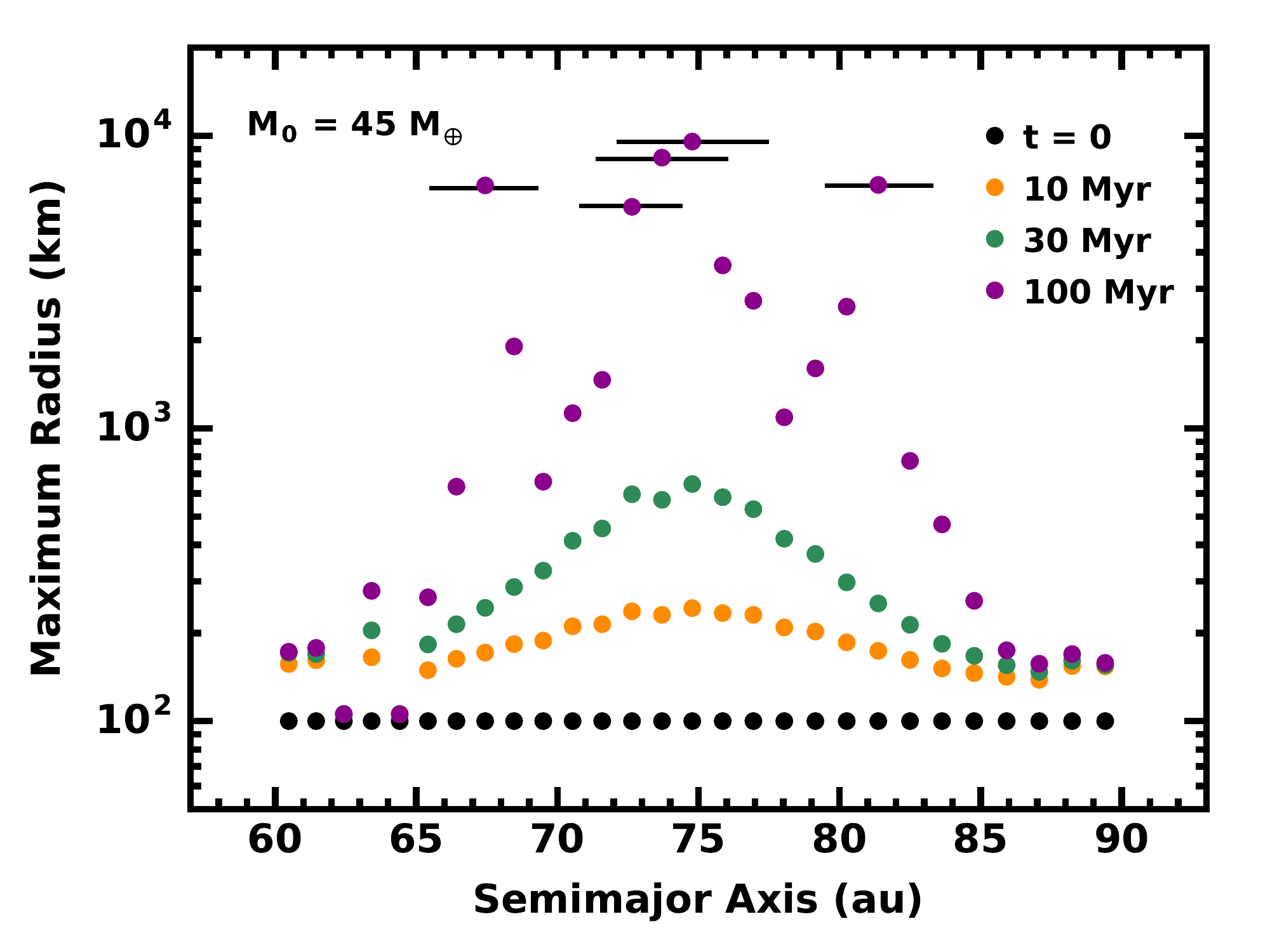}
\vskip -2ex
\caption{
\label{fig: ann0}
Radius of the largest object in each annulus for rings of solids with
\M0\ = 45~\mearth\ and $f = 0.01$ for times listed in the legend. 
Horizontal lines extending from several points in each panel indicate the extent of the `ring of influence', which has a radial extent of 
3.5~\rhill\ around the orbit of each planet. Rings of influence for all protoplanets are omitted for clarity.
At the center of the grid, the largest planetesimals 
slowly grow from 100~km to 200 km (10~Myr) to 700~km (30~Myr). Runaway 
growth leads to the formation of four protoplanets with $M \gtrsim$ 
0.1~\mearth\ and another with $M \approx$ 0.05~\mearth. The overlapping 
rings 
of influence suggest dynamical interactions will begin a phase of chaotic 
growth where the largest protoplanets merge into one or two super-Earths.
}
\end{center}
\end{figure}

Fig.~\ref{fig: ann0} illustrates the evolution for a calculation with 
\M0\ = 45~\mearth\ and $f$ = 0.01. With a surface density maximum in the 
middle of the ring, planets grow fastest (slowest) at 75~au (60--65~au and 
85--90~au). Once 500--600~km protoplanets form at 30~Myr, runaway growth 
begins. Over the next 70~Myr, the five fastest-growing protoplanets each 
surpass the mass of Mars. 
Smaller protoplanets have much smaller masses, $\sim$ 1--30~\mp. The 
horizontal lines in the figure illustrate the extent of the rings of
influence for each of the five largest protoplanets. Rings of 
influence for the two most massive protoplanets 
at 73--75~au overlap and contain a lower mass protoplanet
at 72~au. Rings of influence for the other two protoplanets at 
67~au and at 81~au contain several much smaller planetesimals but no 
other massive protoplanet.

The close proximity of the two largest protoplanets in 
Fig.~\ref{fig: ann0} has two observable outcomes. Initially, the two 
central protoplanets interact dynamically, scattering smaller solids 
out of their orbits and trying to move to a larger separation. This 
interaction involves the protoplanet at 72~au and begins a period 
of chaotic growth, where the three protoplanets at 72--75~au move 
chaotically through the grid and sweep up smaller objects along their 
orbits. Eventually, this process involves the outer two protoplanets 
at 67~au and 81~au. Subsequent collisions and mergers among the 
five largest protoplanets,
the smaller protoplanets with radii $\sim$ 1000--4000~km, and the 
smaller planetesimals are likely to leave behind one or two super-Earth 
mass planets \citep[e.g.,][]{gold2004,kb2006}. Throughout chaotic growth, 
the ring of smaller solids expands radially inward and outward. Thus, at 
$\gtrsim$~200~Myr, ALMA observations would reveal thermal emission from a 
larger ring of small solids with a central depression where the remaining 
super-Earths orbit.

In this example, it seems likely that the size of the gap will be larger
than the standard $4\sqrt{3} \rhill$ expected for a {\it single} massive
planet in a ring of small solids. The five protoplanets at 67--82~au
would experience chaotic growth within an `interaction region' extending 
from $\sim$ 65~au to $\sim$ 83~au. Once chaotic growth ends, this region
would have few small solids. Instead of the $\sim$ 50~\mearth\ single
planet required to create such a large gap, this gap would contain 1--2
super-Earths with a total mass $\sim$ 5~\mearth. 

\begin{figure}[t]
\begin{center}
\includegraphics[width=5.5in]{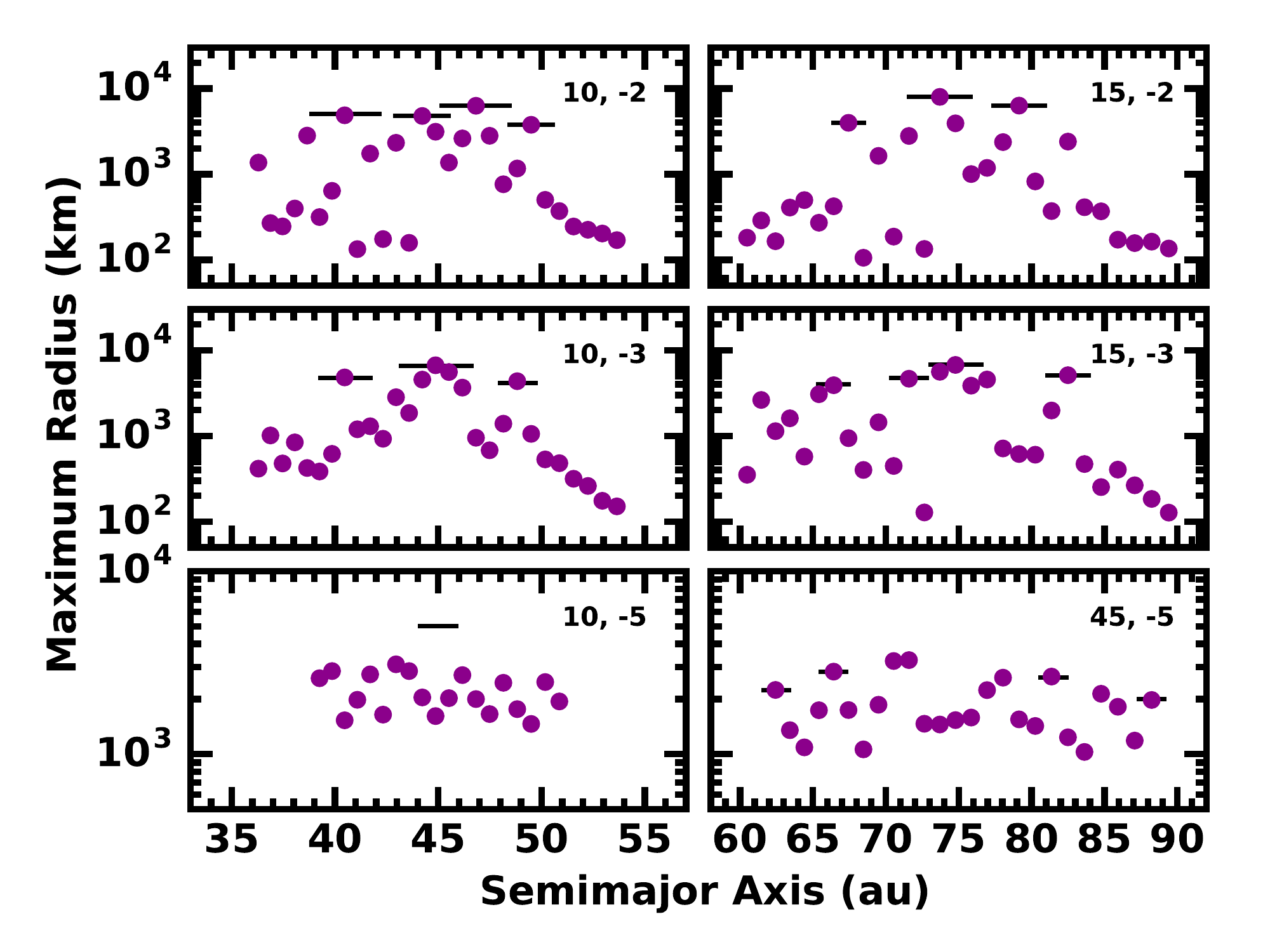}
\vskip -2ex
\caption{
\label{fig: ann1}
Snapshots of the largest protoplanets in each annulus at 1--10 Gyr
for rings at 45~au (left panels) and at 75~au (right panels). The
upper right corner of each panel lists \M0\ and log $f$. Horizontal 
lines indicate rings of influence for several protoplanets in each panel.
Rings of influence for all protoplanets are omitted for clarity.
}
\end{center}
\end{figure}

Among the suite of calculations with \M0\ = 10~\mearth\ at 45~au and
\M0\ = 15--45~\mearth\ at 75~au, many have protoplanets with overlapping 
rings of influence (Fig.~\ref{fig: ann1}). At 45~au (left panel), two 
systems have as many as a half dozen protoplanets within an interaction 
region (top left and middle left panels). Another system has $\sim$ 20 
large objects in an interaction region that takes up most of the ring. 
Calculations at 75~au yield similar systems where a few or many massive
objects will interact chaotically. Some may yield 1-2 super-Earths; others
may produce many Mars-mass planets.

In each of these calculations, we expect that chaotic evolution will
produce gaps and perhaps narrow rings of small solids within broader,
more diffuse rings. With many massive protoplanets evolving within the
interaction region, the final sizes of dark gaps may be much larger than 
expected from the final masses of protoplanets.

Lower mass rings are unlikely to have much structure. The protoplanets in
these systems are not massive enough to scatter small solids out of their
orbits and cannot generate a significant gap in the ring. With little 
scattering, the final extent of the ring should be similar to the initial
extent. 

\section{Discussion} \label{sec:discussion}

The models described in Section 3 illustrate how the collisional evolution of
rings of solids---composed of pebbles and planetesimals---follow diverse evolutionary histories depending on the efficiency of planetesimal formation
($f$) and the initial mass in solids ($\M0$). These parameters establish 
whether rings can grow Pluto-, Mars-, or super-Earth-mass planets in 1 Myr to 
10 Gyr (Section 3.3; Figs.~\ref{fig: mass0}--\ref{fig: mass2}; see also 
Appendix Tables \ref{tab: tab1}--\ref{tab: tab2}). Diverse dust luminosity 
histories result (Section 3.4; Fig.~\ref{fig: ldust0}). The growing planets 
should create gaps in the radial distribution of solids that reflect their 
mass and growth history (Section 3.5; Figs.~\ref{fig: ann0}--\ref{fig: ann1}). 

\begin{figure}[t]
\begin{center}
\includegraphics[width=5in]{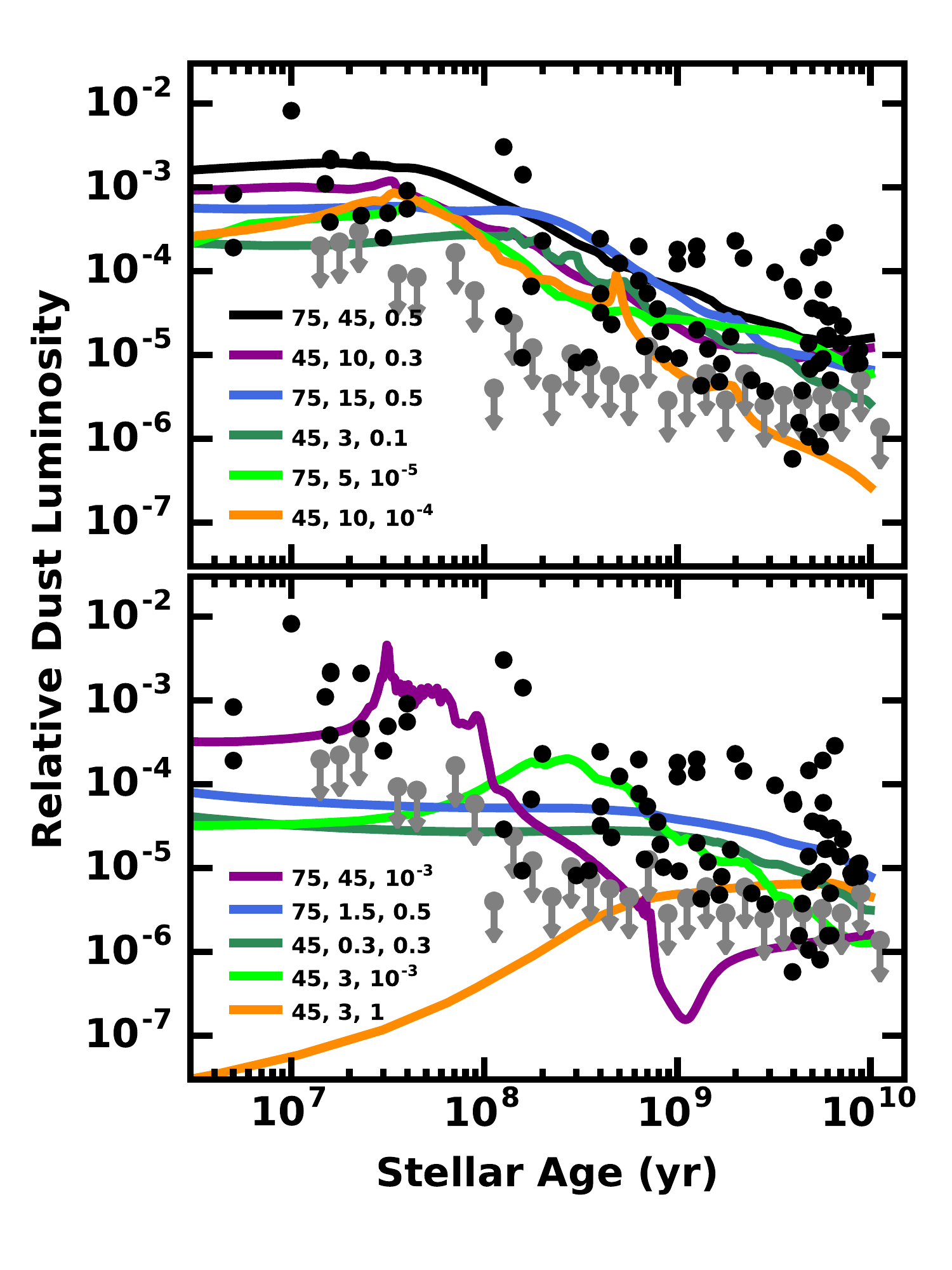}
\vskip -2ex
\caption{
\label{fig: ldustcomp0}
Comparison of the relative dust luminosity derived from the numerical 
calculations (solid lines) with detections (black points, instead of 
blue, green and purple points) and upper limits (grey symbols) from 
Fig.~\ref{fig: alldata}. The legend associates values of $a_0$, \M0, and 
$f$ with each model curve. In the upper panel, a set of `bright stalwart' models 
passes through the ensemble of data points. In the lower panel, `early flare' (purple curve), `steady glow' sources (blue, light green, and dark 
green curves), and `late bloomers' (orange curve) pass through the data 
for young stars (ages $\lesssim$ 100~Myr, early flare) or old stars (ages 
$\gtrsim$ 100~Myr, steady glow sources and late bloomers) but not both. 
}
\end{center}
\end{figure}

Here we compare the model results from Section 3 with observations of debris disks to identify the plausible initial conditions and evolutionary paths that connect protoplanetary disks to debris disks (Fig.~\ref{fig: schema}).
We also associate planet formation outcomes---Pluto,
Mars, or (super-)Earth---with the various paths to debris disks.
Fig.~\ref{fig: ldustcomp0} compares the fractional luminosity \ldlstar\ of the ring models (colored lines) with the observed \ldlstar\ values of debris disks described in Section 2. In each panel, the colored dots from Fig.~\ref{fig: alldata} are reproduced as black dots; grey arrows repeat the median upper limits from Fig.~\ref{fig: alldata}.
The comparison demonstrates that the 
evolution of massive rings of pebbles and planetesimals at 45--75~au 
plausibly explains the observed dust luminosities of known debris disks
with stellar ages of 10~Myr to 10~Gyr. 

\subsection{Bright young disks}

As shown in the two panels of Fig.~\ref{fig: ldustcomp0}, the population of bright young disks ($\ldlstar \gtrsim 10^{-3}$ at 10--100 Myr) is best matched by models of high mass rings (3--45 $\mearth$) with a wide range of planetesimal formation efficiencies ($f$ = 0--0.5). In models with $f \lesssim 10^{-2}$, occasional collisions among large protoplanets create copious debris and pronounced spikes in \ldlstar\ (Fig.~\ref{fig: ldustcomp0}, purple curve
in the lower panel). The largest spikes rival the observed \ldlstar\ of the brightest debris disks with ages of 10--100~Myr (upper panel; 75~au). Fainter debris systems in this age range 
are well matched by rings with smaller masses, smaller radii, or smaller $f$ (Fig.~\ref{fig: ldustcomp0}, upper panel). 

In contrast to these evolutionary paths, rings with similar or larger masses, but completely efficient planetesimal formation ($f$=1) produce very little debris at early times (Fig.~\ref{fig: ldustcomp0}, orange curve in the lower panel). If planetesimals form with such high efficiency, matching the observed fractional luminosities of the bright young disks requires very large initial masses \citep[$\sim 1000$ \mearth;][]{shannon2011,krivov2021}. As shown here, such large masses are unnecessary; more modest planetesimal formation efficiencies and the typical masses of observed protoplanetary disk rings can account for the observed properties of bright young disks.

\subsection{Fainter old disks} 

\begin{figure}[t]
\begin{center}
\includegraphics[width=7.0in]{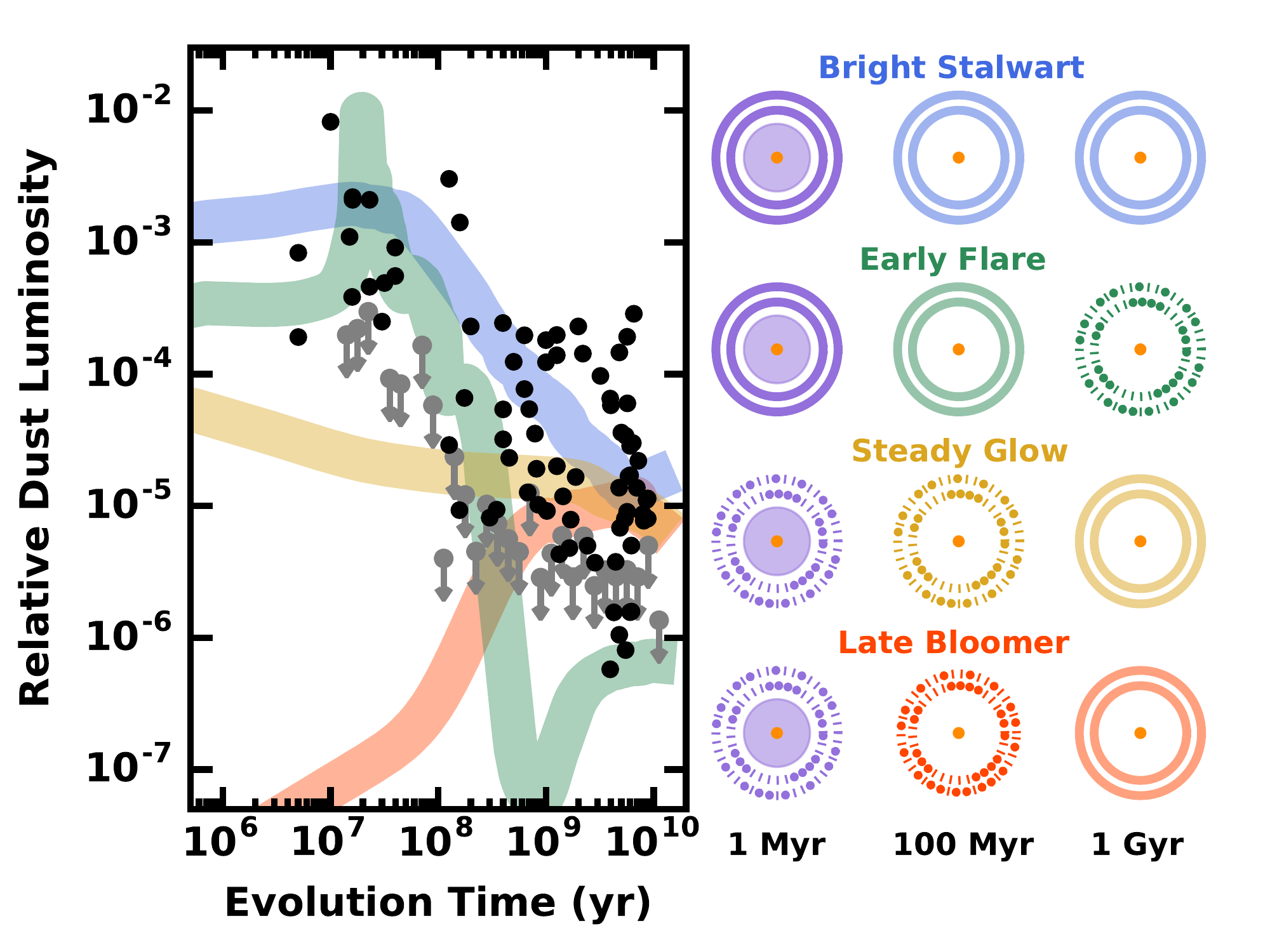}
\vskip -2ex
\caption{
\label{fig: fourpaths}
The four generic pathways to produce cold debris disks 
around solar-type stars with $a \gtrsim$ 40~au.
{\it Left Panel:} filled black circles and grey symbols plot observed dust 
luminosities from  Fig.~\ref{fig: alldata}. Four thick solid curves
illustrate the evolution of \ldlstar\ for four types of models. In the 
`bright stalwart' path, models with $\M0\ \gtrsim$ 5~\mearth\ and $f \approx$ 
0.1--0.5 approximately match debris disk observations at any stellar age. 
Massive disks with $f \lesssim 10^{-2}$ (the `early flares') match 
observations well for young systems with ages $\lesssim$ 50--100~Myr but fade 
into obscurity for older stars. Massive disks with $f$ = 1 (`late 
bloomers') have very small \ldlstar\ until stellar ages $\gtrsim$ 1~Gyr
and gradually rise to an observable level at later times. In the 
`steady glow'
pathway, low mass disks have \ldlstar\ just below detectable levels for
stellar ages $\lesssim$ 100~Myr; their relatively constant \ldlstar\ over
time allows them to match observations for older stars.
{\it Right panel:} schematic evolution in the four pathways. At 1~Myr,
bright stalwarts and early flares have bright rings at $a \gtrsim$ 40~au; 
steady glow sources have low masses of solids in rings at large $a$ that 
are undetectable with current instrumentation. 
Debris rings in late bloomers have much smaller \ldlstar\ than steady
glow sources and are also invisible.
At 100~Myr,
bright stalwarts and early flares have bright debris disks; late bloomers 
and steady glow sources have undetectable debris rings. At 1~Gyr, the debris rings 
in late bloomers, steady glow sources, and Bright stalwarts are bright; 
early flares
have undetectable amounts of debris in rings.
}
\end{center}
\end{figure}

Debris disks trend fainter with age, with $\ldlstar \sim 10^{-5}$ at ages beyond 1 Gyr (Figs.~\ref{fig: excess}, \ref{fig: alldata}, and \ref{fig: ldustcomp0}). The downward trend of $\ldlstar$ with age is readily explained 
by the simple fading of the bright young disks of Section 4.1.
Rings with $f \approx$ 0.01--0.5 follow the classical evolution in dust
luminosity of $\ldlstar \propto t^{-1}$ \citep[see also][]{wyatt2002,
dom2003,krivov2008,lohne2008,wyatt2008,kb2017a} 
via the evolutionary pathway illustrated by nearly all of the curves in 
the upper panel of Fig.~\ref{fig: ldustcomp0}. We identify this evolution
as the ``bright stalwart'' pathway shown in 
Fig.~\ref{fig: fourpaths}.
This interpretation of the known debris disks---that they arise from the 
fading of a population of initially massive disks partially composed of 
pebbles and planetesimals---is consistent with the similar fraction of 
debris disks and protoplanetary disk rings at each evolutionary age ($\sim 
25$\%; Section 2). Thus, the bright stalwart pathway accounts for both 
the incidence rates and luminosities of the known debris disks 
in the age range $\sim$ 50~Myr to 10~Gyr.

The general agreement of this evolutionary path with the observed properties of debris disks 
potentially 
limits the possible role of other evolutionary histories. For example, the fainter old disks could also be explained with a population of low mass rings ($\sim 1~\mearth$) with modest $f$ that are moderately bright at early times ($\ldlstar \sim 10^{-4}$) and evolve more horizontally in 
Fig.~\ref{fig: ldustcomp0}, reaching $\ldlstar \sim 10^{-5}$ at 10 Gyr. This pathway---which we refer to as `steady glow'---is illustrated by the green and blue curves in the lower panel of Fig.~\ref{fig: ldustcomp0} (also the gold curve in Fig.~\ref{fig: fourpaths}). Alternatively, high mass disks (10--40 \mearth) made entirely of planetesimals pursue a stealthy evolutionary path that is extremely faint at early times ($\ldlstar < 10^{-6}$) and brightens at late times to currently observable levels of $\ldlstar \sim 10^{-5}$. 
The orange curve in the lower panel of Fig.~\ref{fig: ldustcomp0} (also
in Fig.~\ref{fig: fourpaths}) shows this `late bloomer' pathway. 

If we assume that $\sim 25$\% of stars start out with massive protoplanetary disk rings (Section 2) and subsequently follow the bright stalwart path, we can account for the observed luminosities and incidence rates of debris disks with age; this assignment leaves little room for significant contribution from the steady glow and late bloomer pathways. If these were important pathways, each populated roughly equally to the 
classical bright stalwart pathway, 
the incidence rate of debris disks at early times (from the 
bright stalwart pathway) would be three times smaller than the rate at late
times when the steady glow and late bloomer populations become detectable with $\ldlstar \sim 10^{-5}$ at $\sim$~1~Gyr.
We can further rule out a significant 
steady glow population, because it would also overproduce sources at ages of 0.1--1~Gyr with $\ldlstar \approx 10^{-4}$. The debris disk incidence rate in this age and luminosity range is restricted by the {\it Herschel} survey upper limits (Section 2). Future debris disk surveys that reach fainter luminosity limits at ages 10--300~Myr can provide additional, direct constraints on the steady glow ($\ldlstar \sim 10^{-5}$) and late bloomer 
($\ldlstar \sim 10^{-6}$) populations.

One can imagine combining the evolutionary tracks from Section 3 in other ways.
For example, if the incidence of cold debris disks at 10--30~Myr is as 
large as the 50\% rate derived for the small population of F stars in the 
20--25~Myr old $\beta$ Pic moving group \citep{pawellek2021}, a significant population of disks must rise to high dust luminosity within a few tens of Myr before fading significantly beyond 100 Myr. This behavior, which we identify as the `early flare' pathway, could
explain the high incidence rate of cold debris disks at young ages and the 
much lower rate among stars with ages $\gtrsim$ 50~Myr. In the calculations, this behavior occurs in rings with high masses and
$f \lesssim 10^{-3}$, where
the combined impact of damping, rapid planet growth, 
and an efficient cascade results in a system where \ldlstar\ rises from 
$\lesssim 10^{-4}$ to $\sim 10^{-2}$ and declines back down to $\lesssim 10^{-4}$ on time
scales of 100~Myr (Fig.~\ref{fig: ldust0}, lower panel, black, purple, and blue curves; Fig.~\ref{fig: fourpaths}, green curve). The late decline in \ldlstar\ in these models is as fast or faster than
the $\ldlstar\ \propto t^{-2}$ required to eliminate the descendants of the
$\beta$ Pic stars from the DEBRIS and DUNES samples at ages of $\sim$ 1~Gyr
\citep{pawellek2021}.

We might further imagine that in addition to the
early flare pathway 
producing the vast majority of the luminous debris disks at ages 
$\lesssim$ 40--50~Myr, 
we also have 
the steady glow and late bloomer pathways 
dominating among much
older stars. With this combination, nearly all of the early flares need to
fall below current detection limits before rings in the steady glow and 
late bloomer pathways begin to contribute to the population. Contributions 
from the classical bright stalwart pathway could `smooth over' the 
transition between these pathways.

Invoking
three pathways---early flare, steady glow, and late bloomer---faces 
several hurdles. Selecting the proper mix to ensure a high incidence rate at
the youngest ages and to maintain a roughly constant rate of $\sim$ 25\% for 
all older stars would require some fine tuning for ages where early flares are
fading away and the other pathways first become detectable. 
In addition,
the required mix 
of initial conditions: massive disks with $f \approx$ 0 for early flares,
massive disks with $f$ = 1 for late bloomers, and intermediate-mass disks with
$f \approx$ 0.1--0.5 for steady glow sources
seems unlikely, unless there are physical mechanisms that can produce such dramatically different planetesimal efficiencies in different disks.
The larger parameter space of $f$ that produces
the classical bright stalwart sources seems more plausible and more in line with the predictions of
simulations of the streaming instability \citep[e.g.,][]{rucska2021}.

We can also distinguish these pathways with a different approach, 
by quantifying the disk substructure created by any massive, embedded planets that form.
Massive rings (10--40 \mearth) that follow the classical bright stalwart pathway ($f$=0.1--0.5; Fig.~\ref{fig: fourpaths}, blue curve) 
or the late bloomer path ($f$ = 1; Fig.~\ref{fig: fourpaths}, orange curve)
would build Mars-mass planets by 10 Gyr.
In contrast, massive rings with lower $f$ that ``burn bright, fade fast'' (the early flare pathway; $f \lesssim 10^{-3}$; Fig.~\ref{fig: fourpaths}, green curve) would create more massive objects---super-Mars to super-Earth mass planets---in 20--30~Myr (Fig.~\ref{fig: mass1}). Modest-mass rings that follow the roughly horizontal steady glow pathway ($\sim$1--5 \mearth, $f$=0.3--0.5; Fig.~\ref{fig: fourpaths}, gold curve) would only build Pluto-mass objects by 10 Gyr (Fig.~\ref{fig: mass2}). 

The gaps created by the more massive planets could be resolved spatially. If a ring with a radius of 75 au creates a planet with the mass of Earth, Mars, or Pluto, the planet would open a gap with a fractional width of at least $2da/a = 4{\sqrt 3}(M_p/M_\odot)^{1/3}$ (Section 3), which corresponds to an angular width of 54 mas, 24 mas, and 8 mas at a distance of 140 pc. In comparison, the ngVLA is anticipated to deliver angular resolutions of 0.5 mas to 50 mas at wavelengths of 2.6 mm to 25 cm
\citep[e.g.,][and references therein]{matthews2018,tobin2018,chalmers2020}.

Thus, future observations could test the hypothesis that the classical 
bright stalwart evolution is the primary pathway for debris disks, with 
few systems pursuing either the late bloomer or the steady glow paths.
In other words, observations could
probe whether protoplanetary disks are born with
massive, initially dark rings of planetesimals (late blooming rings) or initially modest-mass rings of pebbles and planetesimals (steady glow
rings).

\subsection{The Story of Solids in Disks}

If bright stalwarts 
dominate the evolutionary pathways for debris disks, one simple interpretation of the results described here is that 
extended and compact 
protoplanetary disks evolve differently, i.e., disks evolve horizontally in Fig.~\ref{fig: schema}. 
The 25\% of T Tauri 
systems with large ringed protoplanetary disks 
produce detectable debris throughout their lives, from  
$\sim 10$~Myr to 10 Gyr, and are the showy, attention-grabbing celebrities of the debris disk world. In contrast, the majority of T Tauri systems with compact protoplanetary disks live quieter, tidier lives. Few of them are born with `late-blooming' massive, initially dark rings of planetesimals or `steady' modest-mass rings of pebbles and planetesimals. That is, if rings are typically a mixture of pebbles and planetesimals at the end of the protoplanetary disk phase, then compact disks typically leave behind $< 1 \mearth$ of solids at large radii, a small reservoir that produces little debris over the lifetime of the star. 
Instead of each disk evolving along the classical path of 
Class 0/I/II/III sources into debris disks
\citep[e.g.,][]{cieza2007,wahhaj2010,will2011,hardy2015},  
this interpretation
implies that the  `
known debris disks are 
a chapter in the history of only a subset of low mass stars.

While most current observations support this simple picture, the true story may be more complex if the high incidence rate of cold debris disks among the small sample of F stars in the $\beta$ Pic moving group is
typical of all 20--25~Myr old solar-type stars.
As noted earlier, 
a high incidence rate of debris that persists for a 
short time may imply that a significant fraction of the compact protoplanetary disks actually possess massive rings of solids with very small $f$ that are also undetectable because of their very small initial scale height (section 3.2). These disks would evolve into bright early flare sources that appear suddenly on the debris disk stage and quickly fade below current detection limits.

This potential complexity aside,
the divergent debris-production histories of large and small protoplanetary disks connect back to the more fundamental question of what sets the initial distribution of solids in protoplanetary disks. One possibility is that some protoplanetary disks are born large and others small, a consequence of the initial angular momentum of the cloud core and the extent to which angular momentum is shed (or transported) as collapse proceeds \citep[e.g.,][and references therein]{tsc1984,matsumoto1997,
basu1998,yorke1999,nakamura2000,krasnopolsky2002,
tscharnuter2009,joos2012,tomida2015,hennebelle2016,zhao2020}. 
Alternatively, most disks may be born with similar (large) sizes, but some experience greater inward migration of solids and others do not. Disks that create planets (or other disturbances) early on, at large radii, can induce pressure bumps that trap solids and prevent inward migration
\citep[e.g.,][]{pinilla2012,zhu2014,vandermarel2018}. If bright stalwarts
dominate the production of debris disks,  
one or more of these
scenarios are efficient in concentrating or placing a significant solid mass (10--40 \mearth) in rings at large radii in about 20\%--25\% of disks. 

Our analysis and results complement ideas discussed in the literature. 
In a recent examination and interpretation of the dust masses and luminosities of protoplanetary and debris disk sources, \citet{michel2021}
proposed a similar picture to the one described here: they hypothesized that 
debris disks are the descendants of large structured protoplanetary disks, which preserve their solids against inward radial drift through the action of dust trapping in pressure bumps. 
Their compact protoplanetary disk counterparts result from disks without such pressure bumps, whose solids drift inward to small disk radii.
In support of their picture, they 
noted the similar sizes of protoplanetary disk structures (e.g., the central cavity radii of transition disks) and the blackbody radii of debris disks.  
However, they also offered the caveat that their protoplanetary and debris disk samples did not span the same spectral types, with the debris disks skewed to earlier-type stars.
They also asked for modeling that would support their hypothesized scenarios.

Our study complements this work by assembling protoplanetary and debris disk data sets that are matched in stellar mass and by comparing orbital distances (of protoplanetary rings and debris) and their incidence rates as a function of age.  
Our study also provides the detailed modeling that supports the hypothesis of Michel et al. 
We confirm that rings of solids with the properties inferred for protoplanetary disk rings (sizes, solid masses) can indeed evolve to produce the observed luminosities of known debris disks. Moreover, the approximately constant incidence rates of protoplanetary rings and debris disks as a function of age, when combined with our models,  
leaves little room for non-structured (i.e., compact) protoplanetary disks to sustain distant reservoirs of solids $> 1\mearth.$ This result supports the underlying assumption in Michel et al.\ (and many other studies) that ``what you see is what you get'', i.e., that protoplanetary disks with compact millimeter {\it emission} have a compact solid {\it mass} distribution. 

Our results also overlap---but may be less compatible---with the picture
described by \citet{vandermarel2021}.
If they are correct that giant planets create protoplanetary disk rings and eventually migrate in close to the star where they are detected as transiting and radial velocity planets, the radial distribution of the planetesimals that result may not be ring-like. That is, if a giant planet creates a pressure bump
outside its orbit where small solids collect and planetesimals form, the pressure bump moves with the planet as it migrates inward, whereas the planetesimals are poorly coupled to the gas and are left behind. If small solids are continually captured by the pressure bump and planetesimals form, the radial distribution of planetesimals is broadened as the pressure bump migrates. If the giant planets migrate from $\sim 40$ au to $< 1$ au and planetesimals form continuously behind the migrating planet, a very broad disk of planetesimals can result 
\citep[e.g., Figure 2 of][]{shibaike2020}.

Such a broad distribution of planetesimals would eventually generate debris at increasing orbital distance with age \citep[e.g.,][]{kb2008,kb2010}, a 
trend that is not observed in the debris disk population 
\citep[e.g.,][see also Fig.~\ref{fig: extent}]{najita2005,kennedy2010,matthews2014,hughes2018}.
In addition, if small solids migrate inward with the pressure bump, away from the planetesimals they create, the planetesimal-pebble mixture (i.e., the planetesimal mass fraction) will be altered. 
The scenario we explore relies on the longevity of narrow rings (of planetesimals and pebbles with an appropriate mixture) at large radii to account for the properties and demographics of the known debris disk population.

\subsection{Connection to Planetesimal Formation}

\begin{figure}
\begin{center}
\includegraphics[width=6.5in]{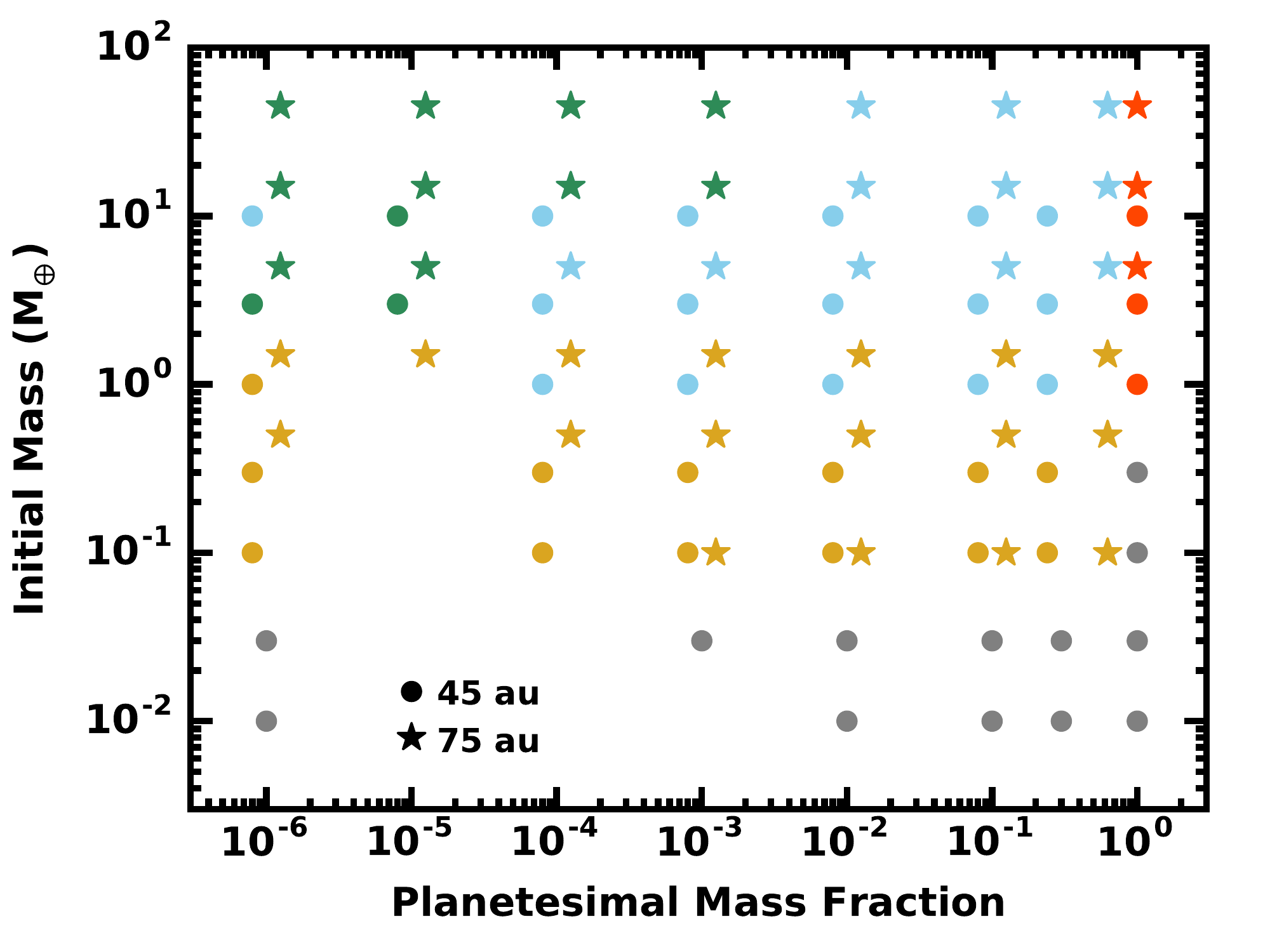}
\vskip -2ex
\caption{
Classification of model results as a function of the planetesimal mass
fraction $f$ and the initial mass in solids \M0. The legend links symbols
to $a_0$ for each calculation. Colors refer to the four debris disk 
pathways summarized in Fig.~\ref{fig: fourpaths}: early flare (green),
bright stalwart (blue), steady glow (gold), and late bloomer (orange). 
Grey symbols indicate models that failed to produce an observable \ldlstar\
with {\it Herschel} (see also Tables~\ref{tab: tab1}--\ref{tab: tab2}).
To avoid overlap, some points have been displaced horizontally.
}
\label{fig: Mf-summary}
\end{center}
\end{figure}

To place the results of Fig.~\ref{fig: fourpaths} in the context of 
planetesimal formation theories, we classify each calculation in terms
of the four evolutionary pathways. Fig.~\ref{fig: Mf-summary} summarizes
the results. Rings with (i) \M0\ $\lesssim$ 0.03~\mearth\ and any $f$ or 
(ii) \M0\ = 0.1-0.3~\mearth\ and $f$ = 1 never generate dust in amounts 
detectable with {\it Herschel} (gray symbols). More massive rings 
(\M0\ $\gtrsim$ 1~\mearth) with $f$ = 1 follow the late bloomer pathway (orange symbols). 
Although the maximum dust luminosity of late bloomers grows with \M0, 
\ldlstar\ never reaches detectable levels before $\sim$ 1~Gyr.

Among the (\M0, $f$) combinations considered here, steady glow and 
bright stalwart outcomes are the most numerous. Rings with \M0\ = 
0.1--1~\mearth\ and any $f \ne$ 1 usually follow the steady glow pathway (gold symbols).
More massive systems (\M0\ $\gtrsim$ 3~\mearth) with $f$ = 0.01--0.5 are
bright stalwarts (blue symbols). The boundary between these two outcomes depends on $a_0$.
When the ring is closer to the host star (e.g., $a_0 \sim$ 45~au), dust
intercepts a larger fraction of stellar radiation and brightens earlier in 
the evolutionary sequence. The dust in more distant rings (e.g., $a_0 \sim$
75~au) intercepts less stellar radiation; this lower dust luminosity falls
below {\it Spitzer} detection limits for stellar ages of 10--100~Myr.
Somewhat more massive rings, $\sim$ 3~\mearth\ at 75~au, are more luminous
at 10--100~Myr and remain detectable throughout their evolution. 

For rings with \M0\ $\gtrsim$ 3--5~\mearth\ and $f \lesssim 10^{-2}$, the 
early evolution of massive solids is stochastic. Sometimes, ensembles of 
1000--2000~km protoplanets undergo a series of mergers that produce 
several super-Earths and generate a rapid rise in \ldlstar\ on time scales
of 10--30~Myr. However, this early flare evolution (green symbols) depletes the system 
of the intermediate mass solids that fuel the collisional cascade; 
\ldlstar\ then drops quickly. When mergers of protoplanets are uncommon, 
the system maintains a plentiful supply of intermediate mass solids, which 
power a more slowly evolving cascade where the dust luminosity rises and
then declines more slowly. On time scales $\lesssim$ 50--100~Myr, this 
bright stalwart evolution generates super-Mars-mass planets but not 
super-Earths.

A preference for the bright stalwart pathway instead of the early flare
and late bloomer tracks is consistent with current understanding of 
planetesimal formation. In recent numerical studies of the streaming 
instability \citep[e.g.,][]{simon2016,li2018,abod2019,carrera2020,
klahr2020,gole2020,rucska2021}, the planetesimal formation efficiencies 
required for bright stalwarts ($f \approx$ 0.1--0.5) are more common than
either $f \lesssim 10^{-2}$ (early flare) or $f \approx 1$ (late bloomer).
Simulations that provide stronger constraints on $f$ would enable better 
estimates of the importance of each of the four pathways outlined in Fig.~\ref{fig: fourpaths}.

The results derived here suggest a way to test numerical simulations
observationally. Although rings with a broad range of $f$ can match a
specific measurement of \ldlstar, robust detection of substructure
within a ring may provide useful limits on $f$. For example, if bright 
debris disks ($\ldlstar \sim 10^{-3}$) at an age of 10--30 Myr commonly 
have a gap that is wide enough to require a several Earth-mass planet,
the numerical calculations discussed here require $f \lesssim$ 0.01 
in order to form so massive an object so quickly (Fig.~\ref{fig: ldust0}, lower 
panel; Fig.~\ref{fig: times} in the Appendix). 

If future observations find a consistent lack of gaps
in bright, young debris disks, the results would point to the larger $f \gtrsim 0.01$ of
bright stalwarts, which can form Earth-mass planets at much later times ($\sim 100$ Myr; Fig.~\ref{fig: mass1}, upper 
panel). As a result,  finding both a lack of gaps in bright, young debris disks and robust
detections of gaps in older debris disks would provide strong support for
$f \gtrsim$ 0.01 and the bright stalwart pathway.

\subsection{Caveats and Open Questions}

The calculations considered here employ standard well-tested techniques and
begin with starting conditions that are consistent with observations (\M0)
and theory ($f$). In a gaussian ring with a constant 
gas to dust ratio, neglect of radial drift from gas drag has little impact 
as drift velocities are small. 
The initial orbital parameters, $e = 10^{-3}$ and 
$\imath = 5 \times 10^{-4}$, are plausible; however, other options are 
possible. If cm-sized pebbles are well-coupled to 
the gas, they would have a large vertical scale-height and a larger 
initial inclination than considered here \citep[e.g.,][]{chiang2010,
riols2018,krapp2020}. As the gas dissipates over 
several Myr, pebbles decouple from the gas; damping from collisions and
residual gas would then act rapidly and reduce $e$ and $\imath$ to values
similar to the starting conditions considered here. The growth of larger
solids and the evolution of \ldlstar\ in the four pathways would differ
little from our description.

If interactions with the gas or other physical 
processes should produce smaller $e$ and $\imath$ than considered here,
outcomes could change dramatically.
Except for models with $f$ = 1, all sequences would 
begin with smaller \ldlstar. Solids in rings following the late bloomer 
and steady glow pathways would adjust on 10--100~Myr time scales and then
follow the evolution described above. Within the bright stalwart and
early flare pathways, planetesimals would have larger gravitational
focusing factors and grow more rapidly. While it would take a little extra
time for \ldlstar\ to begin to rise, these systems might produce larger 
planets and brighter debris disks. The early flare tracks might be more
peaked and fade more rapidly; the bright stalwart tracks might be somewhat
brighter but would fade on time scales similar to those described above.
In both pathways, it might be easier for high resolution observations of 
the debris to identify dark gaps and bright rings due to the more energetic
early stages of planet formation.

Aside from including the \nbody\ component of \orch, 
other changes to the approach (e.g., fragmentation parameters, number
of annuli per ring, or number of mass bins per annulus) are unlikely to
change the outcomes significantly. As discussed in the
Appendix (section B), modifying the fragmentation parameters yields factor 
of two changes in \ldlstar. Previous tests of the coagulation code 
demonstrate that improving the mass and spatial resolution of a calculation
produces similarly small changes in outcomes as a function of evolution
time \cite[e.g.,][]{kb2016a,kb2017a}.

\section{Summary}
\label{sec: summary}

Rings of solids at $a \gtrsim$ 30--40~au offer a way to resolve a long-standing disconnect between
detailed evolutionary models of debris disks and their observed properties.
The lack of a strong trend in the orbital distance of debris with age 
\citep[e.g.,][see also Fig.~\ref{fig: extent}]{najita2005,kennedy2010,matthews2014,hughes2018},
which is predicted by earlier generations of planet formation models, 
is readily explained if the parent bodies that produce the debris are initially distributed in discrete rings rather than over a broad range of orbital radii. 
The prominent rings that are commonly observed in T Tauri (protoplanetary) disks, which have an incidence rate similar to that of cold debris disks around low mass stars ($\sim$ 20--25\%), suggest a compelling starting point for debris disk evolution (Section 2). 

We have explored the potential connection between protoplanetary rings and debris disks, using a new set of evolutionary calculations that follow the evolution of rings of solids spanning a range of initial properties.
The results show that diverse evolutionary histories are possible as a function of planetesimal formation efficiency $f$ and initial solid mass $\M0$ (Section 3). Depending on these parameters, rings of solids can grow Pluto-, Mars-, or super-Earth planets in 0.01--10 Gyr (Section 3.3). The resulting dust luminosity histories are also diverse and fall into four main pathways (Fig.~\ref{fig: fourpaths}): 
an always-bright classical evolutionary pathway that encompasses the known debris disks (``bright stalwart''); tracks that burn bright and fade fast (``early flare''); those that maintain a relatively constant, lower luminosity from 10 Myr to 10 Gyr (``steady glow''); and those that brighten dramatically at late times to a detectable luminosity (``late bloomer''). The largest objects that form via these pathways are expected to clear detectable gaps in the radial distribution of the accompanying debris (Figs.~\ref{fig: ann0}--\ref{fig: ann1}; Section 3.5).

When compared with the model tracks, 
the known population of bright young debris disks ($\ldlstar \gtrsim 10^{-3}$ at 
50--100 Myr)
is well matched by rings that start out with high initial mass ($\M0=$ 5--40 $\mearth$).
As they evolve, these systems pass through the $\ldlstar$ distribution of known debris disks as a function of age, an outcome that is consistent with a large range in planetesimal formation efficiency ($f \lesssim 0.5$).
Thus, if $\sim 25$\% of stars start out with massive protoplanetary rings and follow this classical ``bright stalwart'' evolutionary path, we can readily account for the observed luminosities and incidence rates of debris disks over time (Section 4.2). 

Although most current observations are consistent with this simple picture, the debris disk incidence rate in the 10--50 Myr age range is not well known. The true story may be more complex if the high incidence rate of cold debris disks among the small sample of F stars in the $\beta$ Pic moving group is typical of 10--50 Myr old stars. A large population of such sources may indicate that a significant fraction of disks follow the ``early flare'' pathway, in which massive rings of solids with very small $f$ that are also undetectable because of their very small initial scale height brighten dramatically into observable debris disks, then quickly fade below current detection limits.
Future observations that constrain the debris disk frequency and \ldlstar\ of stars in the 10--50 Myr age range are needed to understand how often protoplanetary disks pursue this evolutionary pathway. 

Constraints from an even earlier stage of evolution, from the Class III phase, are also important to understand the evolutionary pathways of solids in disks. The recent work of \citet{lovell2021}, which studies a small sample of Class III sources in Lupus, is a good start in this direction. Larger samples and deeper observations are needed to understand whether most Class III sources follow the approximately constant luminosity evolution of the ``bright stalwart'' disks at early times, and/or if disks populate the fainter ``steady glow'' evolutionary tracks.

This uncertainty aside, the ability of the classical ``bright stalwart'' pathway to explain the known debris disks appears to limit the role of the 
other generic pathways in producing known debris disks. 
The inferred strong evolutionary connection between the $\sim$ 20--25\% of protoplanetary disks with large rings and the $\sim$ 20--25\% of mature stars with cold debris disks implies that the majority population of compact protoplanetary disks ($\sim$ 75--80\% of all disks) leave behind only modest masses of residual solids at large radii ($\lesssim 1~\mearth$) and evolve primarily into mature low-mass stars without detected debris
at $a \gtrsim$ 30--40~au (Fig.~\ref{fig: schema} and \ref{fig: fourpaths}). In other words, the cold debris disks studied to date are a chapter in the history of a minority of low mass stars (Section 4.3). 

We can test this interpretation by looking for dynamical evidence of the planets 
predicted to form under these conditions. Planets 
should produce gaps in the accompanying debris, which may be resolvable spatially with facilities such as the ngVLA (Section 4.2). Resolving gaps in young disks can also place constraints on the efficiency of planetesimal formation (Section 4.3).  
Improved debris disk demographics can also test this picture. More sensitive surveys that probe $\ldlstar$ down to $10^{-5}-10^{-6}$ in the 10--300 Myr age range can directly constrain the extent to which disks follow the ``steady glow'' and ``late bloomer'' paths. It is also important to study larger samples of young stars ($< 100$ Myr) and to characterize the incidence rate of bright cold debris disks $(\ldlstar \sim 10^{-4}-10^{-2}$) to infer whether they represent $\sim 20$\% of low mass stars 
or a much larger fraction. A much larger fraction (Pawallek et al.\ 2021) would indicate a more complex situation and a possibly significant role for the ``early flare'' pathway.  

Although the picture we have described motivates and awaits new tests, 
our analysis illustrates how models of the evolution of rings of solids, when combined with observational constraints on the demographics of debris disks (e.g., their incidence rate as a function of luminosity and age), can strongly constrain the global evolutionary pathways of debris disks and place constraints on current important unknowns, such as the efficiency of planetesimal formation and the masses of possible dark reservoirs of solids in young disks.
 
\acknowledgements

We thank Meredith Hughes and Gaspard Duch\^ene for sharing 
the data from their ARAA article; Grant Kennedy for sharing summaries of 
the {\it Herschel} survey results; and Jaehan Bae and Andrea Isella for
sharing some results from their upcoming review article.
We also acknowledge a thorough and helpful report from an anonymous 
reviewer. 
Portions of this project were supported by the {\it NASA }
{\it Emerging Worlds} program through grant NNX17AE24G.
Useful data from the numerical calculations are available at a
publicly-accessible repository (https://hive.utah.edu/) with the url
https://doi.org/10.7278/S50d-znrw-973k.
Other portions of this work were supported by {\it NASA} under Agreement No.\ NNX15AD94G to the ``Earths in Other Solar Systems'' program.
The results reported herein  also benefited from collaborations and/or information exchange within NASA’s Nexus for Exoplanet System Science (NExSS) research coordination network sponsored by NASA’s Science Mission Directorate.

\appendix

\section{The ORCHESTRA Code}

To derive the evolution of solid particles within a disk or a ring, 
we perform sets of numerical calculations with \orch, an ensemble of 
computer codes designed to track the accretion, fragmentation, and 
orbital evolution of solid particles ranging in size from a few microns 
to thousands of km \citep{kenyon2002,bk2006, kb2008,bk2011a,bk2013,
kb2016a,kb2016b}.  For the calculations described here, 
we establish a radial grid of 28 or 56 concentric annuli distributed 
in equal intervals of $a^{1/2}$ between $a_{in}$ and $a_{out}$.  Each 
annulus has 140 mass bins; the interval between adjacent bins is 
$\delta = m_{i+1} / m_i$ = 2. The minimum particle radius is 
\rmin\ = 1~\mum; the largest possible object in the grid has a mass 
of roughly 50~\mearth.

Within each calculation, solids have initial mass density $\rho$, 
orbital eccentricity $e_0$, orbital inclination $\imath_0$, and 
total mass \M0.  Each annulus in the grid has initial surface 
density of solids $\Sigma_0 \propto a^{-3/2}$ (disks) or 
$\Sigma_0 \propto e^{-((a - a_0)/\Delta a)^2}$ (rings).
As the calculations proceed, collisions between particles modify 
the number and masses of objects in each mass bin. Collisional 
damping and gravitational interactions between solids change the 
orbital parameters $e$ and $\imath$.  

To evolve the size and velocity distributions of solids in time, 
\orch\ derives collision rates and outcomes with standard 
particle-in-a-box algorithms \citep{kb2012}. Systems start with 
an initial size distribution $n(r)$ . When a pair of solids collides, 
the mass of the merged object is 
\begin{equation}
m = m_1 + m_2 - \mesc ~ ,
\label{eq: msum}
\end{equation}
where $m_1$ and $m_2$ are the masses of the colliding particles.
The mass of debris ejected in a collision is
\begin{equation}
\mesc = 0.5 ~ (m_1 + m_2) \left ( \frac{Q_c}{Q_D^*} \right)^{b_d} ~ .
\label{eq: mesc}
\end{equation}
where $Q_c = m_1 m_2 v^2 / (m_1 + m_2)^2$ is the center-of-mass 
collision energy, $v$ is the collision velocity, and the exponent 
$b_d$ is a constant of order unity \citep[e.g.,][]{davis1985,
weth1993,kl1999a,benz1999,obrien2003,koba2010a,lein2012}.  The 
binding energy of solids, \qdstar, is the energy require to 
disperse half of the combined mass, $m_1 + m_2$, to infinity:
\begin{equation}
\label{eq: qdstar}
\qdstar\ = Q_s r^{e_s} + Q_g \rho r^{e_g} ~ ,
\end{equation}
where $\rho$ = 1.5~\gcmc\ is the mass density and $(Q_s, Q_g, e_s, e_b)$
are model parameters \citep[e.g.,][]{benz1999,lein2012}. 
In this expression, the first (second) term is the strength (gravity) 
component of the binding energy. We choose parameters for normal ice:
$Q_s = 4 \times 10^6$~erg~cm$^{0.4}$~g$^{-1}$, $e_s = -0.4$,
$Q_g$ = 0.3~erg~cm$^{1.65}$~g$^{-1}$, and $e_g$ = 1.35 
\citep{schlicht2013,kb2020}.

To place the debris in the grid of mass bins, we set the mass 
of the largest collision fragment as
\begin{equation}
\mmaxd = m_{l,0} ~ \left ( \frac{Q_c}{Q_D^*} \right)^{-b_l} ~ \mesc ~ ,
\label{eq: mlarge}
\end{equation}
where $m_{l,0} \approx$ 0.01--0.5 and $b_l \approx$ 0--1.25
\citep{weth1993,kb2008,koba2010a,weid2010}. When $b_l$ is large,
catastrophic (cratering) collisions with $Q_c \gtrsim \qdstar$ 
($Q_c \lesssim \qdstar$) crush solids into smaller fragments.  
Lower mass objects have a differential size distribution 
$N(r) \propto r^{-q_d}$ with $q_d \approx$ 3--4. After placing a 
single object with mass \mmaxd\ in an appropriate bin, we place 
material in successively smaller mass bins until 
(i) the mass is exhausted or (ii) mass is placed in the smallest 
mass bin. Any material left over is removed from the grid.

To follow the orbital evolution of solids, we derive collisional 
damping from inelastic collisions and elastic (gravitational) 
interactions.  For inelastic and elastic collisions, we follow 
the statistical, Fokker-Planck approaches of \citet{oht1992} 
and \citet{oht2002}, which treat pairwise interactions (e.g., 
dynamical friction and viscous stirring) between all objects.  
We also compute long-range stirring from distant oligarchs 
\citep{weiden1989}.

We assume the surface density of gas is zero throughout the grid
and ignore interactions between solids and gas.  Previously 
published calculations with \orch\ \citep[e.g.,][]{kb2008,kb2009,
kb2010} demonstrate that gas drag tends to circularize the orbits 
of particles with radii $r \lesssim$ 1~km on time scales of 
several Myr at 30--150~au \citep[see][]{ada1976,weiden1977a,raf2004}.
In parallel, dynamical friction between these particles and much 
larger solids gradually reduces the $e$ and $\imath$ of the larger 
solids. Once runaway growth begins, the largest solids then have 
much larger gravitational focusing factors and grow much more 
rapidly than ensembles of solids where gas drag is neglected 
\citep[see also][]{youdin2013}. 
Because our interest is in the evolution of solids on Gyr time 
scales, the lack of gas drag probably has little impact on our 
results. We consider several comparisons below.

Aside from circularization of orbits, the gas causes 
intermediate-sized solids to drift radially relative to the gas 
\citep{ada1976,weiden1977a,raf2004}. The smallest solids drift 
with the gas; the largest solids are not affected by the gas.  
Here, we avoid the complications of evolving the surface density 
of the gas \citep[e.g.,][]{alex2009,oka2011,bk2011a,martin2012,
martin2014,bitsch2015a,yzhang2015,xiao2017,shadmehri2019} and 
ignore radial drift of the solids. For our focus on long-term 
evolution, this assumption is a reasonable starting point.

Our solutions to the evolution equations conserve mass and energy 
to machine accuracy. Typical calculations require several 12~hr 
runs on a system with 56 cpus; over the $10^6$--$10^8$ timesteps 
in a typical 2--4~Gyr run, calculations conserve mass and energy 
to better than one part in $10^{10}$.

\section{BINDING ENERGY PARAMETERS}

In previous studies, \citet{kb2008,kb2010} demonstrate that 
outcomes of coagulation calculations are remarkably independent 
of initial conditions. Although the growth time for large objects
scales inversely with mass, collisions rapidly erase the initial 
eccentricity and inclination of the solids 
\citep[see also][]{kb2004a,kb2012,kb2016a}. The pace of growth
also depends on the initial size distribution and the maximum 
size of the solids at the start of a calculation. However, the
final radius of the largest object in an annulus is rather 
insensitive to these starting conditions.

Growth of the largest objects is much more sensitive to the 
bulk properties of solids. Solids with larger binding energy 
\qdstar\ generate less debris during a collision and therefore
grow larger with time than solids with smaller \qdstar. In a
suite of coagulation calculations at 15--150~au, weak solids 
reach typical sizes of 3000-7000~km in 10 Gyr \citep{kb2010,kb2012}.
Stronger solids achieve radii somewhat larger than $10^4$~km on 
similar time scales. Despite these differences in final planet 
radii, large changes in the binding energy produce factor of 
$\lesssim$ 2 variations in the maximum surface area of small 
particles \citep{kb2010}.

As an illustration of the sensitivity of the stellar luminosity 
reprocessed by small particles, $L_d$, we consider calculations 
in a single annulus at 30--60~au from the Sun \citep{kb2020}.
Calculations begin with 45~\mearth\ in solids. Particles have
a range of sizes $r \approx$ 100--500km; 100~km particles 
initially contain most of the mass. To initiate a collisional 
cascade at the start of the calculation, particles have initial 
collision velocities of 1~\kms, which guarantees that collisions 
among 100~km particles are destructive. Although there are no 
particles smaller than 100~km at the start of the calculation,
destructive collisions rapidly create them. 

In addition to the parameters appropriate for normal ice listed 
above, we perform calculations with 
`weak' ($Q_s = 2 \times 10^5$~erg~cm$^{0.4}$~g$^{-1}$) and
`strong' ($Q_s = 7 \times 10^7$~erg~cm$^{0.4}$~g$^{-1}$) ice. 
For the gravity component of \qdstar, we examine results for the
\citet{benz1999} approach (BA, $Q_g$ = 2.1~erg~cm$^{1.81}$~g$^{-1}$ 
and $e_g$ = 1.19) and the \citet{lein2012} approach (LS, 
$Q_g$ = 0.3~erg~cm$^{1.81}$~g$^{-1}$ and $e_g$ = 1.35).

To have a sense of how coagulation calculations react to these
choices, we briefly outline how collision outcomes depend on 
\qdstar.  When solids are weak (strong), they are easier (harder) 
to fragment during a high velocity collision. As the largest objects
grow, more (fewer) collision fragments generate a faster (slower) 
rise in the dust luminosity. If collisional damping is important,
a larger mass in small fragments may accelerate the growth of the 
largest objects \citep[e.g.,][]{kb2015a,kb2016b}. Over time, systems 
of weaker solids lose somewhat more mass and contain somewhat less 
mass in small solids than those with stronger solids 
\citep[see also][]{kb2016a}. 

\begin{figure}[t]
\begin{center}
\includegraphics[width=5.5in]{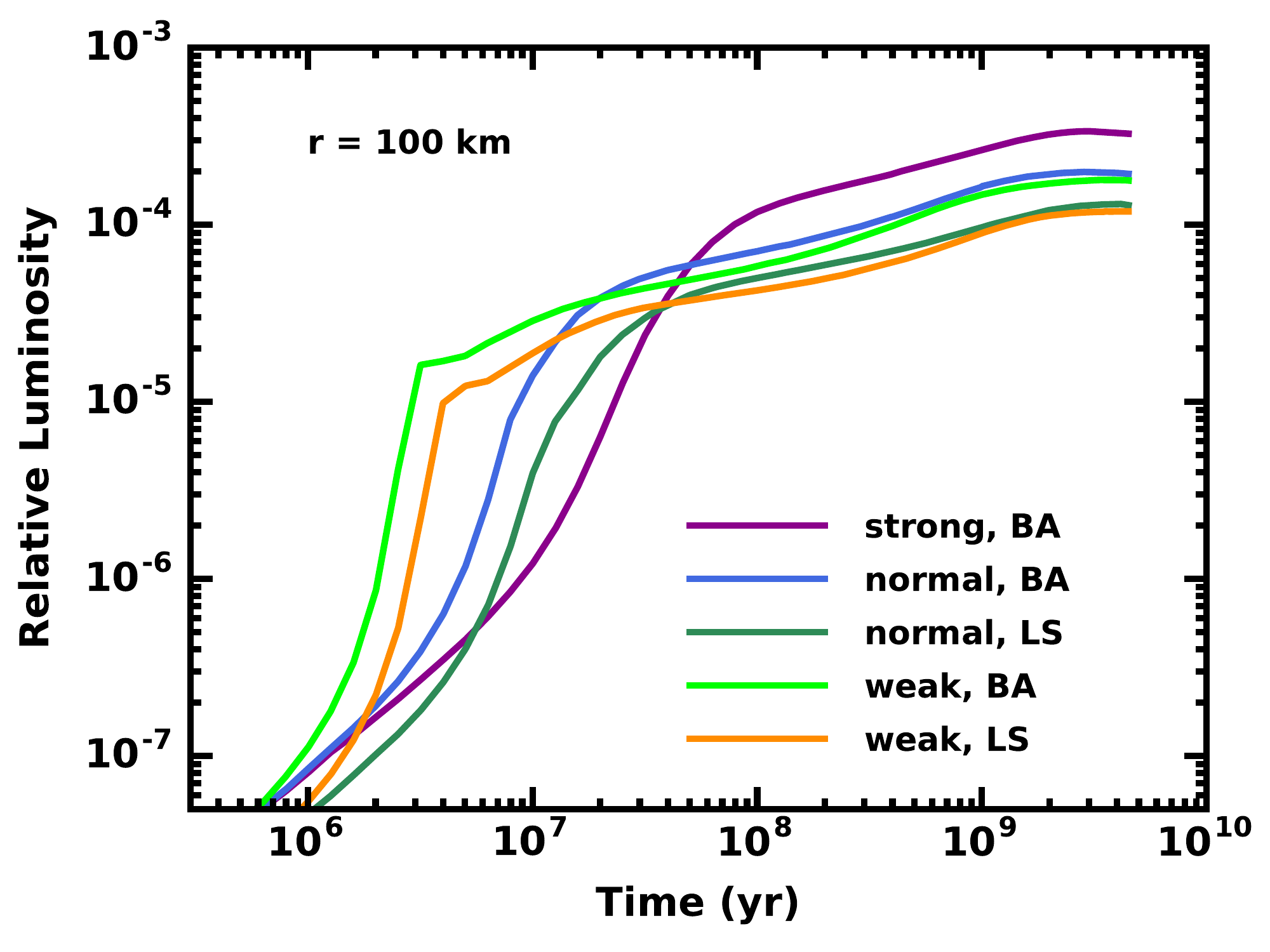}
\vskip -2ex
\caption{
\label{fig: ldust4}
Time evolution of the dust luminosity for single-annulus 
calculations using different approaches for the binding 
energy of solids. In systems with `weak' (`strong') solids,
the dust luminosity rises more rapidly (slowly) than in 
systems with solids composed of `normal' ice. At late times, 
solids with the BA expression for the binding energy 
generate more dust than solids with the LS formulation.
Despite a factor of 300 spread in the strength of small 
objects among these approaches, the range in the dust 
luminosity is only a factor of three.
}
\end{center}
\end{figure}

Fig.~\ref{fig: ldust4} shows the evolution of the dust luminosity 
for the single annulus calculations \citep{kb2020}.  Starting with 
most of the mass in 100~km objects, solids have initial collision 
velocities of 1~\kms\ and no dynamical evolution.  Starting conditions 
are set to initiate a cascade of collisions that gradually grind 
100~km objects into fine dust grains that are ejected by radiation
pressure from the central star \citep[see also][]{wyatt2008}. For 
the chosen parameters, the binding energy of 100~km objects is 
fairly independent of $Q_s$, $Q_g$, and $e_g$. Collisions between 
pairs of these objects produce the same amount of mass in fragments; 
fragments also have the same size distribution.  Across the five 
different calculations, large objects generate fragments at the 
same rate throughout 4.5~Gyr of evolution.  Thus, the mass loss 
rate of the system is the same in all five calculations. After 
4.5~Gyr, all systems have the same final mass in solids.

Differences arise when the 1--10~km fragments collide. At the smallest 
sizes in these calculations (1--10~\mum), the dust luminosity depends 
on the rate mass flows from the fragments to smaller and smaller sizes. 
For systems with the same binding energy at the largest sizes, the mass
flow is largest for `weak' ice systems and is then progressively smaller 
as the bulk strength grows from weak to normal to strong \citep[see also][]
{wyatt2011,kb2016a,kb2017a}. With a constant production rate of 1--10~km 
fragments, strong ice systems retain a larger fraction of this mass 
than weak ice systems and therefore have a larger dust luminosity. 
In addition to the bulk strength, the mass flow down the cascade depends 
on the gravity component of the binding energy. For particle radii 
$r \gtrsim$ 10--100~m, self-gravity dominates bulk strength.  In the BA 
formalism, 0.1--2~km fragments are stronger than in the LS formalism.  
Independent of the bulk strength, the mass flow rate in a system of BA 
solids is then smaller than in a system of LS solids. BA systems retain
more small solids and have larger $L_d$ than LS systems.

In Fig.~\ref{fig: ldust4}, all five systems follow the same evolution. 
During the first 1--10~Myr, the dust luminosity rises rapidly as large 
object collisions produce the first fragments which in turn produce the 
first small dust grains.  Because weaker fragments make more dust, the 
luminosity rises faster in systems of weak particles. At 10--100~Myr, 
the rapid rise in $L_d$ from the initial set of collisions slows. 
Systems enter a plateau phase, where the dust luminosity slowly rises. 
After 3--4~Gyr, collision rates among leftover 100~km objects begin
to drop, dust production slows, and the dust luminosity falls.

At late times, the evolution of the dust luminosity falls into three 
groups with only a factor of $\sim$ 3 range in $L_d$.  Systems with 
the LS gravity component of the binding energy have the largest mass 
flow down the cascade, the smallest dust mass, and the lowest dust 
luminosity.  Although LS systems with `weak' ice initially evolve 
more rapidly than those with normal ice, they converge on nearly the 
same $L_d$ at late times. In these two examples, collisions between
0.1--10~km objects set the mass flow rate down the cascade. For a
collision velocity of 1~\kms, the bulk strength of 1~cm and smaller 
particles has little impact on the population of small particles.
Thus, weak and normal ice calculations with the LS gravity component
yield the same dust luminosity at late times.

In cascades with smaller collision velocities, the difference between
weak and normal bulk strength might lead to different dust luminosities 
at late times. However, we expect these differences to be smaller than
those from adopting different relations for the gravity component of
\qdstar.

Systems with strong ice and the BA gravity component have the smallest 
mass flow rate down the cascade and the largest $L_d$.  At intermediate 
$L_d$, systems with the BA gravity component and either weak or normal 
ice have roughly the same $L_d$ at late times. In these systems, the 
rate of mass flow down the cascade is the same as the strong BA model 
from 100~km to 0.1~km. Below 0.1~km, the mass flow rate grows 
substantially compared to the strong BA model and reaches roughly 
the same level in both calculations. Thus, these models have roughly 
the same $L_d$ at late times.

The small variation in maximum $L_d$ among these calculations 
suggests that the binding energy parameters are not critical 
components for predictions of the luminosity \citep[see also][]{kb2010,
kb2012}. To facilitate comparisons with previous studies, we 
perform calculations with the normal ice and LS binding energy 
parameters. From Fig.~\ref{fig: ldust4}, this choice produces a 
smaller dust luminosity than the BA binding energy parameters.

The maximum sizes of the largest objects are also somewhat smaller.
Expressed another way, the normal ice bulk strength with the LS
gravity component of the binding energy requires the largest mass
to generate an observed maximum $L_d$ and a maximum size for the
largest objects. If the suite of calculations discussed in the main
text explains observations without violating mass budget constraints 
\citep[e.g.,][]{najita2014}, then simulations with other choices for
the binding energy parameters will match observations with lower 
initial masses in solids.

\section{SUPPLEMENTAL RESULTS}

In \S\ref{sec: ringmodelgrowth}--\ref{sec: ringmodelgaps}, we describe 
the time evolution of (i) the largest objects with a broad
range in \M0\ and $f$ (Figs.~\ref{fig: mass0}--\ref{fig: mass2}), 
(ii) \ldlstar\ for selected values of \M0\ and $f$ 
(Fig.~\ref{fig: ldust0}, and 
(iii) the radial distribution of the largest objects within the rings
(Figs.~\ref{fig: ann0}--\ref{fig: ann1}).
To illustrate several other aspects of the calculations, we describe 
the importance of collisional damping on runaway growth and show the 
relation between the growth time, the initial mass in solids, and $f$.
For completeness, this section also includes Tables of results for the
full set of calculations.

{\bf Eccentricity Evolution.}\quad
Collisional damping, dynamical friction, and viscous stirring are central 
features of all calculations with $f < 1$. In a system with a mix of 
pebbles
and planetesimals, collisions among pebbles are inelastic and circularize
the orbits of pebbles \citep[see also][and references therein]{gold2004}.
Gravitational interactions between pebbles and planetesimals try to
equalize the kinetic energy of both species. With the large mass ratio
between pebbles and planetesimals, this process excites the eccentricities
of pebbles and damps the eccentricities of planetesimals. Viscous stirring
transfers angular momentum from planetesimals to pebbles, raising $e$ for
the pebbles. 

\begin{figure}[t]
\begin{center}
\includegraphics[width=5.5in]{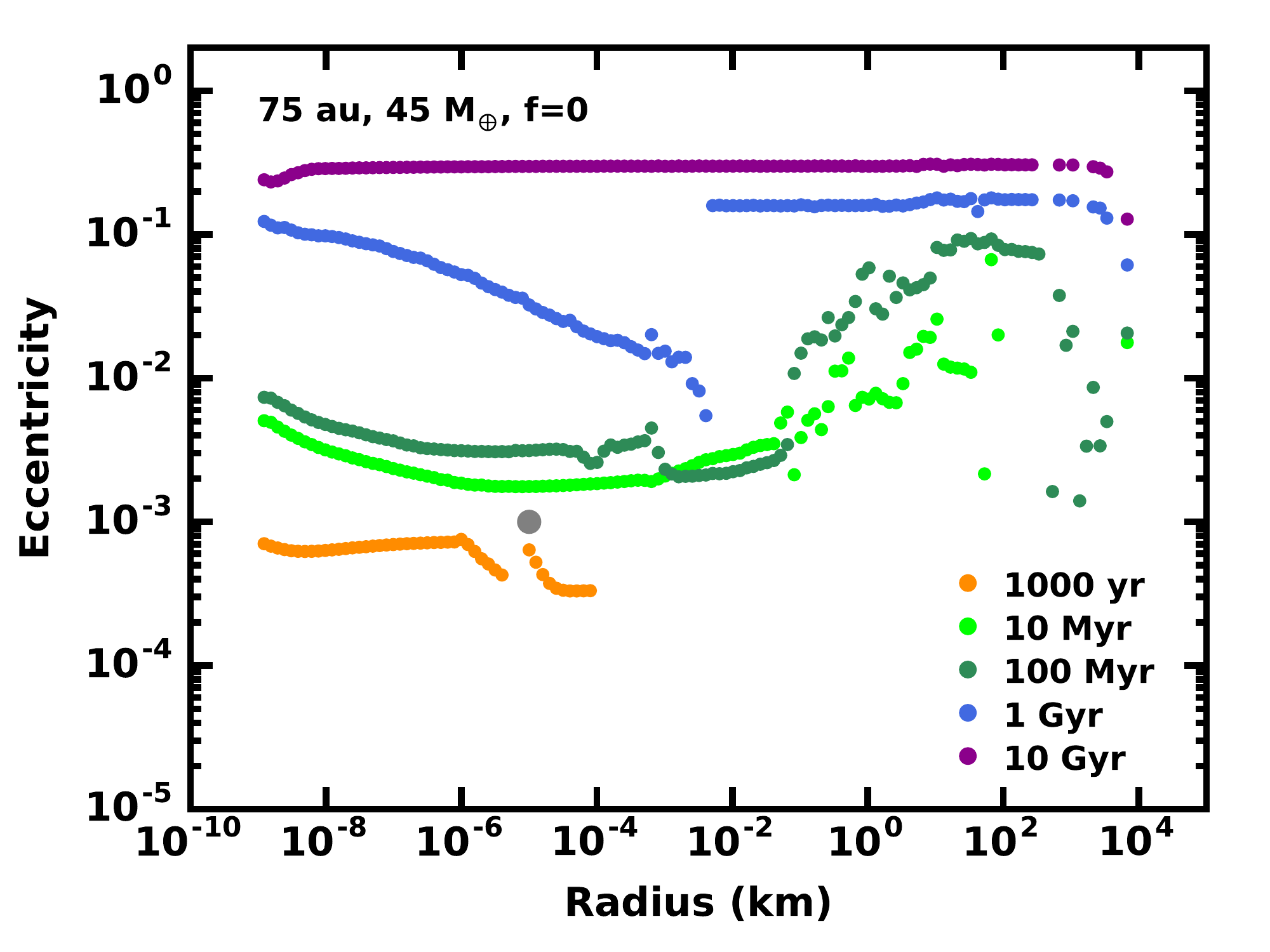}
\vskip -2ex
\caption{
\label{fig: ecc0}
Snapshots of the eccentricity distribution of solids in a model with
$a_0$ = 75~au, \M0\ = 45~\mearth, and $f$ = 0 at five epochs as listed 
in the legend. The large grey dot at $r$ = 1~cm and $e = 10^{-3}$ shows
the starting point for a system composed only of pebbles.  
}
\end{center}
\end{figure}

Fig.~\ref{fig: ecc0} illustrates the impact of these processes for 
a calculation with \m0\ = 45~\mearth\ and $f$ = 0 at 75~au. Initially, 
all pebbles have 
$e = 10^{-3}$. During the first 1000~yr of evolution, collisions among
pebbles create 10~cm rocks and debris with $r \lesssim$ 1~mm. Damping 
reduces the eccentricity of 
1--100~\mum\ (3--10~cm) particles to $\sim 5-6 \times 10^{-4}$
($\sim 3 \times 10^{-4}$).
As pebbles continue to grow, 
modest stirring by m-sized to km-sized planetesimals tries to raise $e$;
pebble damping continues to keep $e$ low. 

This balance between the smallest and largest solids in the ring continues
until the runaway produces large protoplanets. At 10~Myr, several 
0.25~\mearth\ 
planets try to stir all of the lower mass solids. Solids with $R \approx$
100~m to 10~km have little mass; protoplanets effectively stir them to 
larger $e$. Among smaller solids, damping maintains low $e$. At 100~Myr,
the smallest solids continue to resist viscous stirring by the largest
objects. Although 100~m and larger objects now have $e \sim$ 0.01--0.1, 
small solids maintain $e \sim 2 - 4 \times 10^{-3}$. 
Small fluctuations in $e$ for 0.1--1~m solids result from variations
of the mass contained in each mass bin: bins with smaller $e$ have more
mass.

As the evolution proceeds to 1--10~Gyr, the cascade gradually reduces the
population of solids with $r \lesssim$ 1~m. With less mass among pebbles, 
damping is less effective. 
For nearly all solids with $r \gtrsim$ 4~m, 
$e \approx$ 0.1--0.2. Among the 1--4~m solids, damping maintains a small
$e$. The linear rise in the $e$
distribution from 1~m to 1~\mum\ results from the smaller mass in these 
objects: damping maintains small $e$ for 1--4~m objects 
but damping is less and less effective at smaller and smaller sizes.
By 10~Gyr, nearly all solids have roughly the same eccentricity, $e 
\approx$ 0.3; the largest protoplanets have $e \approx$ 0.1.

\begin{figure}[t]
\begin{center}
\includegraphics[width=5.5in]{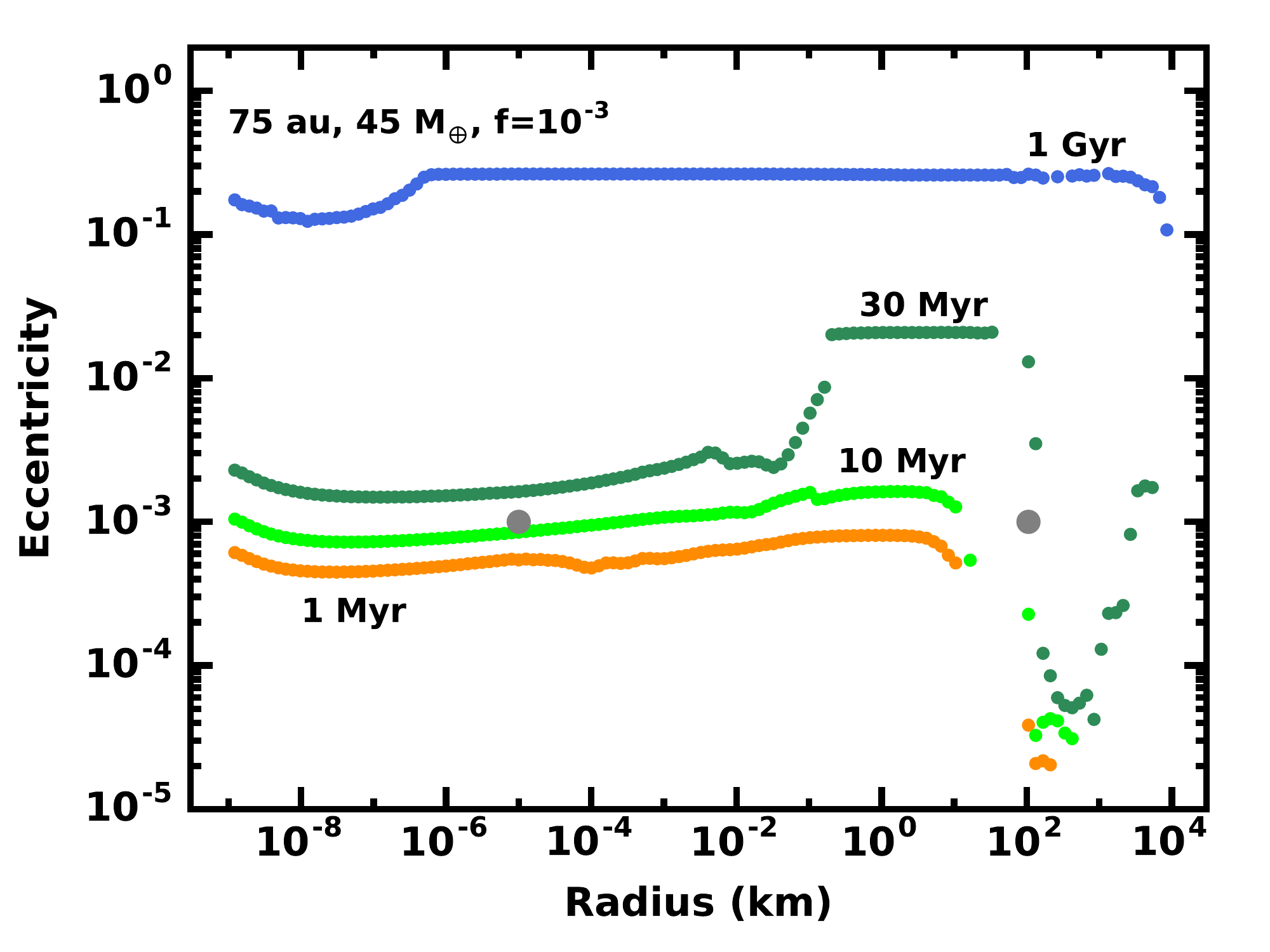}
\vskip -2ex
\caption{
\label{fig: ecc1}
As in Fig.~\ref{fig: ecc0} for a system with $f = 10^{-3}$. The two grey
dots indicate the initial $e$ for pebbles and planetesimals. 
}
\end{center}
\end{figure}

Fig.~\ref{fig: ecc1} shows snapshots of the eccentricity evolution for a
massive ring with planetesimals and pebbles. During the first Myr, growth
is slow: mergers produce a few objects with $r \approx$ 200--300~km and
a swarm of smaller solids with $r \lesssim$ 10~km. Collisional damping
reduces $e$ by 10--20\% for the smaller solids; dynamical friction lowers
$e$ by almost a factor of 100 for the largest solids. 

Over the next 30~Myr, the largest solids grow dramatically to nearly
10,000~km. With most of the mass, protoplanets stir themselves to larger
and larger $e$. Solids with $R \approx$ 0.1--10~km have little mass and
are stirred to even larger $e$. Despite having less mass, pebbles maintain
smaller $e$ through collisional damping. The combined effects of damping
and stirring yield an eccentricity distribution where the largest and 
the smallest solids have $e \approx 2 \times 10^{-3}$. Dynamical friction
between pebbles and 100--300~km solids keep the larger solids at very
small $e \approx 10^{-4}$. 

As the collisional cascade continues, pebbles contain less and less mass.
Damping and dynamical friction are less effective; viscous stirring 
dominates. Eccentricities for all solids grow from $10^{-3}$ to 0.1--0.3.
At either end of the distribution, $e$ drops by a factor of 2--3. For the
large solids, dynamical friction with the rest of the solids reduces $e$.
The smallest solids have just enough mass for damping to reduce and 
maintain the smaller $e$. 

These two examples illustrate how damping among pebbles fuels runaway
growth of the largest planetesimals. Without damping, pebbles have a
factor of 10--100 larger eccentricity. Larger eccentricity lowers 
gravitational focusing factors and reduces the growth rate of the largest
planetesimals. Runaway growth is delayed and is much weaker. Instead of 
reaching super-Earth masses on 1--10~Myr time scales, systems with no 
damping would produce super-Earths on 100~Myr to Gyr time scales. Delaying
the runaway allows smaller planetesimals to accrete more pebbles, reducing
the dust luminosity to levels below observed systems at 10--100~Myr. Once
the runaway begins, a delayed collisional cascade would probably raise 
the dust luminosity to levels well above those observed at 1--10~Gyr. In
this way, the lack of collisional damping would challenge our ability to
match observations with an evolving swarm of pebbles and planetesimals.

\begin{deluxetable}{lccccccc}
\tablecolumns{8}
\tabletypesize{\scriptsize}
\tablenum{1}
\tablecaption{Results for Coagulation Calculations at 36--54~au\tablenotemark{a}}
\tablehead{
   \colhead{$M_0$ (\mearth)} &
   \colhead{$f$} &
   \colhead{$M_f/M_0$} &
   \colhead{$t_{1k}$ (Myr)} &
   \colhead{$r_{max}$ (km)} &
   \colhead{$N_L$} &
   \colhead{$M_L$ (\mearth)} &
   \colhead{$L_{d,max}/L_\star$}
}
\label{tab: tab1}
\startdata
~0.01 &  1.0 &  1.000 & \nodata &  167.5 & 8017 &  0.01 & $<$0.01\\
~0.01 &  0.3 &  0.979 & \nodata &  257.0 &  98 & $<$0.01&  0.50 \\
~0.01 &  0.1 &  0.986 & \nodata &  192.8 &  869 & $<$0.01&  0.58 \\
~0.01 & $10^{-2}$ &  0.993 &  \nodata &  295.1 &  75 & $<$0.01&  0.61 \\
~0.01 &  0.0 &  0.993 & \nodata &  453.9 &   5 & $<$0.01&  0.62 \\
~0.03 &  1.0 &  1.000 & \nodata &  199.5 & 23714 &  0.03 &  0.01 \\
~0.03 &  0.3 &  0.933 & \nodata &  639.7 &   9 & $<$0.01&  1.02 \\
~0.03 &  0.1 &  0.940 & \nodata &  638.3 &  15 & $<$0.01&  1.04 \\
~0.03 & $10^{-2}$ &  0.885 &  \nodata &  540.8 &  178 &  0.01 &  1.04 \\
~0.03 & $10^{-3}$ &  0.545 &  7067.8 & 1049.5 &  17 &  0.01 &  1.05 \\
~0.03 &  0.0 &  0.962 & \nodata &  736.2 &   7 & $<$0.01&  1.05 \\
~0.10 &  1.0 &  1.000 & \nodata &  206.1 & 78705 &  0.10 &  0.01 \\
~0.10 &  0.3 &  0.828 &  8851.2 & 1047.1 &   8 & $<$0.01&  2.15 \\
~0.10 &  0.1 &  0.625 &  3143.6 & 1538.2 &   2 &  0.01 &  2.01 \\
~0.10 & $10^{-2}$ &  0.495 &  5601.9 & 1039.9 &  37 &  0.01 &  1.63 \\
~0.10 & $10^{-3}$ &  0.312 &  2293.1 & 1207.8 &  40 &  0.02 &  2.37 \\
~0.10 &  0.0 &  0.468 &  1402.8 & 1367.7 &  17 &  0.02 &  2.05 \\
~0.30 &  1.0 &  1.000 & \nodata &  434.5 & 1683 &  0.02 &  0.03 \\
~0.30 &  0.3 &  0.774 &  1453.5 & 1552.4 &  10 &  0.01 &  4.95 \\
~0.30 &  0.1 &  0.587 &  820.4 & 2113.5 &   9 &  0.03 &  4.30 \\
~0.30 & $10^{-2}$ &  0.319 &  659.9 & 2387.8 &   3 &  0.02 &  5.16 \\
~0.30 & $10^{-3}$ &  0.254 &  550.4 & 1836.5 &  12 &  0.03 &  6.34 \\
~0.30 & $10^{-4}$ &  0.341 &   95.4 & 2138.0 &  15 &  0.04 &  8.32 \\
~0.30 &  0.0 &  0.752 &  317.4 & 1625.5 &   5 &  0.01 &  2.51 \\
~1.00 &  1.0 &  0.998 & \nodata &  517.6 & 6295 &  0.17 &  0.20 \\
~1.00 &  0.3 &  0.750 &  427.6 & 2128.1 &  11 &  0.05 &  13.30 \\
~1.00 &  0.1 &  0.547 &  229.0 & 3076.1 &   9 &  0.09 &  10.57 \\
~1.00 & $10^{-2}$ &  0.272 &  149.8 & 3639.2 &  5 &  0.12 &  17.22 \\
~1.00 & $10^{-3}$ &  0.217 &  129.5 & 2393.3 &  6 &  0.04 &  20.04 \\
~1.00 & $10^{-4}$ &  0.186 &   30.6 & 3191.5 &  2 &  0.04 &  26.61 \\
~1.00 &  0.0 &  0.267 &   39.4 & 2055.9 &  10 &  0.04 &  7.57 \\
~3.00 &  1.0 &  0.993 &  4818.2 & 2157.7 &  49 &  0.11 &  0.66 \\
~3.00 &  0.3 &  0.676 &   98.7 & 3243.4 &   3 &  0.06 &  32.21 \\
~3.00 &  0.1 &  0.520 &   87.5 & 4385.3 &   4 &  0.16 &  29.65 \\
~3.00 & $10^{-2}$ &  0.234 &   66.6 & 5407.5 &   5 &  0.28 &  65.61 \\
~3.00 & $10^{-3}$ &  0.168 &   51.9 & 5211.9 &   2 &  0.16 &  65.92 \\
~3.00 & $10^{-4}$ &  0.136 &   19.6 & 2747.9 &  46 &  0.22 &  86.50 \\
~3.00 & $10^{-5}$ &  0.094 &   0.6 & 3396.3 &  17 &  0.27 & 110.41 \\
~3.00 &  0.0 &  0.118 &   3.1 & 4335.1 &   6 &  0.15 &  44.16 \\
10.00 &  1.0 &  0.962 &  716.9 & 4375.2 &  22 &  0.66 &  2.18 \\
10.00 &  0.3 &  0.561 &   19.8 & 4168.7 &  20 &  0.50 & 119.40 \\
10.00 &  0.1 &  0.468 &   24.9 & 4405.5 &  13 &  0.48 & 132.74 \\
10.00 & $10^{-2}$ &  0.211 &   14.5 & 6280.6 &   5 &  0.58 & 154.17 \\
10.00 & $10^{-3}$ &  0.143 &   15.8 & 6683.4 &   6 &  0.85 & 350.75 \\
10.00 & $10^{-4}$ &  0.140 &   6.2 & 3784.4 &  22 &  0.28 & 239.88 \\
10.00 & $10^{-5}$ &  0.091 &   0.3 & 3097.4 &  75 &  0.81 & 340.41 \\
10.00 &  0.0 &  0.047 &   1.0 & 3810.7 &   4 &  0.23 & 127.94 \\
\enddata
\tablenotetext{a}{
The columns list the initial mass \M0\ for each calculation,
the initial fraction $f$ of mass in large planetesimals,
the ratio of the final mass in the grid $M_f$ to the initial mass,
the time scale $t_{1k}$ to produce the first protoplanet with 
$r \ge$ 1000~km,
the final radius \rmax\ of the largest planet,
the final number $N_L$ of protoplanets with masses at least as large as
10\% of the mass of the largest planet,
the total mass $M_L$ in these large protoplanets, and
the maximum relative dust luminosity \ldlstar\ in units of $10^{-5}$.
}
\end{deluxetable}

\begin{deluxetable}{lccccccc}
\tablecolumns{8}
\tabletypesize{\scriptsize}
\tablenum{2}
\tablecaption{Results for Coagulation Calculations at 60--90~au\tablenotemark{a}}
\tablehead{
   \colhead{$M_0$ (\mearth)} &
   \colhead{$f$} &
   \colhead{$M_f/M_0$} &
   \colhead{$t_{1k}$ (Myr)} &
   \colhead{$r_{max}$ (km)} &
   \colhead{$N_L$} &
   \colhead{$M_L$ (\mearth)} &
   \colhead{$L_{d,max}/L_\star$}
}
\label{tab: tab2}
\startdata
~0.01 &  1.0 & 1.000 & \nodata & 152.4 &  8035 &  0.01 & $<$0.01 \\
~0.01 &  0.5 & 0.995 & \nodata & 153.8 &  4121 &  0.01 &  0.12 \\
~0.01 &  0.1 & 0.998 & \nodata & 153.1 &   914 &$<$0.01 &  0.21 \\
~0.01 & $10^{-2}$ & 0.998 & \nodata & 195.9 &  81 & $<$0.01 &  0.23 \\
~0.01 &  0.0 & 1.000 & \nodata & 333.4 &   2 & $<$0.01 &  0.23 \\
~0.10 &  1.0 & 1.000 & \nodata & 157.8 & 79616 &  0.10 & $<$0.01 \\
~0.10 &  0.5 & 0.959 & \nodata & 358.9 & 272 & $<$0.01 &  1.20 \\
~0.10 &  0.1 & 0.973 & \nodata & 474.2 &  49 & $<$0.01 &  1.49 \\
~0.10 & $10^{-2}$ & 0.979 & \nodata & 421.7 & 411 &  0.01 &  1.51 \\
~0.10 & $10^{-3}$ & 0.916 & \nodata & 901.6 &  43 &  0.01 &  1.51 \\
~0.10 &  0.0 & 0.984 & \nodata & 772.7 &   2 & $<$0.01 &  1.51 \\
~0.50 &  1.0 & 1.000 & \nodata & 209.4 & 433511 &  0.05 & $<$0.01 \\
~0.50 &  0.5 & 0.883 & \nodata & 863.0 &  56 &  0.01 &  4.46 \\
~0.50 &  0.1 & 0.820 & 4135.2 &  1548.8 &   7 &  0.01 &  4.06 \\
~0.50 & $10^{-2}$ & 0.361 & 3203.7 &  2249.1 &   5 &  0.03 &  3.79 \\
~0.50 & $10^{-3}$ & 0.277 & 2164.2 &  1927.5 &   3 &  0.01 &  3.82 \\
~0.50 & $10^{-4}$ & 0.207 &  685.6 &  2393.3 &   8 &  0.03 &  5.51 \\
~0.50 &  0.0 & 0.962 &  1174.9 &  1188.5 &   6 &  0.01 &  2.65 \\
~1.50 &  1.0 & 1.000 & \nodata & 304.8 &  54576 &  0.29 &  0.07 \\
~1.50 &  0.5 & 0.804 & 2807.4 &  1745.8 &  21 &  0.02 & 10.54 \\
~1.50 &  0.1 & 0.659 & 2629.0 &  2118.4 &   5 &  0.03 &  8.79 \\
~1.50 & $10^{-2}$ & 0.217 &  716.6 &  3372.9 &   2 &  0.04 &  9.33 \\
~1.50 & $10^{-3}$ & 0.194 &  838.4 &  3475.4 &   1 &  0.04 & 10.79 \\
~1.50 & $10^{-4}$ & 0.180 &  601.2 &  2306.7 &  36 &  0.09 & 15.67 \\
~1.50 & $10^{-5}$ & 0.585 &  3.7 &  2685.3 &   6 &  0.06 & 11.04 \\
~1.50 &  0.0 & 1.000 & \nodata &  304.8 &   54576 &  0.29 & 0.07 \\
~5.00 &  1.0 & 0.998 & \nodata &  691.8 & 6209 &  0.34 &  0.33 \\
~5.00 &  0.5 & 0.743 &  401.8 &  1892.3 & 133 &  0.20 & 25.41 \\
~5.00 &  0.1 & 0.521 &  272.1 &  4405.5 &   7 &  0.20 & 20.51 \\
~5.00 & $10^{-2}$ & 0.229 &  245.7 &  5533.5 &   4 &  0.46 & 31.05 \\
~5.00 & $10^{-3}$ & 0.153 &  185.7 &  5546.3 &   2 &  0.22 & 50.47 \\
~5.00 & $10^{-4}$ & 0.106 &   73.5 &  2831.4 &  86 &  0.33 & 43.75 \\
~5.00 & $10^{-5}$ & 0.081 &  1.4 &  3507.5 &  22 &  0.27 & 69.34 \\
~5.00 &  0.0 & 0.256 &  10.1 &  3499.5 &   8 &  0.13 & 20.65 \\
15.00 &  1.0 & 0.995 & 2753.5 &  3404.1 &  11 &  0.27 &  1.14 \\
15.00 &  0.5 & 0.671 &   95.1 &  3715.4 &  10 &  0.25 & 62.95 \\
15.00 &  0.1 & 0.459 &   79.8 &  5081.6 &   5 &  0.22 & 76.74 \\
15.00 & $10^{-2}$ & 0.180 &   79.3 &  7979.9 &   4 &  0.92 &  137.72 \\
15.00 & $10^{-3}$ & 0.116 &   69.2 &  6745.3 &   7 &  0.98 &  233.88 \\
15.00 & $10^{-4}$ & 0.099 &   29.8 &  3435.6 &  45 &  0.40 &  153.82 \\
15.00 & $10^{-5}$ & 0.076 &  0.7 &  3396.3 & 110 &  1.11 &  163.68 \\
15.00 &  1.0 & 0.152 &  4.7 &  4285.5 &   11 &  0.34 &  70.63 \\
45.00 &  1.0 & 0.979 &  853.7 &  4909.1 &   18 &  1.53 &  3.37 \\
45.00 &  0.5 & 0.667 &   28.9 &  5382.7 &  16 &  0.97 &  195.88 \\
45.00 &  0.1 & 0.962 &  \nodata & 861.0 &  3289 &  0.35 &  163.31 \\
45.00 & $10^{-2}$ & 0.986 &   35.7 &  1253.1 &   7 &  0.01 &  103.04  \\
45.00 & $10^{-3}$ & 0.100 &   19.0 &  9397.2 &   7 &  2.23 &  457.09 \\
45.00 & $10^{-4}$ & 0.085 &   10.1 & 12302.7 &   2 &  2.82 &  941.89 \\
45.00 & $10^{-5}$ & 0.083 &  0.3 &  3265.9 & 279 &  3.36 &  613.76 \\
45.00 &  0.0 & 0.025 &  1.5 & 7079.5 &   5 &  0.86 &  309.03 \\
\enddata
\tablenotetext{a}{
The columns list the initial mass \M0\ for each calculation,
the initial fraction $f$ of mass in large planetesimals,
the ratio of the final mass in the grid $M_f$ to the initial mass,
the time scale $t_{1k}$ to produce the first protoplanet with $r \ge$ 1000~km,
the final radius \rmax\ of the largest planet,
the final number $N_L$ of protoplanets with masses at least as large as
10\% of the mass of the largest planet,
the total mass $M_L$ in these large protoplanets, and
the maximum relative dust luminosity \ldlstar\ in units of $10^{-5}$.
}
\end{deluxetable}

{\bf Tables of Results.}\quad
To facilitate comparisons between these calculations and those in other
studies, Tables~\ref{tab: tab1}--\ref{tab: tab2} summarize results for 
each calculation. The first two columns list \M0\ and $f$. In the third
column the ratio of the final mass $M_f$ to \M0\ provides a measure of
the mass ejected in the form of 1~\mum\ and smaller particles. In 
energetic (weak) cascades, $M_f / M_0$ is small (large). Rings with small
$f$ lose much more material than those with large $f$. The next columns
measure the ability of collisional growth to concentrate solids into 
massive planets. The variable $t_{1k}$ in column 4 quantifies the time
scale for the growth of Pluto-mass planets; \rmax\ in column 5 lists the
radius of the largest protoplanet at 10~Gyr. Rings that generate 
super-Earths (Plutos) have large (small) \rmax\ and small (large) 
$t_{1k}$. The next
two columns quantify the amount of mass in the largest objects at the
end of each calculation: $N_L$ ($M_L$) is the number (mass) of objects 
with masses no less than 10\% of the mass of the largest object. Sometimes,
the calculation produces 1--3 large protoplanets; others with similar 
amounts of mass in large objects generate many much smaller protoplanets.
These two quantities allow a comparison of systems with similar initial
masses but very different \rmax. Finally, the last column quantifies the
maximum dust luminosity throughout the calculation. As noted in the main
text, more massive rings have more luminous debris disks. This column
quantifies that statement.

Among the published numerical calculations of pebble accretion, only
\citet{shannon2015} consider the 1--10~Gyr evolution of a swarm of pebbles 
and planetesimals. Starting with \M0\ = 0.01~\mearth\ in a ring at 42--48~au 
with $f = 10^{-3}$ for pebbles and km-sized planetesimals, they follow the 
growth of larger solids with a single annulus coagulation calculation.
With a much smaller initial eccentricity $e_0 = 10^{-6}$ and binding energy
for solids (\qdstar, eq.~\ref{eq: qdstar}), large objects in these models
grow more slowly and fail to reach \rmax\ $\gtrsim$ 1000~km in 10~Gyr. in
contrast, For the initial conditions considered here, results are similar:
\rmax\ $\approx$ 1000~km at 1--2~Gyr and $\sim$ 2000~km at 10~Gyr.

\begin{figure}[t]
\begin{center}
\includegraphics[width=5.5in]{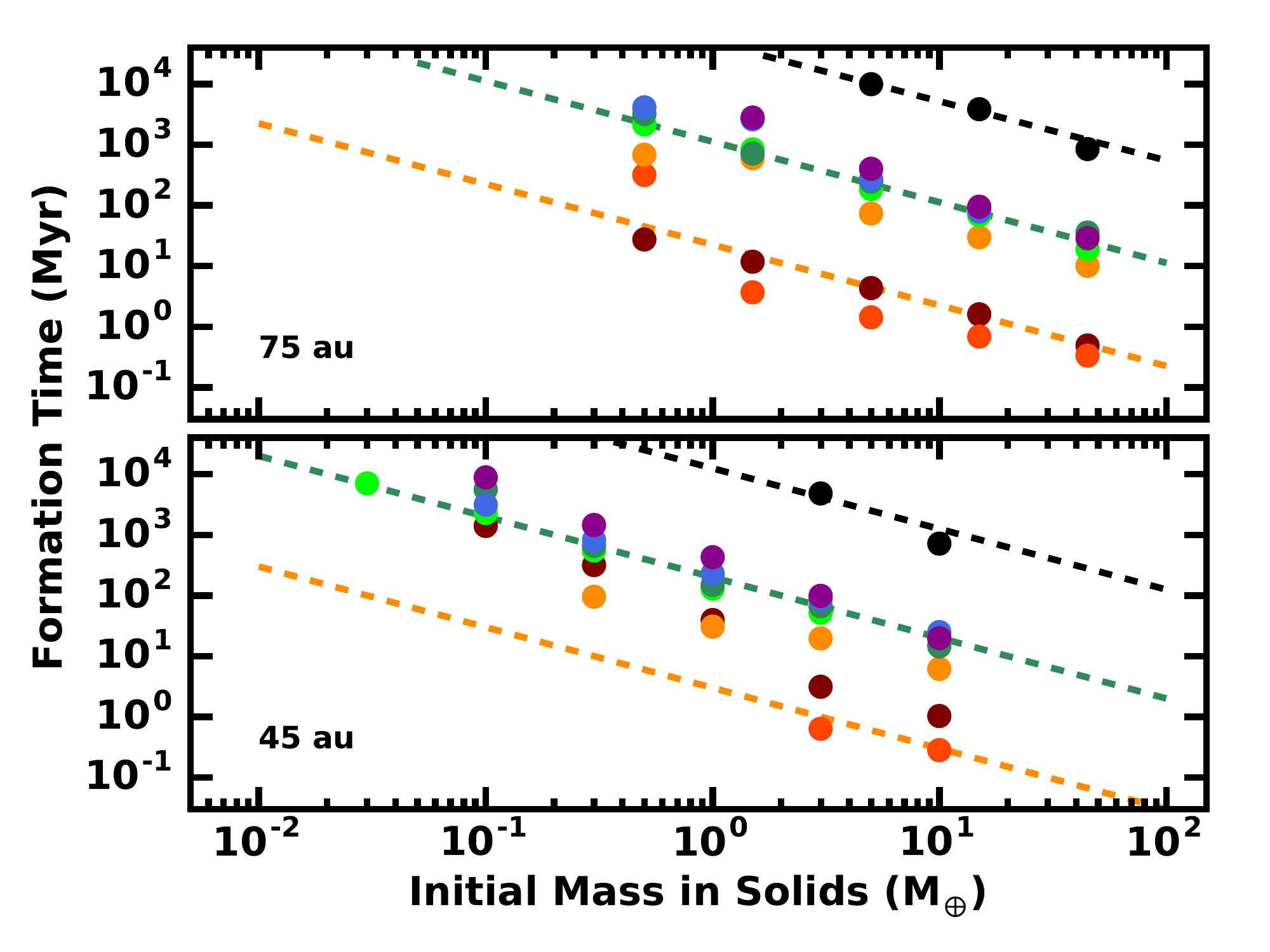}
\vskip -2ex
\caption{
\label{fig: times}
Evolution time $t_{1k}$ to form objects with \rmax\ $\gtrsim$ 1000~km 
for each calculation in Table~\ref{tab: tab1} (lower panel) and 
Table~\ref{tab: tab2} (upper panel). Filled circles indicate results
for $f$ = 1 (black), 0.3 and 0.5 (purple), 0.1 (blue), 
0.01 (dark green), 0.001 (light green), 0.0001 (orange), 
0.00001 (orange-red), and 0 (maroon). Dashed lines show the
relation $t_{1k} \propto \M0^{-1}$ discussed in the text.
}
\end{center}
\end{figure}

{\bf Growth Time as a Function of Initial Mass.}\quad
To conclude this sub-section, Fig.~\ref{fig: times} plots values for
$t_{1k}$ in Tables~\ref{tab: tab1}--\ref{tab: tab2} as a function of
\M0\ for rings at 45~au (lower panel) and 75~au (upper panel). The data
show clear relations between $t_{1k}$ and \M0. For calculations with 
the same $f$, the time scale to form Pluto-mass protoplanets scales
inversely with the initial mass of solids, $t_{1k} \propto \M0^{-1}$. 
Among the full ensemble, there are three trends. Systems with $f$ = 1
($f \lesssim 10^{-5}$) have the longest (shortest) formation times, 
with a three order of magnitude difference independent of \M0. Midway 
between, systems with $10^{-4} \lesssim f \lesssim 0.5$ have a factor of
$\sim$ 30 shorter (longer) formation time than those with $f$ = 1
($f \lesssim 10^{-5}$). The short formation times with small $f$ is a 
hallmark of pebble accretion \citep{gold2004,raf2005}.

The relation for the slow growth rates for systems with $f$ = 1 is 
straightforward to derive. In any 
swarm of solids, the growth time for a solid of radius $r$ is 
$t_c \propto ( r P / \Sigma) [ 1 + (v_{esc} / v)^2 ]^{-1}$, where
$P$ is the orbital period, $\Sigma$ is the surface density of solids, 
$v_{esc}$ is the escape velocity of the pair of colliding solids, 
and $v$ is their collision velocity
\citep[e.g.,][]{saf1969,liss1987,weth1993,gold2004,raf2005,kb2008}. 
When $f$ = 1, 
$v_{esc} \approx v$; $t_c \propto r P / \Sigma$. Compared to the 
prediction in eq.~\ref{eq: t1kp} derived from scaling results in 
\citet{kb2008}, the growth time scales listed in 
Tables \ref{tab: tab1}--\ref{tab: tab2} are a factor of two smaller. 
Analysis of each calculation suggests a simple explanation: in the 
ring-like geometries considered here, planetesimals are concentrated 
in the middle of the ring. Collisions with other planetesimals outside
the central annuli tend to concentrate additional material in these
annuli. This additional material increases the surface density and
lowers the growth time. In the full suite of calculations, the typical
increase in $\Sigma$ is a factor of two, accounting for the factor of
two smaller growth time.

Although the smaller growth times for systems with $f <$ 1 are simple to 
understand, deriving a quantitative relation is a challenge. In a system
of pebbles and planetesimals, damping reduces collision velocities. With
$v \ll v_{esc}$, $t_c \propto (r P / \Sigma) (v/v_{esc})^2$. To evaluate
$(v/v_{esc}^2$, there are two regimes \citep[e.g.,][]{gold2004,raf2005}.
In the `dispersion regime', collision velocities exceed the Hill velocity,
$v_H = \Omega R_H$, where $\Omega$ is the angular velocity. Analytic 
results then yield $(v / v_{esc})^2 \propto f$. In the `shear regime', 
$v < v_H$; then $(v / v_{esc})^2 \propto f^{1/2}$. 

These theories predict the sense of the results in 
Tables~\ref{tab: tab1}--\ref{tab: tab2}, 
but not the clustering of $t_{1k}$ 
for $10^{-4} \lesssim f \lesssim 0.5$ and the second cluster of results 
for smaller $f$. When $f$ is roughly zero, growth is in the shear regime 
and very rapid as summarized in the main text. Larger $f$ places growth 
more in the dispersion regime, where growth is somewhat slower. Our results
suggest that the boundary between rapid and extremely rapid growth is
$f \approx 10^{-5} - 10^{-4}$. However, current theory is insufficient 
to isolate this boundary. At the same time, theory suggests that the 
growth time should scale with $f$ for intermediate $f$ between 0 and 1
rather than a common growth time for a range of $f$. Because relating 
$t_{1k}$ to the initial properties of rings of solids is not a central 
goal of this project, we leave a detailed investigation of this issue
to a separate study.

\section{COMPARISONS WITH PREVIOUS CALCULATIONS}

To illustrate the impact of differences between the calculations in 
this paper and those in \citet{kb2010}, we consider several examples. 
In \citet{kb2010}, solids with radii $r \gtrsim$ 1~m evolve within 64 
annuli extending from an inner radius of 30~au to an outer radius of 
150~au. Solids interact with a gaseous disk; gas drag circularizes orbits 
while radial drift removes smaller particles from the grid. Initially, 
the gas-to-solid ratio is roughly 1:100; during the evolutionary sequence, 
the gas mass declines exponentially with an e-folding time scale of 10~Myr. 
Although \citet{kb2010} consider a broad range of initial conditions, here 
we focus on calculations starting with a mono-disperse set of solids and a 
total mass roughly comparable with the minimum mass solar nebula (solid 
and gas mass).

In the calculations for this paper, the minimum size of solids is 
1~\mum\ instead of 1~m. Interactions with gas are ignored. The smaller 
size allows collisional damping to play a role during the collisional 
cascade \citep[e.g.,][see also 
Figs.~\ref{fig: ecc0}--\ref{fig: ecc1}]{kb2015a,kb2016a,kb2016b}. 
When damping is important, the mass in 0.1~mm to 10~cm particles may 
grow with time and allow a second phase of runaway growth for the 
largest particles. Thus, some solids reach larger masses than in 
calculations without the small particles. The lack of interactions 
with gas tends to slow the growth of the largest particles when 
their initial sizes are 1--10~km. 

\begin{figure}[t]
\begin{center}
\includegraphics[width=5.5in]{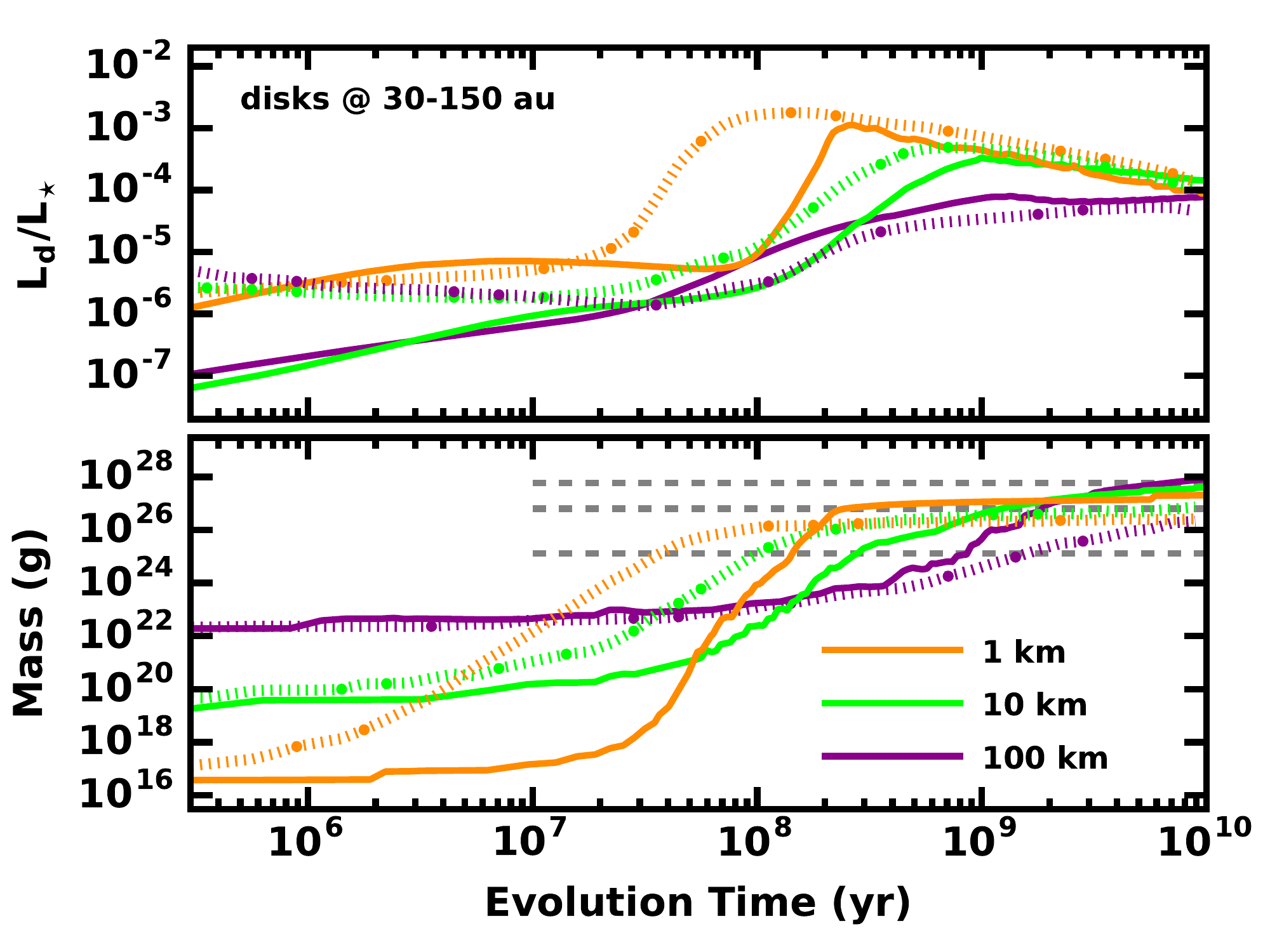}
\vskip -2ex
\caption{
\label{fig: disk0}
Evolution of the largest protoplanets (lower panel) and the relative disk 
luminosity \ldlstar\ (upper panel) in disks of solids at 30--150~au. Dotted 
curves show results from \citet{kb2010}; solid curves follow growth with 
1~\mum\ and larger particles but without gas drag. In the lower panel,
the legend indicates the initial radius for swarms of monodisperse 
solids; dashed lines show the masses of Pluto (lower), Mars (middle), 
and the Earth (upper). Systems with 1--10~km objects evolve faster when 
gas is included in the calculations but reach roughly the same maximum 
\ldlstar. Addition of small particles in the new calculations allows 
planets to reach larger radii.
}
\end{center}
\end{figure}

The lower panel of Fig.~\ref{fig: disk0} compares results for calculations starting with only 
1~km, 10~km, or 100~km objects. At the start of each calculation, 
planetesimals grow slowly. Over 1--3~Myr of evolution, gas drag gradually 
results in larger gravitational focusing factors for the largest objects. 
Despite the loss of collisional debris from radial drift, 1--10~km objects
reach Pluto masses in 30--100~Myr. After this spurt, gravitational 
interactions between the swarm of Plutos and smaller planetesimals 
initiates a collisional cascade, which grinds the leftovers into smaller 
and smaller objects. The loss of material slows growth considerably. The 
largest objects approach the mass of Mars but rarely grow larger.

When gas drag is not included, 1--10 km objects grow more slowly. It takes 
them about three times longer to reach the mass of Pluto. Once they 
initiate the collisional cascade, a modest amount of mass collects in
mm- and cm-sized objects which are stronger and more resistant to the 
cascade than m-sized and larger solids. This reservoir allows Pluto-mass 
objects to grow beyond the mass of Mars and reach the mass of the Earth. 
The time scale to reach Earth-mass is $\sim$ 1--2~Gyr.

Gaseous disks have little impact on the collisional evolution of 100~km 
planetesimals. On time scales of 100--300~Myr, these objects reach masses 
comparable to Charon, Pluto's binary partner. At this point, collisions 
initiate a modest collisional cascade. In calculations with no 
sub-meter-sized particles, the cascade effectively removes debris from 
the grid. Although growth continues, the largest objects reach roughly the 
mass of Mars on 5--10~Gyr time scales. When the small particles are 
included, mass removal is less efficient. Mars (Earth) mass planets 
then form on 1--2~Gyr (5--7~Gyr) time scales.

Despite differences in approach, the evolution of the dust luminosity 
is similar in the two sets of calculations (Fig.~\ref{fig: disk0}, upper
panel). In 
\citet{kb2010}, a second calculation uses the loss rate of sub-meter-sized 
particles from the main coagulation calculation to derive the time 
evolution of 1~\mum\ to 1~m particles; the surface area (and thus 
reprocessed stellar luminosity) of these particles is therefore completely 
distinct from the evolution of the largest particles in the grid. In the 
calculations for this paper, the smallest particles evolve together 
with the largest particles.  This approach should provide a better measure 
of the evolution of the dust luminosity.

The Figure illustrates that the dust luminosity follows the growth of the 
largest objects in each set of calculations. early on, the largest 
particles grow slowly; collisions generate little debris. In turn, the 
dust luminosity is fairly small. As planetesimals grow more rapidly, 
collisions produce copious amounts of dust; the dust luminosity rises in 
step with the growth of the largest planetesimals. As the collisional 
cascade proceeds, there is less and less solid mass in the grid. Growth 
slows; debris production declines; the dust luminosity begins to fade.

For each pair of calculations, the maximum dust luminosity correlates 
well with the initial sizes of planetesimals; 
$L_d / L_\star \approx 10^{-3}$ for 1~km planetesimals,
$L_d / L_\star \approx 3 \times 10^{-4}$ for 10~km planetesimals, and
$L_d / L_\star \approx 10^{-4}$ for 100~km planetesimals,
The timing of this peak is earlier for systems with gas drag and 1--10~km 
planetesimals and is independent of gas drag for 100~km planeteimals.
Independent of the approach, all of the calculations have the same dust 
luminosity at 10~Gyr, \ldlstar $\sim 10^{-4}$. 

Although we do not compare these evolutionary tracks with the data in
Fig.~\ref{fig: alldata} for clarity, a simple comparison shows that the
evolution of full disks provides a poor match to the data. At 100~Myr,
these models have \ldlstar\ $\sim 10^{-5} - 10^{-3}$, as in observed
systems. With \ldlstar\ $\approx 10^{-4}$ at 10~Gyr, all disk models
decline much more slowly than the observations where \ldlstar\ $\sim$
$10^{-5}$ for most sources. As discussed in the main text, rings of
solids evolve in a similar way as the observed systems.

\bibliography{sfpl}{}

\begin{thebibliography}{}
\expandafter\ifx\csname natexlab\endcsname\relax\def\natexlab#1{#1}\fi
\providecommand{\url}[1]{\href{#1}{#1}}
\providecommand{\dodoi}[1]{doi:~\href{http://doi.org/#1}{\nolinkurl{#1}}}
\providecommand{\doeprint}[1]{\href{http://ascl.net/#1}{\nolinkurl{http://ascl.net/#1}}}
\providecommand{\doarXiv}[1]{\href{https://arxiv.org/abs/#1}{\nolinkurl{https://arxiv.org/abs/#1}}}

\bibitem[{{Abod} {et~al.}(2019){Abod}, {Simon}, {Li}, {Armitage}, {Youdin}, \&
  {Kretke}}]{abod2019}
{Abod}, C.~P., {Simon}, J.~B., {Li}, R., {et~al.} 2019, \apj, 883, 192,
  \dodoi{10.3847/1538-4357/ab40a3}

\bibitem[{{Adachi} {et~al.}(1976){Adachi}, {Hayashi}, \& {Nakazawa}}]{ada1976}
{Adachi}, I., {Hayashi}, C., \& {Nakazawa}, K. 1976, Progress of Theoretical
  Physics, 56, 1756

\bibitem[{{Akeson} {et~al.}(2019){Akeson}, {Jensen}, {Carpenter}, {Ricci},
  {Laos}, {Nogueira}, \& {Suen-Lewis}}]{akeson2019}
{Akeson}, R.~L., {Jensen}, E. L.~N., {Carpenter}, J., {et~al.} 2019, \apj, 872,
  158, \dodoi{10.3847/1538-4357/aaff6a}

\bibitem[{{Alexander} \& {Armitage}(2009)}]{alex2009}
{Alexander}, R.~D., \& {Armitage}, P.~J. 2009, \apj, 704, 989,
  \dodoi{10.1088/0004-637X/704/2/989}

\bibitem[{{Alibert} {et~al.}(2018){Alibert}, {Venturini}, {Helled}, {Ataiee},
  {Burn}, {Senecal}, {Benz}, {Mayer}, {Mordasini}, {Quanz}, \&
  {Sch{\"o}nb{\"a}chler}}]{alibert2018}
{Alibert}, Y., {Venturini}, J., {Helled}, R., {et~al.} 2018, Nature Astronomy,
  2, 873, \dodoi{10.1038/s41550-018-0557-2}

\bibitem[{{ALMA Partnership} {et~al.}(2015){ALMA Partnership}, {Brogan},
  {P{\'e}rez}, {Hunter}, {Dent}, {Hales}, {Hills}, {Corder}, {Fomalont},
  {Vlahakis}, {Asaki}, {Barkats}, {Hirota}, {Hodge}, {Impellizzeri}, {Kneissl},
  {Liuzzo}, {Lucas}, {Marcelino}, {Matsushita}, {Nakanishi}, {Phillips},
  {Richards}, {Toledo}, {Aladro}, {Broguiere}, {Cortes}, {Cortes}, {Espada},
  {Galarza}, {Garcia-Appadoo}, {Guzman-Ramirez}, {Humphreys}, {Jung}, {Kameno},
  {Laing}, {Leon}, {Marconi}, {Mignano}, {Nikolic}, {Nyman}, {Radiszcz},
  {Remijan}, {Rod{\'o}n}, {Sawada}, {Takahashi}, {Tilanus}, {Vila Vilaro},
  {Watson}, {Wiklind}, {Akiyama}, {Chapillon}, {de Gregorio-Monsalvo}, {Di
  Francesco}, {Gueth}, {Kawamura}, {Lee}, {Nguyen Luong}, {Mangum}, {Pietu},
  {Sanhueza}, {Saigo}, {Takakuwa}, {Ubach}, {van Kempen}, {Wootten},
  {Castro-Carrizo}, {Francke}, {Gallardo}, {Garcia}, {Gonzalez}, {Hill},
  {Kaminski}, {Kurono}, {Liu}, {Lopez}, {Morales}, {Plarre}, {Schieven},
  {Testi}, {Videla}, {Villard}, {Andreani}, {Hibbard}, \&
  {Tatematsu}}]{alma2015}
{ALMA Partnership}, {Brogan}, C.~L., {P{\'e}rez}, L.~M., {et~al.} 2015, \apjl,
  808, L3, \dodoi{10.1088/2041-8205/808/1/L3}

\bibitem[{{Andrews}(2015)}]{andrews2015}
{Andrews}, S.~M. 2015, \pasp, 127, 961, \dodoi{10.1086/683178}

\bibitem[{{Andrews} {et~al.}(2010){Andrews}, {Czekala}, {Wilner}, {Espaillat},
  {Dullemond}, \& {Hughes}}]{andrews2010}
{Andrews}, S.~M., {Czekala}, I., {Wilner}, D.~J., {et~al.} 2010, \apj, 710,
  462, \dodoi{10.1088/0004-637X/710/1/462}

\bibitem[{{Avenhaus} {et~al.}(2018){Avenhaus}, {Quanz}, {Garufi}, {Perez},
  {Casassus}, {Pinte}, {Bertrang}, {Caceres}, {Benisty}, \&
  {Dominik}}]{avenhaus2018}
{Avenhaus}, H., {Quanz}, S.~P., {Garufi}, A., {et~al.} 2018, \apj, 863, 44,
  \dodoi{10.3847/1538-4357/aab846}

\bibitem[{{Bacciotti} {et~al.}(2018){Bacciotti}, {Girart}, {Padovani}, {Podio},
  {Paladino}, {Testi}, {Bianchi}, {Galli}, {Codella}, {Coffey}, {Favre}, \&
  {Fedele}}]{bacciotti2018}
{Bacciotti}, F., {Girart}, J.~M., {Padovani}, M., {et~al.} 2018, \apjl, 865,
  L12, \dodoi{10.3847/2041-8213/aadf87}

\bibitem[{{Backman} {et~al.}(1995){Backman}, {Dasgupta}, \&
  {Stencel}}]{backman1995}
{Backman}, D.~E., {Dasgupta}, A., \& {Stencel}, R.~E. 1995, \apjl, 450, L35,
  \dodoi{10.1086/309660}

\bibitem[{{Bae} {et~al.}(2018){Bae}, {Pinilla}, \& {Birnstiel}}]{bae2018}
{Bae}, J., {Pinilla}, P., \& {Birnstiel}, T. 2018, \apjl, 864, L26,
  \dodoi{10.3847/2041-8213/aadd51}

\bibitem[{{Basu}(1998)}]{basu1998}
{Basu}, S. 1998, \apj, 509, 229, \dodoi{10.1086/306494}

\bibitem[{{Benz} \& {Asphaug}(1999)}]{benz1999}
{Benz}, W., \& {Asphaug}, E. 1999, Icarus, 142, 5,
  \dodoi{10.1006/icar.1999.6204}

\bibitem[{{Birnstiel} {et~al.}(2016){Birnstiel}, {Fang}, \&
  {Johansen}}]{birnstiel2016}
{Birnstiel}, T., {Fang}, M., \& {Johansen}, A. 2016, \ssr, 205, 41,
  \dodoi{10.1007/s11214-016-0256-1}

\bibitem[{{Bitsch} {et~al.}(2019){Bitsch}, {Izidoro}, {Johansen}, {Raymond},
  {Morbidelli}, {Lambrechts}, \& {Jacobson}}]{bitsch2019}
{Bitsch}, B., {Izidoro}, A., {Johansen}, A., {et~al.} 2019, \aap, 623, A88,
  \dodoi{10.1051/0004-6361/201834489}

\bibitem[{{Bitsch} {et~al.}(2015){Bitsch}, {Johansen}, {Lambrechts}, \&
  {Morbidelli}}]{bitsch2015a}
{Bitsch}, B., {Johansen}, A., {Lambrechts}, M., \& {Morbidelli}, A. 2015, \aap,
  575, A28, \dodoi{10.1051/0004-6361/201424964}

\bibitem[{{Blum} \& {Wurm}(2008)}]{blum2008}
{Blum}, J., \& {Wurm}, G. 2008, \araa, 46, 21,
  \dodoi{10.1146/annurev.astro.46.060407.145152}

\bibitem[{{Bodenheimer} {et~al.}(2018){Bodenheimer}, {Stevenson}, {Lissauer},
  \& {D'Angelo}}]{boden2018}
{Bodenheimer}, P., {Stevenson}, D.~J., {Lissauer}, J.~J., \& {D'Angelo}, G.
  2018, \apj, 868, 138, \dodoi{10.3847/1538-4357/aae928}

\bibitem[{{Brauer} {et~al.}(2008){Brauer}, {Dullemond}, \&
  {Henning}}]{brauer2008}
{Brauer}, F., {Dullemond}, C.~P., \& {Henning}, T. 2008, \aap, 480, 859,
  \dodoi{10.1051/0004-6361:20077759}

\bibitem[{{Brittain} {et~al.}(2019){Brittain}, {Najita}, \&
  {Carr}}]{brittain2019}
{Brittain}, S.~D., {Najita}, J.~R., \& {Carr}, J.~S. 2019, \apj, 883, 37,
  \dodoi{10.3847/1538-4357/ab380b}

\bibitem[{{Bromley} \& {Kenyon}(2006)}]{bk2006}
{Bromley}, B.~C., \& {Kenyon}, S.~J. 2006, \aj, 131, 2737,
  \dodoi{10.1086/503280}

\bibitem[{{Bromley} \& {Kenyon}(2011{\natexlab{a}})}]{bk2011a}
---. 2011{\natexlab{a}}, \apj, 731, 101, \dodoi{10.1088/0004-637X/731/2/101}

\bibitem[{{Bromley} \& {Kenyon}(2011{\natexlab{b}})}]{bk2011b}
---. 2011{\natexlab{b}}, \apj, 735, 29, \dodoi{10.1088/0004-637X/735/1/29}

\bibitem[{{Bromley} \& {Kenyon}(2013)}]{bk2013}
---. 2013, \apj, 764, 192, \dodoi{10.1088/0004-637X/764/2/192}

\bibitem[{{Bromley} \& {Kenyon}(2020)}]{bk2020}
---. 2020, \aj, 160, 85, \dodoi{10.3847/1538-3881/ab9e6c}

\bibitem[{{Bryden} {et~al.}(2009){Bryden}, {Beichman}, {Carpenter}, {Rieke},
  {Stapelfeldt}, {Werner}, {Tanner}, {Lawler}, {Wyatt}, {Trilling}, {Su},
  {Blaylock}, \& {Stansberry}}]{bryden2009}
{Bryden}, G., {Beichman}, C.~A., {Carpenter}, J.~M., {et~al.} 2009, \apj, 705,
  1226, \dodoi{10.1088/0004-637X/705/2/1226}

\bibitem[{{Carpenter} {et~al.}(2009){Carpenter}, {Bouwman}, {Mamajek}, {Meyer},
  {Hillenbrand}, {Backman}, {Henning}, {Hines}, {Hollenbach}, {Kim},
  {Moro-Martin}, {Pascucci}, {Silverstone}, {Stauffer}, \& {Wolf}}]{carp2009a}
{Carpenter}, J.~M., {Bouwman}, J., {Mamajek}, E.~E., {et~al.} 2009, \apjs, 181,
  197, \dodoi{10.1088/0067-0049/181/1/197}

\bibitem[{{Carrera} {et~al.}(2020){Carrera}, {Simon}, {Li}, {Kretke}, \&
  {Klahr}}]{carrera2020}
{Carrera}, D., {Simon}, J.~B., {Li}, R., {Kretke}, K.~A., \& {Klahr}, H. 2020,
  arXiv e-prints, arXiv:2008.01727.
\newblock \doarXiv{2008.01727}

\bibitem[{{Chalmers} {et~al.}(2020){Chalmers}, {Lacy}, {Islam}, {Loewen}, {Van
  Vuuren}, \& {Baker}}]{chalmers2020}
{Chalmers}, D., {Lacy}, G., {Islam}, M., {et~al.} 2020, in Society of
  Photo-Optical Instrumentation Engineers (SPIE) Conference Series, Vol. 11445,
  Society of Photo-Optical Instrumentation Engineers (SPIE) Conference Series,
  114450M, \dodoi{10.1117/12.2562977}

\bibitem[{{Chambers}(2021)}]{chambers2021}
{Chambers}, J. 2021, arXiv e-prints, arXiv:2104.10704.
\newblock \doarXiv{2104.10704}

\bibitem[{{Chambers}(2001)}]{chambers2001a}
{Chambers}, J.~E. 2001, Icarus, 152, 205

\bibitem[{{Chambers} {et~al.}(1996){Chambers}, {Wetherill}, \&
  {Boss}}]{chambers1996}
{Chambers}, J.~E., {Wetherill}, G.~W., \& {Boss}, A.~P. 1996, Icarus, 119, 261,
  \dodoi{10.1006/icar.1996.0019}

\bibitem[{{Chen} \& {Lin}(2020)}]{chen2020}
{Chen}, K., \& {Lin}, M.-K. 2020, \apj, 891, 132,
  \dodoi{10.3847/1538-4357/ab76ca}

\bibitem[{{Chiang} \& {Youdin}(2010)}]{chiang2010}
{Chiang}, E., \& {Youdin}, A.~N. 2010, Annual Review of Earth and Planetary
  Sciences, 38, 493, \dodoi{10.1146/annurev-earth-040809-152513}

\bibitem[{{Chiang} \& {Goldreich}(1997)}]{chiang1997}
{Chiang}, E.~I., \& {Goldreich}, P. 1997, \apj, 490, 368,
  \dodoi{10.1086/304869}

\bibitem[{{Cieza} {et~al.}(2007){Cieza}, {Padgett}, {Stapelfeldt}, {Augereau},
  {Harvey}, {Evans}, {Mer{\'\i}n}, {Koerner}, {Sargent}, {van Dishoeck},
  {Allen}, {Blake}, {Brooke}, {Chapman}, {Huard}, {Lai}, {Mundy}, {Myers},
  {Spiesman}, \& {Wahhaj}}]{cieza2007}
{Cieza}, L., {Padgett}, D.~L., {Stapelfeldt}, K.~R., {et~al.} 2007, \apj, 667,
  308, \dodoi{10.1086/520698}

\bibitem[{{Cieza} {et~al.}(2019){Cieza}, {Ru{\'\i}z-Rodr{\'\i}guez}, {Hales},
  {Casassus}, {P{\'e}rez}, {Gonzalez-Ruilova}, {C{\'a}novas}, {Williams},
  {Zurlo}, {Ansdell}, {Avenhaus}, {Bayo}, {Bertrang}, {Christiaens}, {Dent},
  {Ferrero}, {Gamen}, {Olofsson}, {Orcajo}, {Pe{\~n}a Ram{\'\i}rez},
  {Principe}, {Schreiber}, \& {van der Plas}}]{cieza2019}
{Cieza}, L.~A., {Ru{\'\i}z-Rodr{\'\i}guez}, D., {Hales}, A., {et~al.} 2019,
  \mnras, 482, 698, \dodoi{10.1093/mnras/sty2653}

\bibitem[{{Corder} {et~al.}(2009){Corder}, {Carpenter}, {Sargent}, {Zauderer},
  {Wright}, {White}, {Woody}, {Teuben}, {Scott}, {Pound}, {Plambeck}, {Lamb},
  {Koda}, {Hodges}, {Hawkins}, \& {Bock}}]{corder2009}
{Corder}, S., {Carpenter}, J.~M., {Sargent}, A.~I., {et~al.} 2009, \apjl, 690,
  L65, \dodoi{10.1088/0004-637X/690/1/L65}

\bibitem[{{Cuzzi} {et~al.}(2008){Cuzzi}, {Hogan}, \& {Shariff}}]{cuzzi2008}
{Cuzzi}, J.~N., {Hogan}, R.~C., \& {Shariff}, K. 2008, \apj, 687, 1432,
  \dodoi{10.1086/591239}

\bibitem[{{D'Angelo} {et~al.}(2014){D'Angelo}, {Weidenschilling}, {Lissauer},
  \& {Bodenheimer}}]{dangelo2014}
{D'Angelo}, G., {Weidenschilling}, S.~J., {Lissauer}, J.~J., \& {Bodenheimer},
  P. 2014, \icarus, 241, 298, \dodoi{10.1016/j.icarus.2014.06.029}

\bibitem[{{D'Angelo} {et~al.}(2021){D'Angelo}, {Weidenschilling}, {Lissauer},
  \& {Bodenheimer}}]{dangelo2021}
---. 2021, \icarus, 355, 114087, \dodoi{10.1016/j.icarus.2020.114087}

\bibitem[{{Davis} {et~al.}(1985){Davis}, {Chapman}, {Weidenschilling}, \&
  {Greenberg}}]{davis1985}
{Davis}, D.~R., {Chapman}, C.~R., {Weidenschilling}, S.~J., \& {Greenberg}, R.
  1985, Icarus, 63, 30.
\newblock
  \url{http://adsabs.harvard.edu/cgi-bin/nph-bib_query?bibcode=davis:vesta&db_key=AST}

\bibitem[{{Di Ruscio} {et~al.}(2020){Di Ruscio}, {Fienga}, {Durante}, {Iess},
  {Laskar}, \& {Gastineau}}]{diruscio2020}
{Di Ruscio}, A., {Fienga}, A., {Durante}, D., {et~al.} 2020, \aap, 640, A7,
  \dodoi{10.1051/0004-6361/202037920}

\bibitem[{{Dominik} \& {Decin}(2003)}]{dom2003}
{Dominik}, C., \& {Decin}, G. 2003, \apj, 598, 626, \dodoi{10.1086/379169}

\bibitem[{{Dominik} \& {Tielens}(1997)}]{dominik1997}
{Dominik}, C., \& {Tielens}, A.~G.~G.~M. 1997, \apj, 480, 647,
  \dodoi{10.1086/303996}

\bibitem[{{Dullemond} {et~al.}(2018){Dullemond}, {Birnstiel}, {Huang},
  {Kurtovic}, {Andrews}, {Guzm{\'a}n}, {P{\'e}rez}, {Isella}, {Zhu}, {Benisty},
  {Wilner}, {Bai}, {Carpenter}, {Zhang}, \& {Ricci}}]{dullemond2018}
{Dullemond}, C.~P., {Birnstiel}, T., {Huang}, J., {et~al.} 2018, \apjl, 869,
  L46, \dodoi{10.3847/2041-8213/aaf742}

\bibitem[{{Eiroa} {et~al.}(2013){Eiroa}, {Marshall}, {Mora}, {Montesinos},
  {Absil}, {Augereau}, {Bayo}, {Bryden}, {Danchi}, {del Burgo}, {Ertel},
  {Fridlund}, {Heras}, {Krivov}, {Launhardt}, {Liseau}, {L{\"o}hne},
  {Maldonado}, {Pilbratt}, {Roberge}, {Rodmann}, {Sanz-Forcada}, {Solano},
  {Stapelfeldt}, {Th{\'e}bault}, {Wolf}, {Ardila}, {Ar{\'e}valo}, {Beichmann},
  {Faramaz}, {Gonz{\'a}lez-Garc{\'{\i}}a}, {Guti{\'e}rrez}, {Lebreton},
  {Mart{\'{\i}}nez-Arn{\'a}iz}, {Meeus}, {Montes}, {Olofsson}, {Su}, {White},
  {Barrado}, {Fukagawa}, {Gr{\"u}n}, {Kamp}, {Lorente}, {Morbidelli},
  {M{\"u}ller}, {Mutschke}, {Nakagawa}, {Ribas}, \& {Walker}}]{eiroa2013}
{Eiroa}, C., {Marshall}, J.~P., {Mora}, A., {et~al.} 2013, \aap, 555, A11,
  \dodoi{10.1051/0004-6361/201321050}

\bibitem[{{Esplin} \& {Luhman}(2019)}]{esplin2019}
{Esplin}, T.~L., \& {Luhman}, K.~L. 2019, \aj, 158, 54,
  \dodoi{10.3847/1538-3881/ab2594}

\bibitem[{{Facchini} {et~al.}(2019){Facchini}, {van Dishoeck}, {Manara},
  {Tazzari}, {Maud}, {Cazzoletti}, {Rosotti}, {van der Marel}, {Pinilla}, \&
  {Clarke}}]{facchini2019}
{Facchini}, S., {van Dishoeck}, E.~F., {Manara}, C.~F., {et~al.} 2019, \aap,
  626, L2, \dodoi{10.1051/0004-6361/201935496}

\bibitem[{{Facchini} {et~al.}(2020){Facchini}, {Benisty}, {Bae}, {Loomis},
  {Perez}, {Ansdell}, {Mayama}, {Pinilla}, {Teague}, {Isella}, \&
  {Mann}}]{facchini2020}
{Facchini}, S., {Benisty}, M., {Bae}, J., {et~al.} 2020, \aap, 639, A121,
  \dodoi{10.1051/0004-6361/202038027}

\bibitem[{{Faherty} {et~al.}(2018){Faherty}, {Bochanski}, {Gagn{\'e}},
  {Nelson}, {Coker}, {Smithka}, {Desir}, \& {Vasquez}}]{faherty2018}
{Faherty}, J.~K., {Bochanski}, J.~J., {Gagn{\'e}}, J., {et~al.} 2018, \apj,
  863, 91, \dodoi{10.3847/1538-4357/aac76e}

\bibitem[{{Gagn{\'e}} {et~al.}(2020){Gagn{\'e}}, {David}, {Mamajek}, {Mann},
  {Faherty}, \& {B{\'e}dard}}]{gagne2020}
{Gagn{\'e}}, J., {David}, T.~J., {Mamajek}, E.~E., {et~al.} 2020, \apj, 903,
  96, \dodoi{10.3847/1538-4357/abb77e}

\bibitem[{{Gagn{\'e}} {et~al.}(2018){Gagn{\'e}}, {Faherty}, \&
  {Mamajek}}]{gagne2018}
{Gagn{\'e}}, J., {Faherty}, J.~K., \& {Mamajek}, E.~E. 2018, \apj, 865, 136,
  \dodoi{10.3847/1538-4357/aadaed}

\bibitem[{{Gerbig} {et~al.}(2020){Gerbig}, {Murray-Clay}, {Klahr}, \&
  {Baehr}}]{gerbig2020}
{Gerbig}, K., {Murray-Clay}, R.~A., {Klahr}, H., \& {Baehr}, H. 2020, \apj,
  895, 91, \dodoi{10.3847/1538-4357/ab8d37}

\bibitem[{{Gladman}(1993)}]{gladman1993}
{Gladman}, B. 1993, \icarus, 106, 247, \dodoi{10.1006/icar.1993.1169}

\bibitem[{{Goldreich} {et~al.}(2004){Goldreich}, {Lithwick}, \&
  {Sari}}]{gold2004}
{Goldreich}, P., {Lithwick}, Y., \& {Sari}, R. 2004, \araa, 42, 549

\bibitem[{{Gole} {et~al.}(2020){Gole}, {Simon}, {Li}, {Youdin}, \&
  {Armitage}}]{gole2020}
{Gole}, D.~A., {Simon}, J.~B., {Li}, R., {Youdin}, A.~N., \& {Armitage}, P.~J.
  2020, \apj, 904, 132, \dodoi{10.3847/1538-4357/abc334}

\bibitem[{{Gundlach} \& {Blum}(2015)}]{grundlach2015}
{Gundlach}, B., \& {Blum}, J. 2015, \apj, 798, 34,
  \dodoi{10.1088/0004-637X/798/1/34}

\bibitem[{{Haffert} {et~al.}(2019){Haffert}, {Bohn}, {de Boer}, {Snellen},
  {Brinchmann}, {Girard}, {Keller}, \& {Bacon}}]{haffert2019}
{Haffert}, S.~Y., {Bohn}, A.~J., {de Boer}, J., {et~al.} 2019, Nature
  Astronomy, 3, 749, \dodoi{10.1038/s41550-019-0780-5}

\bibitem[{{Hardy} {et~al.}(2015){Hardy}, {Caceres}, {Schreiber}, {Cieza},
  {Alexander}, {Canovas}, {Williams}, {Wahhaj}, \& {Menard}}]{hardy2015}
{Hardy}, A., {Caceres}, C., {Schreiber}, M.~R., {et~al.} 2015, \aap, 583, A66,
  \dodoi{10.1051/0004-6361/201526504}

\bibitem[{{Hartlep} \& {Cuzzi}(2020)}]{hartlep2020}
{Hartlep}, T., \& {Cuzzi}, J.~N. 2020, \apj, 892, 120,
  \dodoi{10.3847/1538-4357/ab76c3}

\bibitem[{{Hashimoto} {et~al.}(2021){Hashimoto}, {Muto}, {Dong}, {Liu}, {van
  der Marel}, {Francis}, {Hasegawa}, \& {Tsukagoshi}}]{hashimoto2021}
{Hashimoto}, J., {Muto}, T., {Dong}, R., {et~al.} 2021, arXiv e-prints,
  arXiv:2102.05905.
\newblock \doarXiv{2102.05905}

\bibitem[{{Hennebelle} {et~al.}(2016){Hennebelle}, {Commer{\c{c}}on},
  {Chabrier}, \& {Marchand}}]{hennebelle2016}
{Hennebelle}, P., {Commer{\c{c}}on}, B., {Chabrier}, G., \& {Marchand}, P.
  2016, \apjl, 830, L8, \dodoi{10.3847/2041-8205/830/1/L8}

\bibitem[{{Hillenbrand} {et~al.}(2008){Hillenbrand}, {Carpenter}, {Kim},
  {Meyer}, {Backman}, {Moro-Mart{\'{\i}}n}, {Hollenbach}, {Hines}, {Pascucci},
  \& {Bouwman}}]{hillen2008}
{Hillenbrand}, L.~A., {Carpenter}, J.~M., {Kim}, J.~S., {et~al.} 2008, \apj,
  677, 630, \dodoi{10.1086/529027}

\bibitem[{{Homma} {et~al.}(2019){Homma}, {Okuzumi}, {Nakamoto}, \&
  {Ueda}}]{homma2019}
{Homma}, K.~A., {Okuzumi}, S., {Nakamoto}, T., \& {Ueda}, Y. 2019, \apj, 877,
  128, \dodoi{10.3847/1538-4357/ab1de0}

\bibitem[{{Huang} {et~al.}(2018){Huang}, {Andrews}, {Dullemond}, {Isella},
  {P{\'e}rez}, {Guzm{\'a}n}, {{\"O}berg}, {Zhu}, {Zhang}, {Bai}, {Benisty},
  {Birnstiel}, {Carpenter}, {Hughes}, {Ricci}, {Weaver}, \&
  {Wilner}}]{huang2018a}
{Huang}, J., {Andrews}, S.~M., {Dullemond}, C.~P., {et~al.} 2018, \apjl, 869,
  L42, \dodoi{10.3847/2041-8213/aaf740}

\bibitem[{{Hughes} {et~al.}(2018){Hughes}, {Duch{\^e}ne}, \&
  {Matthews}}]{hughes2018}
{Hughes}, A.~M., {Duch{\^e}ne}, G., \& {Matthews}, B.~C. 2018, \araa, 56, 541,
  \dodoi{10.1146/annurev-astro-081817-052035}

\bibitem[{{Hyodo} {et~al.}(2019){Hyodo}, {Ida}, \& {Charnoz}}]{hyodo2019}
{Hyodo}, R., {Ida}, S., \& {Charnoz}, S. 2019, \aap, 629, A90,
  \dodoi{10.1051/0004-6361/201935935}

\bibitem[{{Ida} \& {Guillot}(2016)}]{ida2016}
{Ida}, S., \& {Guillot}, T. 2016, \aap, 596, L3,
  \dodoi{10.1051/0004-6361/201629680}

\bibitem[{{Johansen} \& {Bitsch}(2019)}]{johansen2019}
{Johansen}, A., \& {Bitsch}, B. 2019, \aap, 631, A70,
  \dodoi{10.1051/0004-6361/201936351}

\bibitem[{{Johansen} {et~al.}(2007){Johansen}, {Oishi}, {Mac Low}, {Klahr},
  {Henning}, \& {Youdin}}]{johansen2007b}
{Johansen}, A., {Oishi}, J.~S., {Mac Low}, M.-M., {et~al.} 2007, \nat, 448,
  1022, \dodoi{10.1038/nature06086}

\bibitem[{{Johansen} \& {Youdin}(2007)}]{johansen2007a}
{Johansen}, A., \& {Youdin}, A. 2007, \apj, 662, 627, \dodoi{10.1086/516730}

\bibitem[{{Johansen} {et~al.}(2009){Johansen}, {Youdin}, \& {Mac
  Low}}]{johansen2009}
{Johansen}, A., {Youdin}, A., \& {Mac Low}, M.-M. 2009, \apjl, 704, L75,
  \dodoi{10.1088/0004-637X/704/2/L75}

\bibitem[{{Joos} {et~al.}(2012){Joos}, {Hennebelle}, \& {Ciardi}}]{joos2012}
{Joos}, M., {Hennebelle}, P., \& {Ciardi}, A. 2012, \aap, 543, A128,
  \dodoi{10.1051/0004-6361/201118730}

\bibitem[{{Kataoka} {et~al.}(2013){Kataoka}, {Tanaka}, {Okuzumi}, \&
  {Wada}}]{kataoka2013}
{Kataoka}, A., {Tanaka}, H., {Okuzumi}, S., \& {Wada}, K. 2013, \aap, 557, L4,
  \dodoi{10.1051/0004-6361/201322151}

\bibitem[{{Kelling} {et~al.}(2014){Kelling}, {Wurm}, \&
  {K{\"o}ster}}]{kelling2014}
{Kelling}, T., {Wurm}, G., \& {K{\"o}ster}, M. 2014, \apj, 783, 111,
  \dodoi{10.1088/0004-637X/783/2/111}

\bibitem[{{Kennedy} \& {Wyatt}(2010)}]{kennedy2010}
{Kennedy}, G.~M., \& {Wyatt}, M.~C. 2010, \mnras, 405, 1253,
  \dodoi{10.1111/j.1365-2966.2010.16528.x}

\bibitem[{{Kenyon}(2002)}]{kenyon2002}
{Kenyon}, S.~J. 2002, \pasp, 114, 265, \dodoi{10.1086/339188}

\bibitem[{{Kenyon} \& {Bromley}(2004)}]{kb2004a}
{Kenyon}, S.~J., \& {Bromley}, B.~C. 2004, \aj, 127, 513,
  \dodoi{10.1086/379854}

\bibitem[{{Kenyon} \& {Bromley}(2006)}]{kb2006}
---. 2006, \aj, 131, 1837, \dodoi{10.1086/499807}

\bibitem[{{Kenyon} \& {Bromley}(2008)}]{kb2008}
---. 2008, \apjs, 179, 451, \dodoi{10.1086/591794}

\bibitem[{{Kenyon} \& {Bromley}(2009)}]{kb2009}
---. 2009, \apjl, 690, L140, \dodoi{10.1088/0004-637X/690/2/L140}

\bibitem[{{Kenyon} \& {Bromley}(2010)}]{kb2010}
---. 2010, \apjs, 188, 242, \dodoi{10.1088/0067-0049/188/1/242}

\bibitem[{{Kenyon} \& {Bromley}(2012)}]{kb2012}
---. 2012, \aj, 143, 63, \dodoi{10.1088/0004-6256/143/3/63}

\bibitem[{{Kenyon} \& {Bromley}(2015)}]{kb2015a}
---. 2015, \apj, 806, 42, \dodoi{10.1088/0004-637X/806/1/42}

\bibitem[{{Kenyon} \& {Bromley}(2016{\natexlab{a}})}]{kb2016b}
---. 2016{\natexlab{a}}, \apj, 825, 33, \dodoi{10.3847/0004-637X/825/1/33}

\bibitem[{{Kenyon} \& {Bromley}(2016{\natexlab{b}})}]{kb2016a}
---. 2016{\natexlab{b}}, \apj, 817, 51.
\newblock \doarXiv{1512.01273}

\bibitem[{{Kenyon} \& {Bromley}(2017)}]{kb2017a}
---. 2017, \apj, 839, 38, \dodoi{10.3847/1538-4357/aa6982}

\bibitem[{{Kenyon} \& {Bromley}(2020)}]{kb2020}
---. 2020, The Planetary Science Journal, 1, 40, \dodoi{10.3847/PSJ/aba8a9}

\bibitem[{{Kenyon} \& {Bromley}(2021)}]{kb2021a}
---. 2021, \aj, 161, 211, \dodoi{10.3847/1538-3881/abe858}

\bibitem[{{Kenyon} \& {Hartmann}(1987)}]{kh1987}
{Kenyon}, S.~J., \& {Hartmann}, L. 1987, \apj, 323, 714, \dodoi{10.1086/165866}

\bibitem[{{Kenyon} \& {Luu}(1999)}]{kl1999a}
{Kenyon}, S.~J., \& {Luu}, J.~X. 1999, \aj, 118, 1101, \dodoi{10.1086/300969}

\bibitem[{{Keppler} {et~al.}(2018){Keppler}, {Benisty}, {M{\"u}ller},
  {Henning}, {van Boekel}, {Cantalloube}, {Ginski}, {van Holstein}, {Maire},
  {Pohl}, {Samland}, {Avenhaus}, {Baudino}, {Boccaletti}, {de Boer},
  {Bonnefoy}, {Chauvin}, {Desidera}, {Langlois}, {Lazzoni}, {Marleau},
  {Mordasini}, {Pawellek}, {Stolker}, {Vigan}, {Zurlo}, {Birnstiel},
  {Brandner}, {Feldt}, {Flock}, {Girard}, {Gratton}, {Hagelberg}, {Isella},
  {Janson}, {Juhasz}, {Kemmer}, {Kral}, {Lagrange}, {Launhardt}, {Matter},
  {M{\'e}nard}, {Milli}, {Molli{\`e}re}, {Olofsson}, {P{\'e}rez}, {Pinilla},
  {Pinte}, {Quanz}, {Schmidt}, {Udry}, {Wahhaj}, {Williams}, {Buenzli},
  {Cudel}, {Dominik}, {Galicher}, {Kasper}, {Lannier}, {Mesa}, {Mouillet},
  {Peretti}, {Perrot}, {Salter}, {Sissa}, {Wildi}, {Abe}, {Antichi},
  {Augereau}, {Baruffolo}, {Baudoz}, {Bazzon}, {Beuzit}, {Blanchard}, {Brems},
  {Buey}, {De Caprio}, {Carbillet}, {Carle}, {Cascone}, {Cheetham}, {Claudi},
  {Costille}, {Delboulb{\'e}}, {Dohlen}, {Fantinel}, {Feautrier}, {Fusco},
  {Giro}, {Gluck}, {Gry}, {Hubin}, {Hugot}, {Jaquet}, {Le Mignant}, {Llored},
  {Madec}, {Magnard}, {Martinez}, {Maurel}, {Meyer}, {M{\"o}ller-Nilsson},
  {Moulin}, {Mugnier}, {Orign{\'e}}, {Pavlov}, {Perret}, {Petit}, {Pragt},
  {Puget}, {Rabou}, {Ramos}, {Rigal}, {Rochat}, {Roelfsema}, {Rousset}, {Roux},
  {Salasnich}, {Sauvage}, {Sevin}, {Soenke}, {Stadler}, {Suarez}, {Turatto}, \&
  {Weber}}]{keppler2018}
{Keppler}, M., {Benisty}, M., {M{\"u}ller}, A., {et~al.} 2018, \aap, 617, A44,
  \dodoi{10.1051/0004-6361/201832957}

\bibitem[{{Klahr} \& {Schreiber}(2020)}]{klahr2020}
{Klahr}, H., \& {Schreiber}, A. 2020, \apj, 901, 54,
  \dodoi{10.3847/1538-4357/abac58}

\bibitem[{{Kobayashi} \& {Tanaka}(2010)}]{koba2010a}
{Kobayashi}, H., \& {Tanaka}, H. 2010, \icarus, 206, 735,
  \dodoi{10.1016/j.icarus.2009.10.004}

\bibitem[{{Kobayashi} {et~al.}(2010){Kobayashi}, {Tanaka}, {Krivov}, \&
  {Inaba}}]{koba2010b}
{Kobayashi}, H., {Tanaka}, H., {Krivov}, A.~V., \& {Inaba}, S. 2010, \icarus,
  209, 836, \dodoi{10.1016/j.icarus.2010.04.021}

\bibitem[{{Kokubo} \& {Ida}(1995)}]{kok1995}
{Kokubo}, E., \& {Ida}, S. 1995, Icarus, 114, 247

\bibitem[{{Kokubo} \& {Ida}(1998)}]{kok1998}
---. 1998, Icarus, 131, 171

\bibitem[{{Krapp} {et~al.}(2020){Krapp}, {Youdin}, {Kratter}, \&
  {Ben{\'\i}tez-Llambay}}]{krapp2020}
{Krapp}, L., {Youdin}, A.~N., {Kratter}, K.~M., \& {Ben{\'\i}tez-Llambay}, P.
  2020, \mnras, 497, 2715, \dodoi{10.1093/mnras/staa1854}

\bibitem[{{Krasnopolsky} \& {K{\"o}nigl}(2002)}]{krasnopolsky2002}
{Krasnopolsky}, R., \& {K{\"o}nigl}, A. 2002, \apj, 580, 987,
  \dodoi{10.1086/343890}

\bibitem[{{Krist} {et~al.}(2012){Krist}, {Stapelfeldt}, {Bryden}, \&
  {Plavchan}}]{krist2012}
{Krist}, J.~E., {Stapelfeldt}, K.~R., {Bryden}, G., \& {Plavchan}, P. 2012,
  \aj, 144, 45, \dodoi{10.1088/0004-6256/144/2/45}

\bibitem[{{Krist} {et~al.}(2010){Krist}, {Stapelfeldt}, {Bryden}, {Rieke},
  {Su}, {Chen}, {Beichman}, {Hines}, {Rebull}, {Tanner}, {Trilling}, {Clampin},
  \& {G{\'a}sp{\'a}r}}]{krist2010}
{Krist}, J.~E., {Stapelfeldt}, K.~R., {Bryden}, G., {et~al.} 2010, \aj, 140,
  1051, \dodoi{10.1088/0004-6256/140/4/1051}

\bibitem[{{Krivov} {et~al.}(2008){Krivov}, {M{\"u}ller}, {L{\"o}hne}, \&
  {Mutschke}}]{krivov2008}
{Krivov}, A.~V., {M{\"u}ller}, S., {L{\"o}hne}, T., \& {Mutschke}, H. 2008,
  \apj, 687, 608, \dodoi{10.1086/591507}

\bibitem[{{Krivov} \& {Wyatt}(2021)}]{krivov2021}
{Krivov}, A.~V., \& {Wyatt}, M.~C. 2021, \mnras, 500, 718,
  \dodoi{10.1093/mnras/staa2385}

\bibitem[{{Kruss} {et~al.}(2017){Kruss}, {Teiser}, \& {Wurm}}]{kruss2017}
{Kruss}, M., {Teiser}, J., \& {Wurm}, G. 2017, \aap, 600, A103,
  \dodoi{10.1051/0004-6361/201630251}

\bibitem[{{Kruss} \& {Wurm}(2020)}]{kruss2020}
{Kruss}, M., \& {Wurm}, G. 2020, The Planetary Science Journal, 1, 23,
  \dodoi{10.3847/PSJ/ab93c4}

\bibitem[{{Lambrechts} \& {Johansen}(2012)}]{lamb2012}
{Lambrechts}, M., \& {Johansen}, A. 2012, \aap, 544, A32,
  \dodoi{10.1051/0004-6361/201219127}

\bibitem[{{Lambrechts} {et~al.}(2019){Lambrechts}, {Morbidelli}, {Jacobson},
  {Johansen}, {Bitsch}, {Izidoro}, \& {Raymond}}]{lambrechts2019}
{Lambrechts}, M., {Morbidelli}, A., {Jacobson}, S.~A., {et~al.} 2019, \aap,
  627, A83, \dodoi{10.1051/0004-6361/201834229}

\bibitem[{{Leinhardt} \& {Stewart}(2012)}]{lein2012}
{Leinhardt}, Z.~M., \& {Stewart}, S.~T. 2012, \apj, 745, 79,
  \dodoi{10.1088/0004-637X/745/1/79}

\bibitem[{{Lenz} {et~al.}(2019){Lenz}, {Klahr}, \& {Birnstiel}}]{lenz2019}
{Lenz}, C.~T., {Klahr}, H., \& {Birnstiel}, T. 2019, \apj, 874, 36,
  \dodoi{10.3847/1538-4357/ab05d9}

\bibitem[{{Levison} {et~al.}(2010){Levison}, {Thommes}, \&
  {Duncan}}]{levison2010}
{Levison}, H.~F., {Thommes}, E., \& {Duncan}, M.~J. 2010, \aj, 139, 1297,
  \dodoi{10.1088/0004-6256/139/4/1297}

\bibitem[{{Li} {et~al.}(2018){Li}, {Youdin}, \& {Simon}}]{li2018}
{Li}, R., {Youdin}, A.~N., \& {Simon}, J.~B. 2018, \apj, 862, 14,
  \dodoi{10.3847/1538-4357/aaca99}

\bibitem[{{Li} {et~al.}(2019){Li}, {Youdin}, \& {Simon}}]{li2019}
---. 2019, \apj, 885, 69, \dodoi{10.3847/1538-4357/ab480d}

\bibitem[{{Lin} {et~al.}(2018){Lin}, {Lee}, \& {Chiang}}]{lin2018}
{Lin}, J.~W., {Lee}, E.~J., \& {Chiang}, E. 2018, \mnras, 480, 4338,
  \dodoi{10.1093/mnras/sty2159}

\bibitem[{{Lissauer}(1987)}]{liss1987}
{Lissauer}, J.~J. 1987, Icarus, 69, 249, \dodoi{10.1016/0019-1035(87)90104-7}

\bibitem[{{Lissauer} {et~al.}(2009){Lissauer}, {Hubickyj}, {D'Angelo}, \&
  {Bodenheimer}}]{liss2009}
{Lissauer}, J.~J., {Hubickyj}, O., {D'Angelo}, G., \& {Bodenheimer}, P. 2009,
  \icarus, 199, 338, \dodoi{10.1016/j.icarus.2008.10.004}

\bibitem[{{Liu} \& {Ji}(2020)}]{liu2020}
{Liu}, B., \& {Ji}, J. 2020, Research in Astronomy and Astrophysics, 20, 164,
  \dodoi{10.1088/1674-4527/20/10/164}

\bibitem[{{Liu} {et~al.}(2019){Liu}, {Ormel}, \& {Johansen}}]{lei2019}
{Liu}, B., {Ormel}, C.~W., \& {Johansen}, A. 2019, \aap, 624, A114,
  \dodoi{10.1051/0004-6361/201834174}

\bibitem[{{Lodato} {et~al.}(2019){Lodato}, {Dipierro}, {Ragusa}, {Long},
  {Herczeg}, {Pascucci}, {Pinilla}, {Manara}, {Tazzari}, {Liu}, {Mulders},
  {Harsono}, {Boehler}, {M{\'e}nard}, {Johnstone}, {Salyk}, {van der Plas},
  {Cabrit}, {Edwards}, {Fischer}, {Hendler}, {Nisini}, {Rigliaco}, {Avenhaus},
  {Banzatti}, \& {Gully-Santiago}}]{lodato2019}
{Lodato}, G., {Dipierro}, G., {Ragusa}, E., {et~al.} 2019, \mnras, 486, 453,
  \dodoi{10.1093/mnras/stz913}

\bibitem[{{L{\"o}hne} {et~al.}(2008){L{\"o}hne}, {Krivov}, \&
  {Rodmann}}]{lohne2008}
{L{\"o}hne}, T., {Krivov}, A.~V., \& {Rodmann}, J. 2008, \apj, 673, 1123,
  \dodoi{10.1086/524840}

\bibitem[{{Long} {et~al.}(2018){Long}, {Pinilla}, {Herczeg}, {Harsono},
  {Dipierro}, {Pascucci}, {Hendler}, {Tazzari}, {Ragusa}, {Salyk}, {Edwards},
  {Lodato}, {van de Plas}, {Johnstone}, {Liu}, {Boehler}, {Cabrit}, {Manara},
  {Menard}, {Mulders}, {Nisini}, {Fischer}, {Rigliaco}, {Banzatti}, {Avenhaus},
  \& {Gully-Santiago}}]{long2018}
{Long}, F., {Pinilla}, P., {Herczeg}, G.~J., {et~al.} 2018, \apj, 869, 17,
  \dodoi{10.3847/1538-4357/aae8e1}

\bibitem[{{Long} {et~al.}(2019){Long}, {Herczeg}, {Harsono}, {Pinilla},
  {Tazzari}, {Manara}, {Pascucci}, {Cabrit}, {Nisini}, {Johnstone}, {Edwards},
  {Salyk}, {Menard}, {Lodato}, {Boehler}, {Mace}, {Liu}, {Mulders}, {Hendler},
  {Ragusa}, {Fischer}, {Banzatti}, {Rigliaco}, {van de Plas}, {Dipierro},
  {Gully-Santiago}, \& {Lopez-Valdivia}}]{long2019}
{Long}, F., {Herczeg}, G.~J., {Harsono}, D., {et~al.} 2019, \apj, 882, 49,
  \dodoi{10.3847/1538-4357/ab2d2d}

\bibitem[{{Loomis} {et~al.}(2017){Loomis}, {{\"O}berg}, {Andrews}, \&
  {MacGregor}}]{loomis2017}
{Loomis}, R.~A., {{\"O}berg}, K.~I., {Andrews}, S.~M., \& {MacGregor}, M.~A.
  2017, \apj, 840, 23, \dodoi{10.3847/1538-4357/aa6c63}

\bibitem[{{Lovell} {et~al.}(2021){Lovell}, {Wyatt}, {Ansdell}, {Kama},
  {Kennedy}, {Manara}, {Marino}, {Matr{\`a}}, {Rosotti}, {Tazzari}, {Testi}, \&
  {Williams}}]{lovell2021}
{Lovell}, J.~B., {Wyatt}, M.~C., {Ansdell}, M., {et~al.} 2021, \mnras, 500,
  4878, \dodoi{10.1093/mnras/staa3335}

\bibitem[{{Luhman}(2018)}]{luhman2018}
{Luhman}, K.~L. 2018, \aj, 156, 271, \dodoi{10.3847/1538-3881/aae831}

\bibitem[{{Luhman} {et~al.}(2010){Luhman}, {Allen}, {Espaillat}, {Hartmann}, \&
  {Calvet}}]{luhman2010}
{Luhman}, K.~L., {Allen}, P.~R., {Espaillat}, C., {Hartmann}, L., \& {Calvet},
  N. 2010, \apjs, 186, 111, \dodoi{10.1088/0067-0049/186/1/111}

\bibitem[{{Mamajek} \& {Hillenbrand}(2008)}]{mamajek2008}
{Mamajek}, E.~E., \& {Hillenbrand}, L.~A. 2008, \apj, 687, 1264,
  \dodoi{10.1086/591785}

\bibitem[{{Marino} {et~al.}(2018){Marino}, {Carpenter}, {Wyatt}, {Booth},
  {Casassus}, {Faramaz}, {Guzman}, {Hughes}, {Isella}, {Kennedy}, {Matr{\`a}},
  {Ricci}, \& {Corder}}]{marino2018}
{Marino}, S., {Carpenter}, J., {Wyatt}, M.~C., {et~al.} 2018, \mnras, 479,
  5423, \dodoi{10.1093/mnras/sty1790}

\bibitem[{{Marino} {et~al.}(2020){Marino}, {Zurlo}, {Faramaz}, {Milli},
  {Henning}, {Kennedy}, {Matr{\`a}}, {P{\'e}rez}, {Delorme}, {Cieza}, \&
  {Hughes}}]{marino2020}
{Marino}, S., {Zurlo}, A., {Faramaz}, V., {et~al.} 2020, \mnras, 498, 1319,
  \dodoi{10.1093/mnras/staa2386}

\bibitem[{{Marshall} {et~al.}(2011){Marshall}, {L{\"o}hne}, {Montesinos},
  {Krivov}, {Eiroa}, {Absil}, {Bryden}, {Maldonado}, {Mora}, {Sanz-Forcada},
  {Ardila}, {Augereau}, {Bayo}, {Del Burgo}, {Danchi}, {Ertel}, {Fedele},
  {Fridlund}, {Lebreton}, {Gonz{\'a}lez-Garc{\'{\i}}a}, {Liseau}, {Meeus},
  {M{\"u}ller}, {Pilbratt}, {Roberge}, {Stapelfeldt}, {Th{\'e}bault}, {White},
  \& {Wolf}}]{marshall2011}
{Marshall}, J.~P., {L{\"o}hne}, T., {Montesinos}, B., {et~al.} 2011, \aap, 529,
  A117, \dodoi{10.1051/0004-6361/201116673}

\bibitem[{{Martin} \& {Livio}(2012)}]{martin2012}
{Martin}, R.~G., \& {Livio}, M. 2012, \mnras, 425, L6,
  \dodoi{10.1111/j.1745-3933.2012.01290.x}

\bibitem[{{Martin} \& {Livio}(2014)}]{martin2014}
---. 2014, \apjl, 783, L28, \dodoi{10.1088/2041-8205/783/2/L28}

\bibitem[{{Matr{\`a}} {et~al.}(2018){Matr{\`a}}, {Marino}, {Kennedy}, {Wyatt},
  {{\"O}berg}, \& {Wilner}}]{matra2018}
{Matr{\`a}}, L., {Marino}, S., {Kennedy}, G.~M., {et~al.} 2018, \apj, 859, 72,
  \dodoi{10.3847/1538-4357/aabcc4}

\bibitem[{{Matsumoto} {et~al.}(1997){Matsumoto}, {Hanawa}, \&
  {Nakamura}}]{matsumoto1997}
{Matsumoto}, T., {Hanawa}, T., \& {Nakamura}, F. 1997, \apj, 478, 569,
  \dodoi{10.1086/303822}

\bibitem[{{Matsumura} {et~al.}(2017){Matsumura}, {Brasser}, \&
  {Ida}}]{matsumura2017}
{Matsumura}, S., {Brasser}, R., \& {Ida}, S. 2017, \aap, 607, A67,
  \dodoi{10.1051/0004-6361/201731155}

\bibitem[{{Matthews} {et~al.}(2018){Matthews}, {Greaves}, {Kennedy}, {Matra},
  {Wilner}, \& {Wyatt}}]{matthews2018}
{Matthews}, B., {Greaves}, J., {Kennedy}, G., {et~al.} 2018, arXiv e-prints,
  arXiv:1810.06719.
\newblock \doarXiv{1810.06719}

\bibitem[{{Matthews} {et~al.}(2014){Matthews}, {Krivov}, {Wyatt}, {Bryden}, \&
  {Eiroa}}]{matthews2014}
{Matthews}, B.~C., {Krivov}, A.~V., {Wyatt}, M.~C., {Bryden}, G., \& {Eiroa},
  C. 2014, in Protostars and Planet VI, ed. {Beuther, H., Klessen, R. S.,
  Dullemond, C.~P., \& Henning, T.} (The University of Arizona Press, Tucson,
  AZ), 521--544

\bibitem[{{McNally} {et~al.}(2021){McNally}, {Lovascio}, \&
  {Paardekooper}}]{mcnally2021}
{McNally}, C.~P., {Lovascio}, F., \& {Paardekooper}, S.-J. 2021, \mnras, 502,
  1469, \dodoi{10.1093/mnras/stab112}

\bibitem[{{Michel} {et~al.}(2021){Michel}, {van der Marel}, \&
  {Matthews}}]{michel2021}
{Michel}, A., {van der Marel}, N., \& {Matthews}, B. 2021, arXiv e-prints,
  arXiv:2104.05894.
\newblock \doarXiv{2104.05894}

\bibitem[{{Montesinos} {et~al.}(2016){Montesinos}, {Eiroa}, {Krivov},
  {Marshall}, {Pilbratt}, {Liseau}, {Mora}, {Maldonado}, {Wolf}, {Ertel},
  {Bayo}, {Augereau}, {Heras}, {Fridlund}, {Danchi}, {Solano}, {Kirchschlager},
  {del Burgo}, \& {Montes}}]{montesinos2016}
{Montesinos}, B., {Eiroa}, C., {Krivov}, A.~V., {et~al.} 2016, \aap, 593, A51,
  \dodoi{10.1051/0004-6361/201628329}

\bibitem[{{Morbidelli}(2020)}]{morbidelli2020}
{Morbidelli}, A. 2020, \aap, 638, A1, \dodoi{10.1051/0004-6361/202037983}

\bibitem[{{Mordasini} {et~al.}(2015){Mordasini}, {Molli{\`e}re}, {Dittkrist},
  {Jin}, \& {Alibert}}]{mordasini2015}
{Mordasini}, C., {Molli{\`e}re}, P., {Dittkrist}, K.-M., {Jin}, S., \&
  {Alibert}, Y. 2015, International Journal of Astrobiology, 14, 201,
  \dodoi{10.1017/S1473550414000263}

\bibitem[{{Najita} \& {Williams}(2005)}]{najita2005}
{Najita}, J., \& {Williams}, J.~P. 2005, \apj, 635, 625, \dodoi{10.1086/497159}

\bibitem[{{Najita} \& {Kenyon}(2014)}]{najita2014}
{Najita}, J.~R., \& {Kenyon}, S.~J. 2014, \mnras, 445, 3315,
  \dodoi{10.1093/mnras/stu1994}

\bibitem[{{Nakamura}(2000)}]{nakamura2000}
{Nakamura}, F. 2000, \apj, 543, 291, \dodoi{10.1086/317113}

\bibitem[{{Nederlander} {et~al.}(2021){Nederlander}, {Hughes}, {Fehr},
  {Flaherty}, {Su}, {Moor}, {Chiang}, {Andrews}, {Wilner}, \&
  {Marino}}]{nederlander2021}
{Nederlander}, A., {Hughes}, A.~M., {Fehr}, A.~J., {et~al.} 2021, arXiv
  e-prints, arXiv:2101.08849.
\newblock \doarXiv{2101.08849}

\bibitem[{{Nimmo} {et~al.}(2018){Nimmo}, {Kretke}, {Ida}, {Matsumura}, \&
  {Kleine}}]{nimmo2018}
{Nimmo}, F., {Kretke}, K., {Ida}, S., {Matsumura}, S., \& {Kleine}, T. 2018,
  \ssr, 214, 101, \dodoi{10.1007/s11214-018-0533-2}

\bibitem[{{O'Brien} \& {Greenberg}(2003)}]{obrien2003}
{O'Brien}, D.~P., \& {Greenberg}, R. 2003, Icarus, 164, 334

\bibitem[{{Ohtsuki}(1992)}]{oht1992}
{Ohtsuki}, K. 1992, Icarus, 98, 20, \dodoi{10.1016/0019-1035(92)90202-I}

\bibitem[{{Ohtsuki} {et~al.}(2002){Ohtsuki}, {Stewart}, \& {Ida}}]{oht2002}
{Ohtsuki}, K., {Stewart}, G.~R., \& {Ida}, S. 2002, Icarus, 155, 436,
  \dodoi{10.1006/icar.2001.6741}

\bibitem[{{Oka} {et~al.}(2011){Oka}, {Nakamoto}, \& {Ida}}]{oka2011}
{Oka}, A., {Nakamoto}, T., \& {Ida}, S. 2011, \apj, 738, 141,
  \dodoi{10.1088/0004-637X/738/2/141}

\bibitem[{{Okuzumi} {et~al.}(2012){Okuzumi}, {Tanaka}, {Kobayashi}, \&
  {Wada}}]{okuzumi2012}
{Okuzumi}, S., {Tanaka}, H., {Kobayashi}, H., \& {Wada}, K. 2012, \apj, 752,
  106, \dodoi{10.1088/0004-637X/752/2/106}

\bibitem[{{Ormel} {et~al.}(2010){Ormel}, {Dullemond}, \& {Spaans}}]{ormel2010c}
{Ormel}, C.~W., {Dullemond}, C.~P., \& {Spaans}, M. 2010, \apjl, 714, L103,
  \dodoi{10.1088/2041-8205/714/1/L103}

\bibitem[{{Pan} {et~al.}(2011){Pan}, {Padoan}, {Scalo}, {Kritsuk}, \&
  {Norman}}]{pan2011}
{Pan}, L., {Padoan}, P., {Scalo}, J., {Kritsuk}, A.~G., \& {Norman}, M.~L.
  2011, \apj, 740, 6, \dodoi{10.1088/0004-637X/740/1/6}

\bibitem[{{Pan} \& {Yu}(2020)}]{pan2020}
{Pan}, L., \& {Yu}, C. 2020, \apj, 898, 7, \dodoi{10.3847/1538-4357/ab9cab}

\bibitem[{{Pawellek} {et~al.}(2021){Pawellek}, {Wyatt}, {Matr{\`a}}, {Kennedy},
  \& {Yelverton6}}]{pawellek2021}
{Pawellek}, N., {Wyatt}, M., {Matr{\`a}}, L., {Kennedy}, G., \& {Yelverton6},
  B. 2021, \mnras, 502, 5390, \dodoi{10.1093/mnras/stab269}

\bibitem[{{Pinilla} {et~al.}(2012){Pinilla}, {Benisty}, \&
  {Birnstiel}}]{pinilla2012}
{Pinilla}, P., {Benisty}, M., \& {Birnstiel}, T. 2012, \aap, 545, A81,
  \dodoi{10.1051/0004-6361/201219315}

\bibitem[{{Pitjeva} \& {Pitjev}(2018)}]{pitjeva2018}
{Pitjeva}, E.~V., \& {Pitjev}, N.~P. 2018, Celestial Mechanics and Dynamical
  Astronomy, 130, 57, \dodoi{10.1007/s10569-018-9853-5}

\bibitem[{{Rafikov}(2001)}]{raf2001}
{Rafikov}, R.~R. 2001, \aj, 122, 2713, \dodoi{10.1086/323451}

\bibitem[{{Rafikov}(2004)}]{raf2004}
---. 2004, \aj, 128, 1348, \dodoi{10.1086/423216}

\bibitem[{{Rafikov}(2005)}]{raf2005}
---. 2005, \apjl, 621, L69, \dodoi{10.1086/428899}

\bibitem[{{Ricci} {et~al.}(2015){Ricci}, {Carpenter}, {Fu}, {Hughes}, {Corder},
  \& {Isella}}]{ricci2015}
{Ricci}, L., {Carpenter}, J.~M., {Fu}, B., {et~al.} 2015, \apj, 798, 124,
  \dodoi{10.1088/0004-637X/798/2/124}

\bibitem[{{Riols} \& {Lesur}(2018)}]{riols2018}
{Riols}, A., \& {Lesur}, G. 2018, \aap, 617, A117,
  \dodoi{10.1051/0004-6361/201833212}

\bibitem[{{Rucska} \& {Wadsley}(2021)}]{rucska2021}
{Rucska}, J.~J., \& {Wadsley}, J.~W. 2021, \mnras, 500, 520,
  \dodoi{10.1093/mnras/staa3295}

\bibitem[{{Safronov}(1969)}]{saf1969}
{Safronov}, V.~S. 1969, {Evoliutsiia doplanetnogo oblaka. (Evolution of the
  Protoplanetary Cloud and Formation of the Earth and Planets, Nauka, Moscow
  [Translation 1972, NASA TT F-677]} (1969.).
\newblock
  \url{http://adsabs.harvard.edu/cgi-bin/nph-bib_query?bibcode=1969QB981.S26......&db_key=AST}

\bibitem[{{Sallum} {et~al.}(2015){Sallum}, {Eisner}, {Close}, {Hinz}, {Skemer},
  {Bailey}, {Briguglio}, {Follette}, {Males}, {Morzinski}, {Puglisi},
  {Rodigas}, {Weinberger}, \& {Xompero}}]{sallum2015}
{Sallum}, S., {Eisner}, J.~A., {Close}, L.~M., {et~al.} 2015, \apj, 801, 85,
  \dodoi{10.1088/0004-637X/801/2/85}

\bibitem[{{Sch{\"a}fer} {et~al.}(2017){Sch{\"a}fer}, {Yang}, \&
  {Johansen}}]{schafer2017}
{Sch{\"a}fer}, U., {Yang}, C.-C., \& {Johansen}, A. 2017, \aap, 597, A69,
  \dodoi{10.1051/0004-6361/201629561}

\bibitem[{{Schlichting} {et~al.}(2013){Schlichting}, {Fuentes}, \&
  {Trilling}}]{schlicht2013}
{Schlichting}, H.~E., {Fuentes}, C.~I., \& {Trilling}, D.~E. 2013, \aj, 146,
  36, \dodoi{10.1088/0004-6256/146/2/36}

\bibitem[{{Sekiya} \& {Onishi}(2018)}]{sekiya2018}
{Sekiya}, M., \& {Onishi}, I.~K. 2018, \apj, 860, 140,
  \dodoi{10.3847/1538-4357/aac4a7}

\bibitem[{{Shadmehri} \& {Ghoreyshi}(2019)}]{shadmehri2019}
{Shadmehri}, M., \& {Ghoreyshi}, S.~M. 2019, \mnras, 488, 4623,
  \dodoi{10.1093/mnras/stz2025}

\bibitem[{{Shannon} \& {Wu}(2011)}]{shannon2011}
{Shannon}, A., \& {Wu}, Y. 2011, \apj, 739, 36,
  \dodoi{10.1088/0004-637X/739/1/36}

\bibitem[{{Shannon} {et~al.}(2015){Shannon}, {Wu}, \& {Lithwick}}]{shannon2015}
{Shannon}, A., {Wu}, Y., \& {Lithwick}, Y. 2015, \apj, 801, 15,
  \dodoi{10.1088/0004-637X/801/1/15}

\bibitem[{{Shibaike} \& {Alibert}(2020)}]{shibaike2020}
{Shibaike}, Y., \& {Alibert}, Y. 2020, \aap, 644, A81,
  \dodoi{10.1051/0004-6361/202039086}

\bibitem[{{Sibthorpe} {et~al.}(2018){Sibthorpe}, {Kennedy}, {Wyatt},
  {Lestrade}, {Greaves}, {Matthews}, \& {Duch{\^e}ne}}]{sibthorpe2018}
{Sibthorpe}, B., {Kennedy}, G.~M., {Wyatt}, M.~C., {et~al.} 2018, \mnras, 475,
  3046, \dodoi{10.1093/mnras/stx3188}

\bibitem[{{Simon} {et~al.}(2016){Simon}, {Armitage}, {Li}, \&
  {Youdin}}]{simon2016}
{Simon}, J.~B., {Armitage}, P.~J., {Li}, R., \& {Youdin}, A.~N. 2016, \apj,
  822, 55, \dodoi{10.3847/0004-637X/822/1/55}

\bibitem[{{Simon} {et~al.}(2017){Simon}, {Guilloteau}, {Di Folco}, {Dutrey},
  {Grosso}, {Pi{\'e}tu}, {Chapillon}, {Prato}, {Schaefer}, {Rice}, \&
  {Boehler}}]{simon2017}
{Simon}, M., {Guilloteau}, S., {Di Folco}, E., {et~al.} 2017, \apj, 844, 158,
  \dodoi{10.3847/1538-4357/aa78f1}

\bibitem[{{Squire} \& {Hopkins}(2018)}]{squire2018}
{Squire}, J., \& {Hopkins}, P.~F. 2018, \mnras, 477, 5011,
  \dodoi{10.1093/mnras/sty854}

\bibitem[{{Squire} \& {Hopkins}(2020)}]{squire2020}
---. 2020, \mnras, 498, 1239, \dodoi{10.1093/mnras/staa2311}

\bibitem[{{Steinpilz} {et~al.}(2019){Steinpilz}, {Joeris}, {Jungmann}, {Wolf},
  {Brendel}, {Teiser}, {Shinbrot}, \& {Wurm}}]{steinpilz2019}
{Steinpilz}, T., {Joeris}, K., {Jungmann}, F., {et~al.} 2019, Nature Physics,
  16, 225, \dodoi{10.1038/s41567-019-0728-9}

\bibitem[{{Teiser} {et~al.}(2021){Teiser}, {Kruss}, {Jungmann}, \&
  {Wurm}}]{teiser2021}
{Teiser}, J., {Kruss}, M., {Jungmann}, F., \& {Wurm}, G. 2021, \apjl, 908, L22,
  \dodoi{10.3847/2041-8213/abddc2}

\bibitem[{{Terebey} {et~al.}(1984){Terebey}, {Shu}, \& {Cassen}}]{tsc1984}
{Terebey}, S., {Shu}, F.~H., \& {Cassen}, P. 1984, \apj, 286, 529,
  \dodoi{10.1086/162628}

\bibitem[{{Tobin} {et~al.}(2018){Tobin}, {Sheehan}, {Johnstone}, \&
  {Sharma}}]{tobin2018}
{Tobin}, J., {Sheehan}, P., {Johnstone}, D., \& {Sharma}, R. 2018, arXiv
  e-prints, arXiv:1810.07174.
\newblock \doarXiv{1810.07174}

\bibitem[{{Tomida} {et~al.}(2015){Tomida}, {Okuzumi}, \&
  {Machida}}]{tomida2015}
{Tomida}, K., {Okuzumi}, S., \& {Machida}, M.~N. 2015, \apj, 801, 117,
  \dodoi{10.1088/0004-637X/801/2/117}

\bibitem[{{Tscharnuter} {et~al.}(2009){Tscharnuter}, {Sch{\"o}nke}, {Gail},
  {Trieloff}, \& {L{\"u}ttjohann}}]{tscharnuter2009}
{Tscharnuter}, W.~M., {Sch{\"o}nke}, J., {Gail}, H.~P., {Trieloff}, M., \&
  {L{\"u}ttjohann}, E. 2009, \aap, 504, 109,
  \dodoi{10.1051/0004-6361/200912120}

\bibitem[{{Ujjwal} {et~al.}(2020){Ujjwal}, {Kartha}, {Mathew}, {Manoj}, \&
  {Narang}}]{ujjwal2020}
{Ujjwal}, K., {Kartha}, S.~S., {Mathew}, B., {Manoj}, P., \& {Narang}, M. 2020,
  \aj, 159, 166, \dodoi{10.3847/1538-3881/ab76d6}

\bibitem[{{Umurhan} {et~al.}(2020){Umurhan}, {Estrada}, \&
  {Cuzzi}}]{umurhan2020}
{Umurhan}, O.~M., {Estrada}, P.~R., \& {Cuzzi}, J.~N. 2020, \apj, 895, 4,
  \dodoi{10.3847/1538-4357/ab899d}

\bibitem[{{van der Marel} {et~al.}(2018){van der Marel}, {Matthews}, {Dong},
  {Birnstiel}, \& {Isella}}]{vandermarel2018}
{van der Marel}, N., {Matthews}, B., {Dong}, R., {Birnstiel}, T., \& {Isella},
  A. 2018, in Astronomical Society of the Pacific Conference Series, Vol. 517,
  Science with a Next Generation Very Large Array, ed. E.~{Murphy}, 199

\bibitem[{{van der Marel} \& {Mulders}(2021)}]{vandermarel2021}
{van der Marel}, N., \& {Mulders}, G.~D. 2021, \aj, 162, 28,
  \dodoi{10.3847/1538-3881/ac0255}

\bibitem[{{Vican}(2012)}]{vican2012}
{Vican}, L. 2012, \aj, 143, 135, \dodoi{10.1088/0004-6256/143/6/135}

\bibitem[{{Vitense} {et~al.}(2012){Vitense}, {Krivov}, {Kobayashi}, \&
  {L{\"o}hne}}]{vitense2012}
{Vitense}, C., {Krivov}, A.~V., {Kobayashi}, H., \& {L{\"o}hne}, T. 2012, \aap,
  540, A30, \dodoi{10.1051/0004-6361/201118551}

\bibitem[{{Voelkel} {et~al.}(2020){Voelkel}, {Klahr}, {Mordasini},
  {Emsenhuber}, \& {Lenz}}]{voelkel2020}
{Voelkel}, O., {Klahr}, H., {Mordasini}, C., {Emsenhuber}, A., \& {Lenz}, C.
  2020, \aap, 642, A75, \dodoi{10.1051/0004-6361/202038085}

\bibitem[{{Wahhaj} {et~al.}(2010){Wahhaj}, {Cieza}, {Koerner}, {Stapelfeldt},
  {Padgett}, {Case}, {Keller}, {Mer{\'\i}n}, {Evans}, {Harvey}, {Sargent}, {van
  Dishoeck}, {Allen}, {Blake}, {Brooke}, {Chapman}, {Mundy}, \&
  {Myers}}]{wahhaj2010}
{Wahhaj}, Z., {Cieza}, L., {Koerner}, D.~W., {et~al.} 2010, \apj, 724, 835,
  \dodoi{10.1088/0004-637X/724/2/835}

\bibitem[{{Weidenschilling}(1977)}]{weiden1977a}
{Weidenschilling}, S.~J. 1977, \mnras, 180, 57

\bibitem[{{Weidenschilling}(1989)}]{weiden1989}
---. 1989, Icarus, 80, 179, \dodoi{10.1016/0019-1035(89)90166-8}

\bibitem[{{Weidenschilling}(2010)}]{weid2010}
---. 2010, \apj, 722, 1716, \dodoi{10.1088/0004-637X/722/2/1716}

\bibitem[{{Weidenschilling} {et~al.}(1997){Weidenschilling}, {Spaute}, {Davis},
  {Marzari}, \& {Ohtsuki}}]{weiden1997b}
{Weidenschilling}, S.~J., {Spaute}, D., {Davis}, D.~R., {Marzari}, F., \&
  {Ohtsuki}, K. 1997, Icarus, 128, 429, \dodoi{10.1006/icar.1997.5747}

\bibitem[{{Weidling} {et~al.}(2009){Weidling}, {G{\"u}ttler}, {Blum}, \&
  {Brauer}}]{weidling2009}
{Weidling}, R., {G{\"u}ttler}, C., {Blum}, J., \& {Brauer}, F. 2009, \apj, 696,
  2036, \dodoi{10.1088/0004-637X/696/2/2036}

\bibitem[{{Wetherill} \& {Stewart}(1993)}]{weth1993}
{Wetherill}, G.~W., \& {Stewart}, G.~R. 1993, Icarus, 106, 190,
  \dodoi{10.1006/icar.1993.1166}

\bibitem[{{Williams} \& {Cieza}(2011)}]{will2011}
{Williams}, J.~P., \& {Cieza}, L.~A. 2011, \araa, 49, 67,
  \dodoi{10.1146/annurev-astro-081710-102548}

\bibitem[{{Windmark} {et~al.}(2012){Windmark}, {Birnstiel}, {Ormel}, \&
  {Dullemond}}]{windmark2012}
{Windmark}, F., {Birnstiel}, T., {Ormel}, C.~W., \& {Dullemond}, C.~P. 2012,
  \aap, 544, L16, \dodoi{10.1051/0004-6361/201220004}

\bibitem[{{Wurm} \& {Blum}(1998)}]{wurm1998}
{Wurm}, G., \& {Blum}, J. 1998, \icarus, 132, 125,
  \dodoi{10.1006/icar.1998.5891}

\bibitem[{{Wyatt}(2008)}]{wyatt2008}
{Wyatt}, M.~C. 2008, \araa, 46, 339,
  \dodoi{10.1146/annurev.astro.45.051806.110525}

\bibitem[{{Wyatt} {et~al.}(2011){Wyatt}, {Clarke}, \& {Booth}}]{wyatt2011}
{Wyatt}, M.~C., {Clarke}, C.~J., \& {Booth}, M. 2011, Celestial Mechanics and
  Dynamical Astronomy, 111, 1, \dodoi{10.1007/s10569-011-9345-3}

\bibitem[{{Wyatt} \& {Dent}(2002)}]{wyatt2002}
{Wyatt}, M.~C., \& {Dent}, W.~R.~F. 2002, \mnras, 334, 589,
  \dodoi{10.1046/j.1365-8711.2002.05533.x}

\bibitem[{{Xiao} {et~al.}(2017){Xiao}, {Niu}, \& {Zhang}}]{xiao2017}
{Xiao}, L., {Niu}, R., \& {Zhang}, H. 2017, \mnras, 467, 2869,
  \dodoi{10.1093/mnras/stx278}

\bibitem[{{Yang} {et~al.}(2017){Yang}, {Johansen}, \& {Carrera}}]{yang2017}
{Yang}, C.~C., {Johansen}, A., \& {Carrera}, D. 2017, \aap, 606, A80,
  \dodoi{10.1051/0004-6361/201630106}

\bibitem[{{Yorke} \& {Bodenheimer}(1999)}]{yorke1999}
{Yorke}, H.~W., \& {Bodenheimer}, P. 1999, \apj, 525, 330,
  \dodoi{10.1086/307867}

\bibitem[{{Youdin}(2010)}]{youdin2010}
{Youdin}, A.~N. 2010, in EAS Publications Series, Vol.~41, EAS Publications
  Series, ed. {T.~Montmerle, D.~Ehrenreich, \& A.-M.~Lagrange}, 187--207,
  \dodoi{10.1051/eas/1041016}

\bibitem[{{Youdin} \& {Goodman}(2005)}]{youdin2005}
{Youdin}, A.~N., \& {Goodman}, J. 2005, \apj, 620, 459, \dodoi{10.1086/426895}

\bibitem[{{Youdin} \& {Kenyon}(2013)}]{youdin2013}
{Youdin}, A.~N., \& {Kenyon}, S.~J. 2013, {From Disks to Planets}, ed. T.~D.
  {Oswalt}, L.~M. {French}, \& P.~{Kalas} (Dordrecht: Springer Science \&
  Business Media), 1, \dodoi{10.1007/978-94-007-5606-9_1}

\bibitem[{{Zhang} \& {Jin}(2015)}]{yzhang2015}
{Zhang}, Y., \& {Jin}, L. 2015, \apj, 802, 58,
  \dodoi{10.1088/0004-637X/802/1/58}

\bibitem[{{Zhao} {et~al.}(2020){Zhao}, {Tomida}, {Hennebelle}, {Tobin},
  {Maury}, {Hirota}, {S{\'a}nchez-Monge}, {Kuiper}, {Rosen}, {Bhandare},
  {Padovani}, \& {Lee}}]{zhao2020}
{Zhao}, B., {Tomida}, K., {Hennebelle}, P., {et~al.} 2020, \ssr, 216, 43,
  \dodoi{10.1007/s11214-020-00664-z}

\bibitem[{{Zhu} \& {Stone}(2014)}]{zhu2014}
{Zhu}, Z., \& {Stone}, J.~M. 2014, \apj, 795, 53,
  \dodoi{10.1088/0004-637X/795/1/53}

\bibitem[{{Zsom} {et~al.}(2010){Zsom}, {Ormel}, {G{\"u}ttler}, {Blum}, \&
  {Dullemond}}]{zsom2010}
{Zsom}, A., {Ormel}, C.~W., {G{\"u}ttler}, C., {Blum}, J., \& {Dullemond},
  C.~P. 2010, \aap, 513, A57, \dodoi{10.1051/0004-6361/200912976}

\bibitem[{{Zurlo} {et~al.}(2021){Zurlo}, {Garufi}, {P{\'e}rez}, {Alves},
  {Girart}, {Zhu}, {Franco}, \& {Cleeves}}]{zurlo2021}
{Zurlo}, A., {Garufi}, A., {P{\'e}rez}, S., {et~al.} 2021, \apj, 912, 64,
  \dodoi{10.3847/1538-4357/abec42}

\end{thebibliography}
\bibliographystyle{aasjournal}

\end{document}